%% file: texte.tex
\documentclass[graybox]{svmult}

\usepackage{mathptmx}       
\usepackage{helvet}         
\usepackage{courier}        
\usepackage{type1cm}        
\usepackage{makeidx}         
\usepackage{graphicx}        
\usepackage{multicol}        
\usepackage[bottom]{footmisc}
\usepackage{verbatim}

\makeindex             

\usepackage{amssymb}
\usepackage{amsmath} 

\newcommand{\bea}{\begin{eqnarray}}
\newcommand{\eea}{\end{eqnarray}}
\newcommand{\be}{\begin{equation}}
\newcommand{\ee}{\end{equation}}
\newcommand{\rr}{\mathbf{r}}
\newcommand{\YY}{\mathbf{Y}}
\newcommand{\yy}{\mathbf{y}}
\newcommand{\kk}{\mathbf{k}}
\newcommand{\pp}{\mathbf{p}}
\newcommand{\KK}{\mathbf{K}}
\newcommand{\qq}{\mathbf{q}}
\newcommand{\uu}{\mathbf{u}}
\newcommand{\RR}{\mathbf{R}}
\newcommand{\CC}{\mathbf{C}}
\newcommand{\PP}{\mathbf{P}}
\newcommand{\XX}{\mathbf{X}}
\newcommand{\nn}{\mathbf{n}}
\newcommand{\vn}{\mathbf{0}}
\newcommand{\vv}{\mathbf{v}}
\newcommand{\Omegab}{\mbox{\boldmath$\Omega$}}
\newcommand{\epsk}{\epsilon_{\mathbf{k}}}
\newcommand{\up}{\uparrow}
\newcommand{\down}{\downarrow}
\newcommand{\gr}{\mathbf{\nabla}}

\begin{document}

\title*{The Unitary Gas and its Symmetry Properties}
\author{Yvan Castin and F\'elix Werner}
\institute{Yvan Castin \at Laboratoire Kastler Brossel, 
Ecole normale sup\'erieure, CNRS and UPMC, Paris (France),
\email{yvan.castin@lkb.ens.fr}
\and
F\'elix Werner \at Department of Physics, 
University of Massachusetts, Amherst (USA), \email{werner@lkb.ens.fr}}
%
%
\maketitle

\abstract*{
\input abstract.txt
}

\abstract{
\input abstract.txt
}

The chapter is organized as follows.
In section \ref{sec:sf}, we introduce useful concepts, and we 
present a simple definition and basic
properties of the unitary gas, related to its scale invariance.
In section \ref{sec:vm}, we describe various models that may be used
to describe the BEC-BCS crossover, and in particular the unitary gas,
each model having its own advantage and shedding some particular
light on the unitary gas properties:
scale invariance and a virial theorem hold within the zero-range model,
relations between the derivative of the energy with respect to the inverse scattering length
and the
short range pair correlations or the tail of the  momentum distribution are easily derived
 using the lattice model, and
 the same derivative is immediately related to the
number of molecules in the closed channel (recently measured at Rice)
using the two-channel model.
In section \ref{sec:ds}, we describe the dynamical symmetry properties
of the unitary gas in a harmonic trap, and we extract their physical 
consequences for many-body and few-body problems.

\bigskip
\setcounter{minitocdepth}{5}
\dominitoc
\newpage

\section{Simple facts about the unitary gas}
\label{sec:sf}

\subsection{What is the unitary gas ?}

First, the unitary gas is $\ldots$ a gas. As opposed to a liquid, 
it is a dilute system with respect to the interaction range $b$: its
mean number density $\rho$ satisfies the constraint
\be
\label{eq:gaz}
\rho b^3 \ll 1.
\ee
For a rapidly decreasing interaction potential $V(r)$, $b$ is the spatial
width of $V(r)$. In atomic physics, where $V(r)$ may be viewed
as a strongly repulsive core and a Van der Waals attractive tail $-C_6/r^6$,
one usually assimilates $b$ to the Van der Waals length
$(m C_6/\hbar^2)^{1/4}$.

The intuitive picture of a gas is that the particles mainly experience
binary scattering, the probability that more than two particles 
are within a volume $b^3$ being negligible. As a consequence,
what should really matter is the knowledge of the scattering amplitude
$f_k$ of two particles, where $k$ is the relative momentum,
rather than the $r$ dependence of the interaction potential $V(r)$. 
This expectation has 
guided essentially all many-body works on the BEC-BCS crossover: 
One uses convenient models for $V(r)$ that are very different
from the true atomic interaction potential, but that reproduce
correctly the momentum dependence of $f_k$ at the relevant low values
of $k$, such as the Fermi momentum or the inverse thermal de Broglie wavelength,
these relevant low values of $k$ having to satisfy $k b \ll 1$ for this
modelization to be acceptable.

Second, the unitary gas is such that, for the relevant values of 
the relative momentum $k$, the modulus of $f_k$ reaches the maximal
value allowed by quantum mechanics, the so-called unitary limit
\cite{livre_collisions}. Here, we consider $s$-wave scattering 
between two opposite-spin fermions, so that $f_k$ depends
only on the modulus of the relative momentum. The optical
theorem, a consequence of the unitarity of the quantum evolution operator
\cite{livre_collisions}, then implies
\be
\mathrm{Im}\, f_k = k |f_k|^2.
\ee
Dividing by $|f_k|^2$, and using $f_k/|f_k|^2=1/f_k^*$,
one sees that this fixes the value of the imaginary part of
$1/f_k$, so that it is strictly equivalent to the requirement that
there exists a real function $u(k)$ such that
\be
\label{eq:u(k)}
f_k = - \frac{1}{ik+u(k)}
\ee
for all values of $k$. We then obtain the upper bound $|f_k| \leq 1/k$.
Ideally, the unitary gas saturates this inequality for all values
of $k$:
\be
\label{eq:fku}
f_k^{\rm unitary} = -\frac{1}{ik}.
\ee

\begin{figure}[b]
\centerline{\includegraphics[width=9cm,clip=]{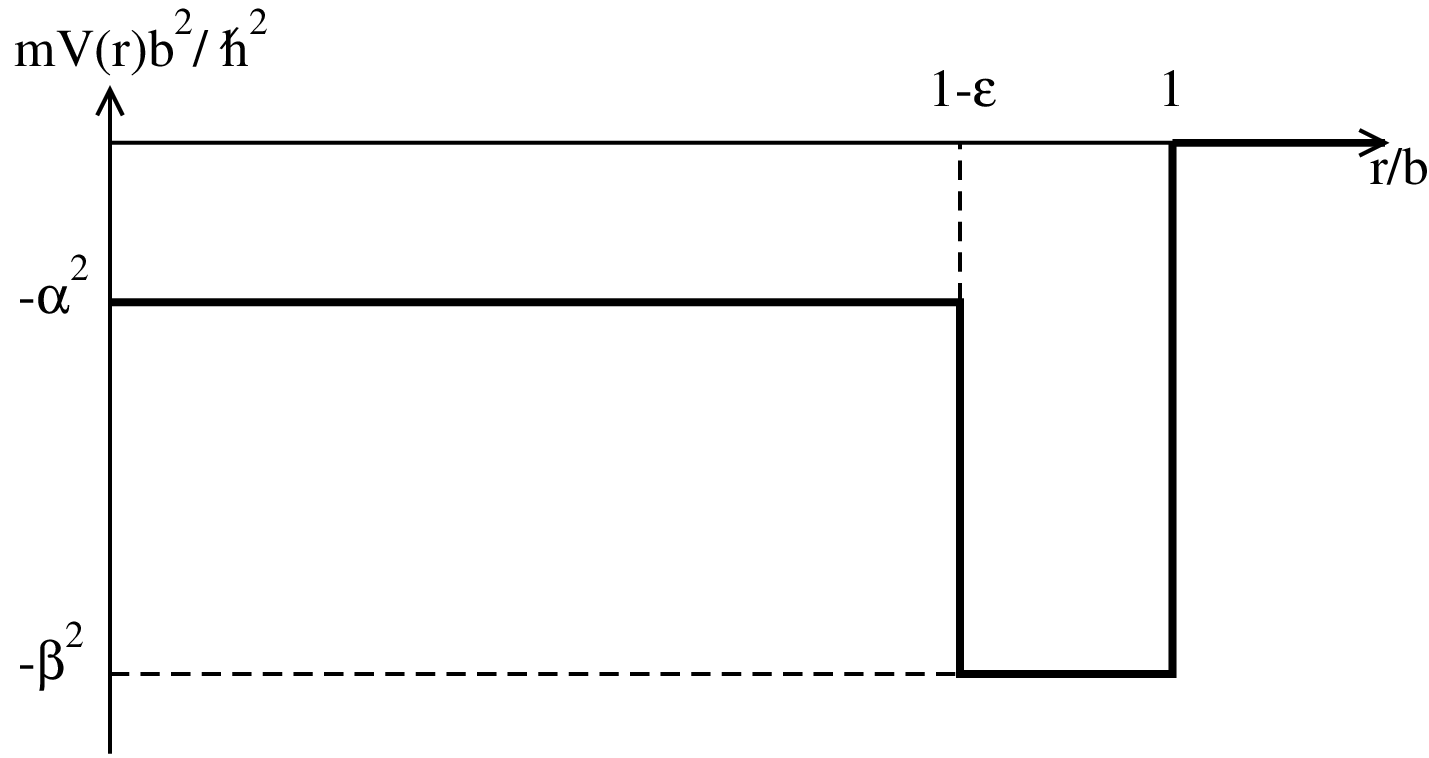}}
\caption{A class of non-positive
potentials (of compact support of radius $b$)
that may lead to a negative effective range in the resonant
case $a^{-1}=0$. The resonant case is achieved when the three parameters
$\alpha,\beta$ and $\epsilon$ satisfy
$\tan[(1-\epsilon)\alpha] \tan(\epsilon \beta) = \alpha/\beta.$
Then from Smorodinskii's formula, see Problem 1 in \S 131 of \cite{Landau},
one sees that $r_e/b \leq 2$. One also finds that
$r_e/b\sim -\cos^2\theta/(\pi\epsilon)^2
\to -\infty$ when $\epsilon\to 0$ with
$\alpha=\pi$, $\beta\epsilon \to \theta$,  where $\theta=2.798386\ldots$
solves $1+\theta\tan\theta=0$.
}
\label{fig:contre}
\end{figure}

In reality, Eq.(\ref{eq:fku}) cannot hold for all $k$. 
It is thus important to understand over which range of $k$ 
Eq.(\ref{eq:fku}) should hold to have a unitary gas, and to estimate
the deviations from Eq.(\ref{eq:fku}) in that range in a real 
experiment.
To this end, we use the usual low-$k$ expansion of the denominator
of the scattering amplitude \cite{livre_collisions}, under validity
conditions specified in \cite{math_re}:
\be
\label{eq:lke}
u(k) = \frac{1}{a} - \frac{1}{2} r_e k^2 + \ldots
\ee
The length $a$ is the scattering length, the length $r_e$ is the effective
range of the interaction. Both $a$ and $r_e$ can be of arbitrary sign.
Even for $1/a=0$, even for an everywhere non-positive interaction potential,
$r_e$ can be of arbitrary sign. As this last property seems to contradict 
a statement in the solution of problem 1 
in \S 131 of \cite{Landau},
we have constructed an explicit example depicted in Fig.~\ref{fig:contre},
which even shows that the effective range may be very
different in absolute value from the true potential range $b$, i.e.\
$r_e/b$ for $a^{-1}=0$
may be in principle an arbitrarily large and negative number.
Let us assume that the $\ldots$ in Eq.(\ref{eq:lke}) are negligible if 
$k b \ll 1$, an assumption that will be revisited in \S\ref{subsubsec:nccm}.
Noting $k_{\rm typ}$ a typical relative momentum in the gas, we thus
see that the unitary gas is in practice obtained as a double limit,
a {\sl zero range} limit 
\be
\label{eq:zrl}
k_{\rm typ} b \ll 1, k_{\rm typ} |r_e| \ll 1
\ee
and an {\sl infinite scattering length} limit:
\be
\label{eq:isll}
k_{\rm typ} |a| \gg 1.
\ee

At zero temperature, we assume that $k_{\rm typ}= k_F$, where
the Fermi momentum is conventionally defined in terms of the gas
total density $\rho$ as for the ideal spin-1/2 Fermi gas:
\be
\label{eq:kf}
k_F = (3\pi^2\rho)^{1/3}.
\ee
In a trap, $\rho$ and thus $k_F$ are position dependent.
Condition~(\ref{eq:isll}) is well satisfied experimentally, thanks
to the Feshbach resonance. The condition $k_F b \ll 1$
is also well satisfied at the per cent level, because
$b \approx$ the Van der Waals length is in the nanometer range.
Up to now, there is no experimental tuning of the effective range $r_e$, and there
are cases where $k_F |r_e|$ is not small. However, to study
the BEC-BCS crossover, one uses in practice
the so-called broad Feshbach resonances, which do not require
a too stringent control of the spatial homogeneity of the 
magnetic field, and where $|r_e|\sim b$; then Eq.(\ref{eq:zrl}) is also
satisfied.

We note that the assumption $k_{\rm typ}=k_F$, although quite intuitive,
is not automatically correct. For example, for bosons, as shown
by Efimov \cite{Efimov70}, an effective three-body attraction takes place,
leading to the occurrence of the Efimov trimers; this attraction leads
to the so-called problem of {\sl fall to the center} \cite{Landau},
and one has $1/k_{\rm typ}$ of the order of the largest of the 
two ranges $b$ and $|r_e|$. Eq.(\ref{eq:zrl}) is then violated,
and an extra length scale, the three-body parameter, has to be introduced,
breaking the scale invariance of the unitary gas.
Fortunately, for three fermions, 
there is no Efimov attraction, except for the case
 of different masses for the two spin components:
If two fermions of mass $m_\uparrow$ interact with a lighter particle
of mass $m_\downarrow$, the Efimov effect takes place
for $m_\uparrow / m _\downarrow$ larger than $\simeq 13.607$
\cite{Efimov73,Petrov}.
If a third fermion of mass $m_\uparrow$ is added, a four-body Efimov effect appears at a 
slightly lower mass ratio $m_\uparrow / m _\downarrow\simeq13.384$~\cite{CMP}.
In what follows we consider the case of equal masses, unless specified otherwise.

At non-zero temperature $T>0$, 
another length scale appears in the unitary gas properties,
the thermal de Broglie wavelength $\lambda_{\rm dB}$, defined as
\be
\label{eq:ldb}
\lambda_{\rm dB}^2 = \frac{2\pi \hbar^2}{m k_B T}.
\ee
At temperatures larger than the Fermi temperature $T_F
= \hbar^2 k_F^2/(2mk_B)$, one has to take $k_{\rm typ}\sim 1/\lambda_{\rm dB}$
in the conditions~(\ref{eq:zrl},\ref{eq:isll}).
In practice, the most interesting regime is however the degenerate
regime $T<T_F$, where the non-zero temperature does not bring
new conditions for unitarity.

\subsection{Some simple properties of the unitary gas}
\label{subsec:ssfp}

As is apparent in the expression of the two-body
scattering amplitude Eq.(\ref{eq:fku}), there is no parameter or
length scales issuing from the interaction.
As a consequence, for a gas in the trapping potential $U(\rr)$, the
eigenenergies $E_i$ of the $N$-body problem only depend on $\hbar^2/m$
and on the spatial dependence of $U(\rr)$:
the length scale required to get an energy out
of $\hbar^2/m$ is obtained from the shape of the container.

This is best formalized in terms of a spatial scale invariance.
Qualitatively, if one changes the volume of the container, even if the gas
becomes arbitrarily dilute, it remains
at unitarity and strongly interacting. This is of course
not true for a finite value of the scattering length $a$: 
If one reduces the gas density, $\rho^{1/3} a$ drops eventually to small
values, and the gas becomes weakly interacting.

Quantitatively, if one applies to the container 
a similarity factor $\lambda$ in all directions, which
changes its volume from $V$ to $\lambda^3 V$, we expect 
that each eigenenergy scales as
\be
\label{eq:sca_ener}
E_i \rightarrow \frac{E_i}{\lambda^2}
\ee
and each eigenwavefunction scales as
\be
\label{eq:sca_psi}
\psi_i(\XX) \rightarrow \frac{\psi_i(\XX/\lambda)}{\lambda^{3N/2}}.
\ee
Here $\XX=(\rr_1,\ldots,\rr_N)$ is the set of all coordinates of the particles,
and the $\lambda$-dependent factor ensures that the wavefunction
remains normalized. The properties~(\ref{eq:sca_ener},\ref{eq:sca_psi}),
which are at the heart of what the unitary gas really is,
will be put on mathematical grounds  
in section \ref{sec:vm} by replacing the interaction 
with contact conditions on $\psi$.
Simple consequences may be obtained from these scaling properties, as we now discuss.

In a harmonic isotropic trap, where a single particle has an
oscillation angular frequency $\omega$, taking as the scaling factor
the harmonic oscillator length $a_{\rm ho}=[\hbar/(m\omega)]^{1/2}$,
one finds that
\be
\label{eq:simple}
\frac{E_i}{\hbar\omega} = \mathcal{F}_i(N)
\ee
where the functions $\mathcal{F}_i$ are universal functions, ideally independent
of the fact that one uses lithium 6 or potassium 40 atoms, and depending
only on the particle number.

In free space, the unitary gas cannot have a $N$-body bound state (an eigenstate
of negative energy), whatever the value of $N\geq 2$.
If there was such a bound state, which corresponds to a square integrable
eigenwavefunction of the relative (Jacobi) coordinates of the particles,
one could generate a continuum of such square integrable eigenstates
using Eqs.(\ref{eq:sca_ener},\ref{eq:sca_psi}). This would violate
a fundamental property of self-adjoint Hamiltonians \cite{analyse_spectrale}.
Another argument is that the energy of a discrete universal bound state would depend only on $\hbar$ and $m$, 
which is impossible by dimensional analysis.

At thermal equilibrium in the canonical ensemble in a box, 
say a cubic box of volume $V=L^3$ with periodic boundary conditions, 
several relations may be obtained if one takes the thermodynamic 
limit $N\to +\infty$, $L^3\to +\infty$ with a fixed density $\rho$
and temperature $T$, and if one assumes that the free energy $F$
is an extensive quantity. Let us consider for simplicity
the case of equal population of the two spin states,
$N_\uparrow=N_\downarrow$. Then, in the thermodynamic
limit, the free energy per
particle $F/N\equiv f$ is a function of the density $\rho$ and
temperature $T$. If one applies a similarity of factor $\lambda$
and if one change $T$ to $T/\lambda^2$ so as to keep a constant
ratio $E_i/(k_B T)$, that is a constant occupation probability
for each eigenstate, one obtains from Eq.(\ref{eq:sca_ener}) that
\be
\label{eq:sca_f}
f(\rho/\lambda^3, T/\lambda^2) = f(\rho, T)/\lambda^2.
\ee
At zero temperature, $f$ reduces to the ground state energy per
particle $e_0(\rho)$. From Eq.(\ref{eq:sca_f}) it appears that 
$e_0(\rho)$ scales as $\rho^{2/3}$, exactly as the ground state energy
of the ideal Fermi gas.  One thus simply has
\be
\label{eq:e0}
e_0(\rho) = \xi \, e_0^{\rm ideal}(\rho) = \frac{3\xi}{5} 
\frac{\hbar^2 k_F^2}{2m}
\ee
where $k_F$ is defined by Eq.(\ref{eq:kf}) and
$\xi$ is a universal number. This is also a simple consequence
of dimensional analysis \cite{Ho}. Taking the derivative with respect
to $N$ or to the volume, 
this shows that the same type of relation holds for the zero temperature
chemical potential, $\mu_0(\rho) = \xi \mu_0^{\rm ideal}(\rho)$,
and for the zero temperature total pressure, 
$P_0(\rho) = \xi P_0^{\rm ideal}(\rho)$,
so that
\bea
\label{eq:mu0}
\mu_0(\rho) &=& \xi \frac{\hbar^2 k_F^2}{2m}\\
P_0(\rho) &=& \frac{2\xi}{5}  \rho \frac{\hbar^2 k_F^2}{2m}.
\label{eq:P0}
\eea

At non-zero temperature, taking the derivative of Eq.(\ref{eq:sca_f})
with respect to $\lambda$ in $\lambda=1$, and using $F=E-TS$, where
$E$ is the mean energy and $S=-\partial_T F$ is the entropy, 
as well as $\mu=\partial_N F$, one obtains
\be
\frac{5}{3} E - \mu N = T S.
\ee
From the Gibbs-Duhem relation, the grand potential
$\Omega=F-\mu N$ is equal to $-P V$, where $P$ is the pressure
of the gas. This gives finally the useful relation
\be
\label{eq:pv}
PV = \frac{2}{3} E,
\ee
that can also be obtained from dimensional analysis \cite{Ho}, and that
of course also holds at zero temperature (see above).
All these properties actually also apply to the ideal Fermi gas, 
which is obviously scaling invariant. The relation (\ref{eq:pv})
for example was established for the ideal gas in \cite{ideal}.

Let us finally describe at a macroscopic level, i.e.\ in a hydrodynamic picture,
the effect of the similarity Eq.(\ref{eq:sca_psi}) on the quantum state of a unitary gas,
assuming that it was initially at thermal equilibrium in a trap.
In the initial state of the gas, consider a small (but still macroscopic) element, enclosed in a volume $dV$ around point $\rr$.
It is convenient to assume that $dV$ is a fictitious cavity with periodic boundary conditions.
In the hydrodynamic picture, this small element is assumed to be at local thermal equilibrium
with a temperature $T$. Then one performs the spatial scaling transform Eq.(\ref{eq:sca_ener}) 
on each many-body eigenstate $\psi$ of the thermal statistical mixture,
which does not change the statistical weigths.
How will the relevant physical quantities be transformed in the hydrodynamic approach~?

The previously considered small element is now at position $\lambda \rr$, and occupies a volume $\lambda^3 dV$,
with the same number of particles. The hydrodynamic mean density profile after rescaling,
$\rho_\lambda$, is thus related to the mean density profile $\rho$ before scaling as
\be
\label{eq:resca_macro_dens}
\rho_\lambda(\lambda \rr) = \rho(\rr)/\lambda^3.
\ee
Second, is the small element still at (local) thermal equilibrium after scaling~?
Each eigenstate of energy $E_{\rm loc}$ of the locally homogeneous unitary gas within the initial cavity of
volume $dV$ is transformed by the scaling
into an eigenstate within the cavity of volume $\lambda^3 dV$, with the eigenenergy $E_{\rm loc}/\lambda^2$.
Since the occupation probabilities of each local eigenstate
are not changed, the local statistical mixture remains thermal provided that one rescales
the temperature as
\be
\label{eq:resca_macro_temp}
T_\lambda =  T/\lambda^2.
\ee
A direct consequence is that the entropy of the small element of the gas is unchanged by the scaling,
so that the local entropy {\sl per particle} $s$ in the hydrodynamic approach obeys
\be
\label{eq:resca_macro_entropie_par_particule}
s_\lambda(\lambda \rr) = s(\rr).
\ee
Also, since the mean energy of the small element is reduced by the factor $\lambda^2$ due to the scaling,
and the volume of the small element is multiplied by $\lambda^3$, the equilibrium relation Eq.(\ref{eq:pv}) imposes
that the local pressure is transformed by the scaling as
\be
\label{eq:resca_macro_pression}
p_\lambda(\lambda\rr) = p(\rr)/\lambda^5.
\ee

\subsection{Application: Inequalities on $\xi$ and finite-temperature 
quantities}

Using the previous constraints imposed by scale invariance of the unitary
gas on thermodynamic quantities, in addition to standard
thermodynamic inequalities, we show that one can produce constraints
involving both the zero-temperature quantity $\xi$ and finite-temperature
quantities of the gas.

Imagine that, at some temperature $T$, the energy $E$ and the chemical
potential $\mu$ of the non-polarized unitary Fermi gas have been
obtained, in the thermodynamic limit. 
If one introduces the
 Fermi momentum
Eq.(\ref{eq:kf}) and the corresponding Fermi energy
$E_F=\hbar^2 k_F^2/(2m)$, this means that on has at hand the two
dimensionless quantities
\bea
A &\equiv& \frac{E}{N E_F} \\
B &\equiv& \frac{\mu}{E_F}.
\eea
As a consequence of Eq.(\ref{eq:pv}), one also has access to the 
 pressure $P$.
We now show that the following inequalities hold at any temperature $T$:
\be
\label{eq:xib}
\left(\frac{3}{5A}\right)^{2/3}  B^{5/3} \leq
\xi \leq \frac{5A}{3}.
\ee

In the canonical ensemble, the 
 mean energy
$E(N,T,V)$ is an increasing function of temperature for fixed
volume $V$ and atom number $N$.
Indeed one has the well-known relation
$k_B T^2 \partial_T E(N,T,V) = \mathrm{Var}\, H$,
and the variance of the Hamiltonian is non-negative.
As a consequence, for any temperature $T$:
\be
E(N,T,V) \ge E(N,0,V).
\ee
From Eq.(\ref{eq:e0}) we then reach the upper bound on $\xi$ given
in Eq.(\ref{eq:xib}).

In the grand canonical ensemble, the 
 pressure $P(\mu,T)$ is
an increasing function of temperature for a fixed chemical potential.
This results from the Gibbs-Duhem relation
$\Omega(\mu,T,V)=-V P(\mu,T)$ where $\Omega$ is the grand potential
and $V$ the volume, and from the differential relation 
$\partial_T \Omega(\mu,T)=-S$ where $S\geq 0$ is the
 entropy.
As a consequence, for any temperature $T$:
\be
P(\mu,T) \geq P(\mu,0).
\ee
For the unitary gas,
the left hand side can be expressed in terms of $A$ using (\ref{eq:pv}).
Eliminating the density between Eq.(\ref{eq:mu0}) and Eq.(\ref{eq:P0})
we obtain the zero temperature pressure
\be
P(\mu,0) = \frac{1}{15\pi^2\xi^{3/2}} \frac{\hbar^2}{m} 
\left(\frac{2m\mu}{\hbar^2}\right)^{5/2}.
\ee
This leads to the lower bound on $\xi$ given in Eq.(\ref{eq:xib}).

Let us apply Eq.(\ref{eq:xib}) to the Quantum Monte Carlo results
of \cite{Burovski}: At the critical temperature $T=T_c$,
$A=0.310(10)$ and $B=0.493(14)$, so that
\be
0.48(3) \leq \xi_{\cite{Burovski}}  \leq 0.52(2).
\ee
This deviates by two standard deviations from  the fixed node
result $\xi \leq 0.40(1)$ \cite{Carlson}.
The Quantum Monte Carlo results of \cite{Bulgac},
if one takes a temperature equal to the critical temperature of \cite{Burovski},
  give
$A=0.45(1)$ and $B=0.43(1)$; these values, in clear disagreement 
with \cite{Burovski}, lead to the non-restrictive bracketing
$0.30(2)\leq \xi_{\cite{Bulgac}}\leq 0.75(2)$.
The more recent work~\cite{Goulko} finds
$k_BT_c/E_F=0.171(5)$
and at this critical temperature,
$A=0.276(14)$ and $B=0.429(9)$,
leading to
\be
0.41(3) \leq \xi_{\cite{Goulko}}  \leq 0.46(2).
\ee

\begin{figure}[b]
\centerline{\includegraphics[width=9cm,clip=]{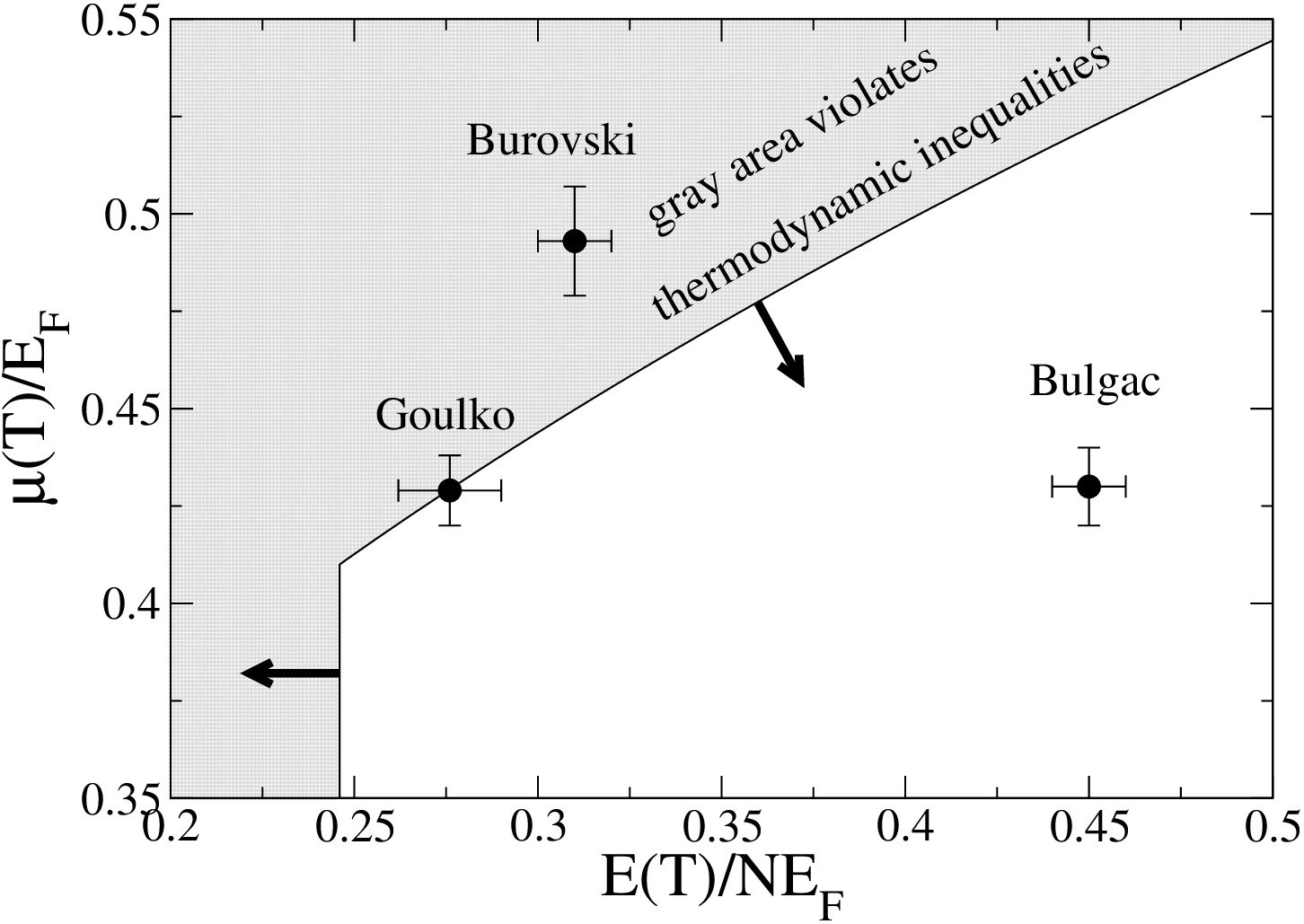}}
\caption{For the spin balanced uniform unitary gas at thermal equilibrium: Assuming $\xi=0.41$ in Eq.(\ref{eq:xib})
defines a zone (shaded in gray) in the plane energy-chemical potential that is forbidden
at all temperatures.
The black disks correspond to the unbiased Quantum Monte Carlo results
of Burovski et al. \cite{Burovski}, of Bulgac et al. \cite{Bulgac},
and of Goulko et al. \cite{Goulko} at the critical temperature. Taking the unknown exact value of $\xi$, which is below the
fixed node upper bound 0.41 \cite{Carlson}, will shift the forbidden zone boundaries
as indicated by the arrows.
}
\label{fig:exclue}
\end{figure}

Another, more graphical application of our simple bounds
is to assume some reasonable value of $\xi$, and then to use
Eq.(\ref{eq:xib}) to construct a zone in the energy-chemical potential
plane that is forbidden at all temperatures. In Fig.\ref{fig:exclue},  we took $\xi=0.41$, inspired
by the fixed node upper bound on the exact value of $\xi$ \cite{Carlson}:
The shaded area is the resulting forbidden zone, and the filled disks with error bars represent the in principle
exact Quantum Monte Carlo results of various groups at $T=T_c$. The prediction
of \cite{Burovski} lies within the forbidden zone.
The prediction of \cite{Bulgac} is well within the allowed zone, whereas the most recent
prediction of \cite{Goulko} is close to the boundary between the forbidden and the allowed zones.
If one takes a smaller value for $\xi$,
the boundaries of the forbidden zone will shift as indicated by the arrows on the figure.
All this shows that simple reasonings may be useful to test and guide
numerical studies of the unitary gas.

\subsection{Is the unitary gas attractive or repulsive ?}
\label{subsec:aor}

According to a common saying, a weakly interacting Fermi gas
($k_F |a|\ll 1$)
experiences an effective repulsion for a positive scattering length $a>0$,
and an effective attraction for a negative scattering length $a<0$.
Another common fact is that, in the unitary limit
$|a|\to +\infty$, the gas properties do not depend on the sign
of $a$. As the unitary limit may be apparently equivalently obtained
by taking the limit $a\to +\infty$ or the limit $a\to -\infty$, 
one reaches a paradox, considering the fact that the unitary
gas does not have the same ground state energy than the ideal gas
and cannot be at the same time an attractive and repulsive state of matter.

This paradox may be resolved by considering the case of two
particles in an isotropic harmonic trap.
After elimination of the center of mass motion, and restriction
to a zero relative angular momentum to have $s$-wave interaction,
one obtains the radial Schr\"odinger equation
\be
\label{eq:schr3d}
-\frac{\hbar^2}{2\mu} \left[\psi''(r) + \frac{2}{r} \psi'(r)\right]
+\frac{1}{2} \mu \omega^2 r^2 \psi(r) = E_{\rm rel} \psi(r),
\ee
with the relative mass $\mu=m/2$. The interactions
are included in the zero range limit by the $r=0$ boundary conditions,
the so-called Wigner-Bethe-Peierls contact conditions described in section
\ref{sec:vm}:
\be
\label{eq:bp}
\psi(r) = A [r^{-1}-a^{-1}] + O(r)
\ee
that correctly reproduce the free space scattering amplitude
\be
\label{eq:fkzr}
f_k^{\rm zero\ range} = -\frac{1}{a^{-1} + ik}.
\ee
The general solution of Eq.(\ref{eq:schr3d}) may be expressed in
terms of Whittaker $M$ et $W$ functions. For an energy $E_{\rm rel}$
not belonging to the non-interacting spectrum $\{(\frac{3}{2}+2n)\hbar
\omega,n\in \mathbb{N}\}$, the Whittaker function $M$ diverges
exponentially for large $r$ and has to be disregarded. The small $r$
behavior of the Whittaker function $W$, together with the Wigner-Bethe-Peierls
contact condition, leads to the implicit equation for the relative
energy, in accordance with \cite{Wilkens}:
\be
\label{eq:impl}
\frac{\Gamma(\frac{3}{4}- \frac{E_{\rm rel}}{2\hbar\omega})}
{\Gamma(\frac{1}{4}- \frac{E_{\rm rel}}{2\hbar\omega})} 
= \frac{a_{\rm ho}^{\rm rel}}{2a}
\ee
with the harmonic oscillator length of the relative motion,
$a_{\rm ho}^{\rm rel} = [\hbar/(\mu\omega)]^{1/2}$.

The function $\Gamma(x)$ is different from zero $\forall x\in \mathbb{R}$
and diverges on each non-positive integers.
Thus Eq.(\ref{eq:impl}) immediately leads in the unitary case to the spectrum
$E_{\rm rel}\in \{(2n+1/2)\hbar \omega,n\in\mathbb{N}\}$.
This can be readily obtained by setting in Eq.(\ref{eq:schr3d})
$\psi(r)=f(r)/r$, so that $f$ obeys Schr\"odinger's equation
for a 1D harmonic oscillator, with the constraint issuing
from Eq.(\ref{eq:bp}) that $f(r=0)\neq 0$, which selects the even 1D
states.

\begin{figure}[b]
\sidecaption
\includegraphics[width=7cm,clip=]{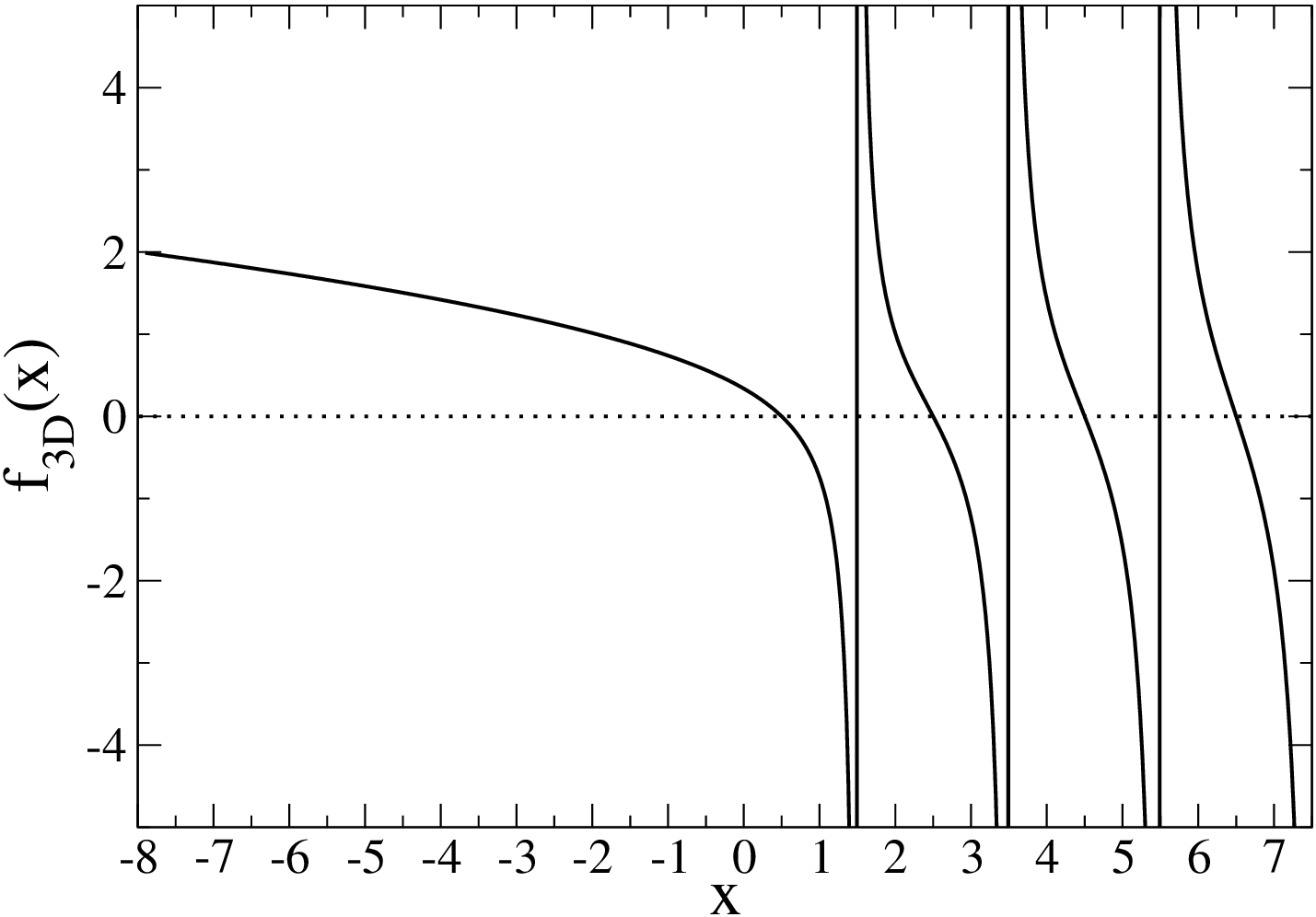}
\caption{For the graphical solution of Eq.(\ref{eq:impl}),
which gives the spectrum for two particles in a three-dimensional
isotropic harmonic trap, plot of the function
$f_{3D}(x) = 
\Gamma(\frac{3}{4}- \frac{x}{2})/
\Gamma(\frac{1}{4}- \frac{x}{2})$, where $x$ 
stands for $E_{\rm rel}/(\hbar\omega)$.
}
\label{fig:spec3d}
\end{figure}

The graphical solution of Eq.(\ref{eq:impl}), see Fig.~\ref{fig:spec3d},
allows to resolve the paradox about the attractive or repulsive
nature of the unitary gas. E.g. starting with the ground state
wavefunction of the ideal gas case, of relative energy $E_{\rm rel}
=\frac{3}{2}\hbar \omega$, it appears that the two adiabatic followings
(i) $a=0^+\rightarrow a= +\infty$ and (ii) $a=0^-\rightarrow -\infty$ 
lead to {\sl different} final eigenstates of the unitary case,
to an excited state $E_{\rm rel}=\frac{5}{2}\hbar\omega$ for the procedure
(i), and to the ground state $E_{\rm rel}=\frac{1}{2}\hbar\omega$
for procedure (ii). 

\begin{figure}[b]
\sidecaption
\includegraphics[width=7cm,clip=]{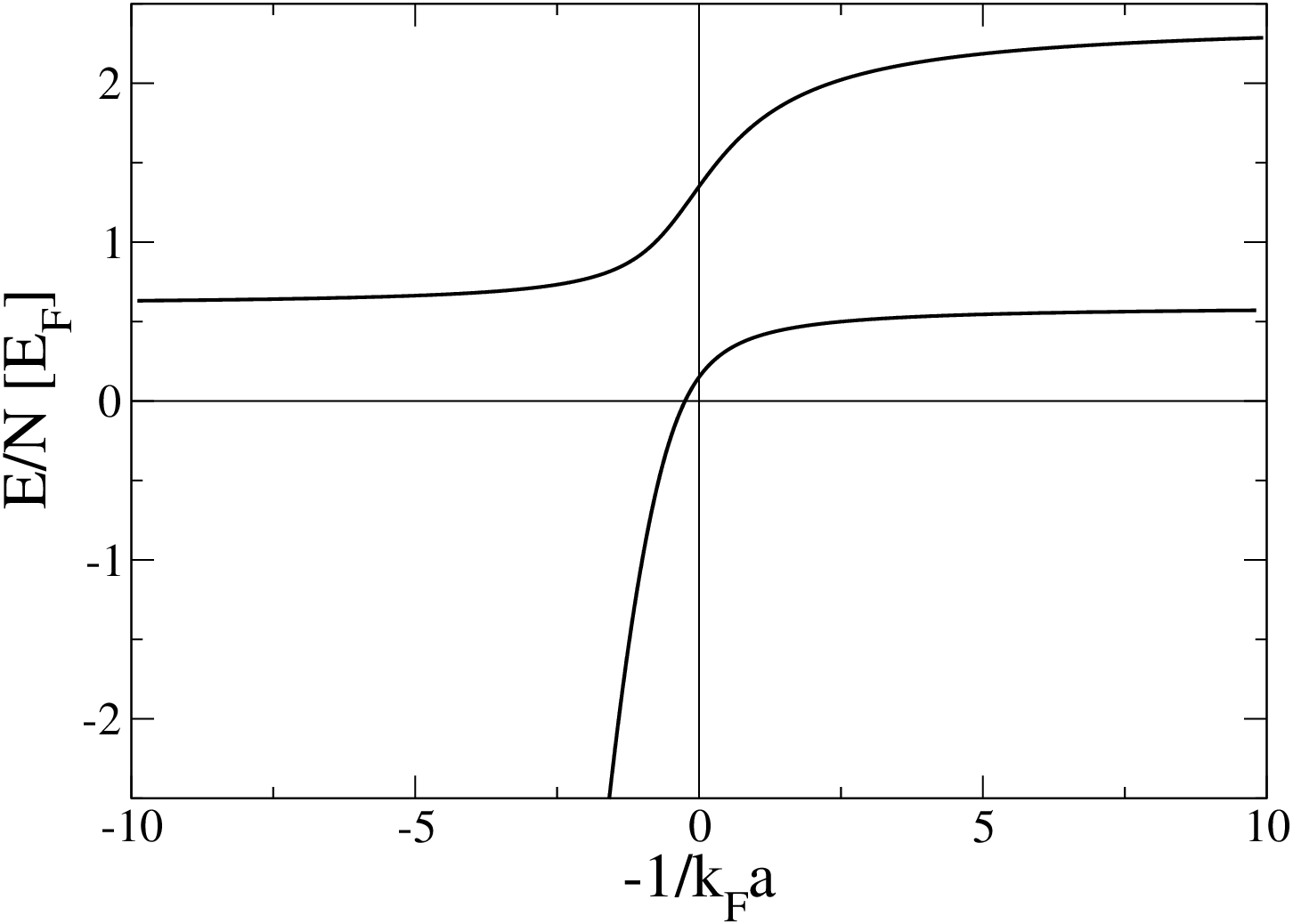}
\caption{
In the toy model of \cite{PricoupenkoToy},
for the homogeneous two-component unpolarized Fermi gas,
energy per particle on the ground branch and the first excited branch
as a function of the inverse scattering length.
The Fermi wavevector is defined in Eq.(\ref{eq:kf}), $E_F=\hbar^2 k_F^2/(2m)$
is the Fermi energy, and $a$ is the scattering length.
}
\label{fig:toy}
\end{figure}

The same explanation holds for the many-body case: The interacting
gas has indeed several energy branches in the BEC-BCS crossover,
as suggested by the toy model 
\footnote{This toy model replaces the many-body
problem with the one of a matterwave interacting with a single scatterer
in a hard wall cavity of radius $\propto 1/k_F$.}
of \cite{PricoupenkoToy}, see Fig.~\ref{fig:toy}. 
Starting from the weakly attractive Fermi gas 
and ramping the scattering length down to $-\infty$ one explores
a part of the ground energy branch, where the unitary gas is attractive;
this ground branch continuously evolves into a weakly repulsive
condensate of dimers \cite{PetrovShlyapSalomon}
if $1/a$ further moves from $0^-$ to $0^+$ and then to $+\infty$.
The attractive nature of the unitary gas on the ground energy branch
will become apparent in the lattice model of section \ref{sec:vm}.
On the other hand, starting from the weakly repulsive Fermi gas and 
ramping the scattering up to $+\infty$, one explores an effectively 
repulsive excited branch.

In the first experiments on the BEC-BCS crossover, 
the ground branch was explored by adiabatic
variations of the scattering length and was found to be stable.
The first excited energy branch
was also investigated in the early work~\cite{Bourdel_Eint}, and more recently in~ \cite{Ketterle_excited} looking for a Stoner demixing instability
of the strongly repulsive two-component Fermi gas.
A difficulty for the study of this excited branch is its metastable
character: Three-body collisions gradually transfer
the gas to the ground branch, leading e.g. to the formation
of dimers if $0 < k_F a \lesssim 1$.

\subsection{Other partial waves, other dimensions}

We have previously considered the two-body scattering amplitude
in the $s$-wave channel. What happens for example in the $p$-wave
channel~? This channel is relevant for the interaction between
fermions in the same internal state, where a Feshbach resonance
technique is also available \cite{Salomon_p,Jin_p}. Can one
also reach the unitarity limit Eq.(\ref{eq:fku}) in the $p$-wave channel~?

Actually the optical theorem shows that
relation Eq.(\ref{eq:u(k)}) also holds for the $p$-wave scattering
amplitude $f_k$. What differs is the low-$k$ expansion of $u(k)$,
that is now given by
\be
\label{eq:ukp}
u(k)= \frac{1}{k^2\mathcal{V}_s} + \alpha + \ldots,
\ee
where $\mathcal{V}_s$ is the scattering volume (of arbitrary sign)
and $\alpha$ has the dimension of the inverse of
a length.
The unitary limit would require $u(k)$ negligible as compared
to $k$. One can in principle tune $\mathcal{V}_s$ to infinity
with a Feshbach resonance. Can one then have a small value of $\alpha$
at resonance~? A theorem for a compact support interaction
potential of radius $b$ shows however that \cite{LudoPRL_ondeP,Jona}
\be
\lim_{|\mathcal{V}_s|\to +\infty} \alpha \geq 1/b.
\ee
A similar conclusion holds using two-channel models of the Feshbach
resonance \cite{Jona,Chevy}. $\alpha$ thus assumes a huge positive
value on resonance, which breaks the scale invariance and precludes
the existence of a $p$-wave unitary gas.
This does not prevent however to reach the unitary
limit in the {\sl vicinity} of a particular value of $k$.
For $\mathcal{V}_s$ large and negative, neglecting
the $\ldots$ in Eq.(\ref{eq:ukp}) under the condition $k b \ll 1$,
one indeed has $|u(k)|\ll k$,
so that $f_k \simeq -1/(ik)$, in a vicinity of
\be
k_0 = \frac{1}{(\alpha |\mathcal{V}_s|)^{1/2}}.
\ee

Turning back to the interaction in the $s$-wave channel, an interesting
question is whether the unitary gas exists in reduced dimensions.

In a one-dimensional system the zero range interaction may be
modeled by a Dirac potential $V(x)=g\delta(x)$. If $g$ is finite,
it introduces a length scale $\hbar^2/(m g)$ that breaks the scaling 
invariance. Two cases are thus scaling invariant, the ideal gas
$g=0$ and the impenetrable case $1/g=0$. The impenetrable case
however is mappable to an ideal gas in one dimension, it has in particular
the same energy spectrum and thermodynamic properties \cite{Gaudin}.

In a two-dimensional system, the scattering amplitude for a zero
range interaction potential is given by \cite{Olshanii2D}
\be
\label{eq:fk2D}
f_k^{2D} = \frac{1}{\gamma+\ln(k a_{2D}/2) -i\pi/2}
\ee
where $\gamma=0.57721566\ldots$ 
is Euler's constant and $a_{2D}$ is the scattering length.
For a finite value of $a_{2D}$, there is no scale invariance.
The case $a_{2D}\to 0$ corresponds to the ideal gas limit. At first sight,
the opposite limit $a_{2D}\to +\infty$ is a good candidate for a
two-dimensional unitary gas; however this limit also corresponds
to an ideal gas. This appears in the 2D version of the lattice
model of section \ref{sec:vm} \cite{Tonini}.
This can also be checked for two particles in an isotropic 
harmonic trap. Separating out the center of mass motion,
and taking a zero angular momentum state for the relative motion,
to have interaction in the $s$-wave channel, one has to solve
the radial Schr\"odinger equation:
\be
-\frac{\hbar^2}{2\mu} [\psi''(r) + \frac{1}{r} \psi'(r)]
+\frac{1}{2}\mu \omega^2 r^2 \psi(r) = E_{\rm rel} \psi(r)
\label{eq:rad_schr}
\ee
where $\mu=m/2$ is the reduced mass of the two particles,
$E_{\rm rel}$ is an eigenenergy of the relative motion,
and $\omega$ is the single particle angular oscillation frequency.
The interactions are included by the boundary condition in $r=0$:
\be
\psi(r) = A\ln(r/a_{2D}) + O(r),
\ee
which is constructed to reproduce the expression of the 
scattering amplitude Eq.(\ref{eq:fk2D}) for the free space problem.

\begin{figure}[b]
\sidecaption
\includegraphics[width=7cm,clip=]{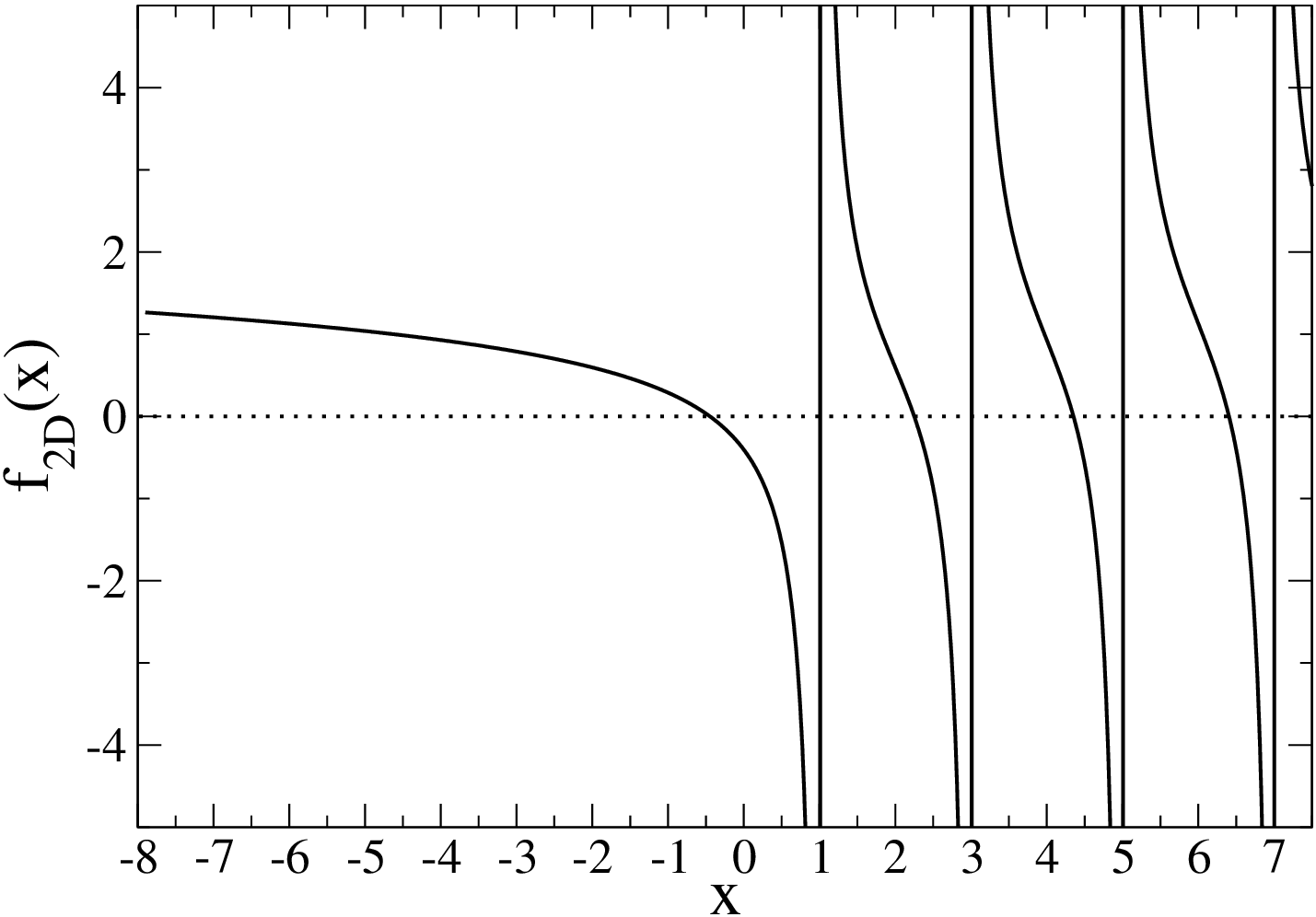}
\caption{For the graphical solution of Eq.(\ref{eq:impl2d}),
which gives the spectrum for two interacting particles in a two-dimensional
isotropic harmonic trap, plot of the function
$f_{2D}(x) = \frac{1}{2} 
\psi[(1-x)/2] +\gamma$
where $x$ stands for $E_{\rm rel}/(\hbar\omega)$ and the special
function $\psi$
is the logarithmic derivative of the $\Gamma$ function.
}
\label{fig:spec2d}
\end{figure}

The general solution of Eq.(\ref{eq:rad_schr}) may be expressed
in terms of Whittaker functions $M$ and $W$.
Assuming that $E_{\rm rel}$ does not belong to the ideal gas spectrum
$\{(2n+1)\hbar \omega,n\in \mathbb{N}\}$, one finds that the 
$M$ solution has to be disregarded because it diverges 
exponentially for $r\to +\infty$. From the small $r$ behavior
of the $W$ solution, one obtains the implicit equation
\be
\label{eq:impl2d}
\frac{1}{2} \psi\left(\frac{\hbar\omega-E_{\rm rel}}{2\hbar\omega}\right)
+\gamma = - \ln(a_{2D}/a_{\rm ho}^{\rm rel})
\ee
where the relative harmonic oscillator length is $a_{\rm ho}^{\rm rel}
=[\hbar/(\mu\omega)]^{1/2}$ and the digamma function $\psi$
is the logarithmic derivative of the $\Gamma$ function.
If $a_{2D}\to +\infty$, one then finds that $E_{\rm rel}$ 
tends to the ideal gas spectrum $\{(2n+1)\hbar \omega,n\in \mathbb{N}\}$
from below, see Fig.~\ref{fig:spec2d}, in agreement with
the lattice model result that the 2D gas with a large and finite
$a_{2D}$ is a weakly attractive gas.

\section{Various models and general relations}
\label{sec:vm}

There are basically two approaches to model the interaction 
between particles for the unitary gas (and more generally for
the BEC-BCS crossover).

In the first approach, see subsections \ref{subsec:lm} and \ref{subsec:tcm},
one takes a model with a finite range $b$
and a fixed (e.g. infinite) scattering length $a$. This model
may be in continuous space or on a lattice, with one or several channels. 
Then one tries to calculate the eigenenergies, the thermodynamic properties from
the thermal density operator $\propto \exp(-\beta H)$,
etc, and the zero range limit $b\to 0$ should be taken at the end of the 
calculation. Typically, this approach is followed in numerical
many-body methods, such as the approximate fixed node Monte Carlo
method \cite{Carlson,Panda,Giorgini} or unbiased Quantum Monte Carlo
methods \cite{Burovski,Bulgac,Juillet}.
A non-trivial question however is whether each eigenstate of the model
is universal in the zero range limit, that is if the eigenenergy
$E_i$ and the corresponding wavefunction $\psi_i$ converge
for $b\to 0$. In short, the challenge is to prove
that the ground state energy of the system does not tend
to $-\infty$ when $b\to 0$. 

In the second approach,  see subsection \ref{subsec:bp},
one directly considers the zero range limit,
and one replaces the interaction by the so-called Wigner-Bethe-Peierls
contact conditions on the $N$-body
wavefunction. This constitutes what we shall call the {\it zero-range model}.
The advantage is that only the scattering length
appears in the problem, without unnecessary details on the interaction,
which simplifies the problem and allows to obtain analytical results.
E.g. the scale invariance of the unitary gas becomes clear.
A non-trivial question however is to know whether the zero-range model
leads to a self-adjoint Hamiltonian, with a spectrum then necessarily
bounded from below for the unitary gas (see Section \ref{subsec:ssfp}), without having to add 
extra boundary conditions.
For $N=3$ bosons,
due to the Efimov effect,
 the Wigner-Bethe-Peierls or zero-range model becomes self-adjoint
only if one adds an extra three-body contact condition,
involving a
so-called three-body parameter.
  In an isotropic
harmonic trap, at unitarity, there exists however a non-complete
family of bosonic universal states, independent from
the three-body parameter
and to which the restriction of the Wigner-Bethe-Peierls
model is hermitian \cite{Jonsell,WernerPRL}.
For equal mass two-component fermions, it is hoped in the physics
literature that the zero-range model is self-adjoint 
for an arbitrary number of particles $N$.
Surprisingly, there exist works in mathematical physics 
predicting
 that this is
{\sl not} the case when $N$ is large enough \cite{Teta,Minlos};
however
the critical mass ratio for the appearance of an Efimov effect in the unequal-mass $3+1$ body problem given without proof in~\cite{Minlos}
was not confirmed by the numerical study~\cite{CMP},
and
the variational ansatz used in~\cite{Teta} 
to show that the energy is unbounded below does not have the proper fermionic exchange symmetry. 
This mathematical problem thus remains open.

\subsection{Lattice models and
general relations}
\label{subsec:lm}

\subsubsection{The lattice models}

The model that we consider here assumes that the spatial positions are discretized on a cubic lattice,
of lattice constant that we call $b$ as the interaction range. 
It is quite appealing in its simplicity and generality.
It naturally allows to consider a contact interaction potential, opposite spin fermions
interacting only when they are on the same lattice site. Formally, this constitutes
a separable potential for the interaction (see subsection \ref{subsec:tcm} for a reminder),
a feature known to simplify diagrammatic calculations \cite{NSR}.
Physically, it belongs to the same class as the Hubbard model, so that it may truly
be realized with ultracold atoms in optical lattices \cite{BlochMott}, and it allows
to recover the rich lattice physics of condensed matter physics and the
corresponding theoretical tools such as Quantum Monte Carlo methods
\cite{Burovski,Juillet}.

The spatial coordinates $\rr$ of the particles are thus discretized on a cubic
grid of step $b$. As a consequence, the components of the wavevector of a particle have 
a meaning modulo $2\pi/b$ only, since the plane wave function $\rr\rightarrow \exp(i\kk\cdot\rr\,)$
defined on the grid is not changed if a component of $\kk$ is shifted by an integer
multiple of $2\pi/b$. We shall therefore restrict the wavevectors to the first Brillouin
zone of the lattice:
\be
\kk \in {\mathcal D}\equiv \left[-\frac{\pi}{b},\frac{\pi}{b}\right[^3.
\ee
This shows that the lattice structure in real space automatically provides a cut-off in
momentum space.
In the absence of interaction and of confining potential, eigenmodes of the system are plane waves 
with a dispersion relation $\kk\to \epsk$, supposed to be an even and non-negative function of $\kk$. 
We assume that this dispersion relation is independent of the spin state, which is a natural choice
since the $\uparrow$ and $\downarrow$ particles have the same mass. To recover
the correct continuous space physics in the zero lattice spacing limit $b\to 0$, we further impose
that it reproduces the free space dispersion relation in that limit, so that
\be
\label{eq:rdavdz}
\epsk \sim \frac{\hbar^2 k^2}{2m} \ \ \ \mbox{for}\ \ \ kb\to 0.
\ee
The interaction between opposite spin particles takes place when two particles are
on the same lattice site, as in the Hubbard model. In first quantized form, 
it is represented by a discrete delta potential:
\be
V= \frac{g_0}{b^3} \delta_{\rr_1,\rr_2}.
\ee
The factor $1/b^3$ is introduced because $b^{-3} \delta_{\rr,\mathbf{0}}$ is 
equivalent to the Dirac distribution $\delta(\rr)$ in the continuous space limit.
To summarize, the lattice Hamiltonian in second quantized form in the general trapped case is
\bea
H &=& \sum_{\sigma=\uparrow,\downarrow}
\int_{\mathcal{D}} \frac{d^3k}{(2\pi)^3}\,\epsk c_\sigma^\dagger(\kk)c_\sigma(\kk) + 
\sum_{\sigma=\uparrow,\downarrow} 
\sum_{\rr} 
b^3 U(\rr) (\psi_\sigma^\dagger \psi_\sigma)(\rr) \nonumber \\
&& + g_0 \sum_\rr b^3 (\psi^\dagger_\uparrow\psi^\dagger_\downarrow\psi_\downarrow\psi_\uparrow)(\rr).
\eea
The plane wave annihilation operators $c_\sigma(\kk)$ in spin state $\sigma$ 
obey the usual continuous space anticommutation relations $\{c_\sigma(\kk),c_{\sigma'}^\dagger(\kk')\}=(2\pi)^3
\delta(\kk-\kk')\delta_{\sigma\sigma'}$ if $\kk$ and $\kk'$ are in the first Brillouin zone
\footnote{In the general case, $\delta(\kk-\kk')$ has to be replaced with $\sum_{\KK} \delta(\kk-\kk'-\KK)$
where $\KK\in (2\pi/b)\mathbb{Z}^3$ is any vector in the reciprocal lattice.},
and the field operators $\psi_{\sigma}(\rr)$ obey the usual discrete space
anticommutation relations $\{\psi_{\sigma}(\rr),\psi_{\sigma'}^\dagger(\rr')\}=
b^{-3}\delta_{\rr\rr'} \delta_{\sigma\sigma'}$.
In the absence of trapping potential, in a cubic box with size $L$ integer multiple of $b$,
with periodic boundary conditions, the integral in the kinetic energy term is replaced by the sum
$\sum_{\kk\in\mathcal{D}} \epsk \tilde{c}_{\kk\sigma}^\dagger \tilde{c}_{\kk\sigma}$ 
where the annihilation operators
then obey the discrete anticommutation relations $\{\tilde{c}_{\kk\sigma},\tilde{c}^\dagger_{\kk'\sigma'}\}=
\delta_{\kk\kk'} \delta_{\sigma\sigma'}$ for $\kk,\kk'\in \mathcal{D}$.

The coupling constant $g_0$ is a function of the grid spacing $b$. It is adjusted
to reproduce the scattering length of the true interaction.
The scattering amplitude of two atoms on the lattice with vanishing total momentum,
that is with incoming particles of opposite spin and opposite momenta $\pm\kk_0$,
reads
\be
\label{eq:fklm}
f_{k_0} = -\frac{m}{4\pi\hbar^2}
\left[g_0^{-1}-\int_{\mathcal D}\frac{d^3k}{(2\pi)^3}
\frac{1}{E+i0^+-2\epsk}\right]^{-1}
\ee
as derived in details in \cite{Houches03} for a quadratic dispersion relation and in~\cite{Tangen} for a general dispersion relation.
Here the scattering state energy 
$E=2\epsilon_{\kk_0}$ actually introduces
a dependence of the scattering amplitude 
on the direction of $\kk_0$ when the dispersion relation $\epsilon_\kk$
is not parabolic.
If one is only interested in the expansion of $1/f_{k_0}$ 
up to second order in $k_0$, e.g.\ for an effective range calculation,
one may conveniently use the isotropic approximation $E=\hbar^2 k_0^2/m$
thanks to~(\ref{eq:rdavdz}).
Adjusting $g_0$ to recover the correct scattering length gives
from Eq.(\ref{eq:fklm}) for $k_0\to 0$:
\be
\frac{1}{g_0}= \frac{1}{g} -\int_{\mathcal D} \frac{d^3k}{(2\pi)^3} \,
\frac{1}{2 \epsk},
\label{eq:g0_gen}
\ee
with $g=4\pi\hbar^2 a/m$.  The above formula Eq.(\ref{eq:g0_gen}) is reminiscent of the technique of renormalization of the coupling
constant \cite{Randeria,Randeria2}. 
A natural case to consider is the one of the usual parabolic dispersion relation,
\be
\label{eq:pdr}
\epsk=\frac{\hbar^2 k^2}{2m}.
\ee
A more explicit form of Eq.(\ref{eq:g0_gen}) is then \cite{Mora,LudoVerif}:
\be
g_0= \frac{4\pi\hbar^2 a/m}{1-Ka/b}
\label{eq:g0_exp}
\ee
with a numerical constant given by
\be
K=\frac{12}{\pi}\int_{0}^{\pi/4} 
d\theta\, \ln(1+1/\cos^{2}\theta)
= 2.442\ 749\ 607\ 806\ 335\ldots,
\ee
and that may be expressed analytically in terms of the dilog special function.

\subsubsection{Simple variational upper bounds}

The relation Eq.(\ref{eq:g0_exp}) is quite instructive in the zero range limit $b\to 0$, for fixed non-zero
scattering length $a$ and atom numbers $N_\sigma$: In this limit, the lattice filling factor tends to zero, and the lattice model
is expected to converge to the continuous space zero-range model, that is to the Wigner-Bethe-Peierls
model described in subsection \ref{subsec:bp}. For each of the eigenenergies this means that
\be
\label{eq:lat_to_bp}
\lim_{b\to 0} E_i(b) = E_i,
\ee
where in the right hand side the set of $E_i$'s are the energy spectrum of the zero range model.
On the other hand, for a small enough value of $b$, the denominator in the
right-hand side of Eq.(\ref{eq:g0_exp}) is dominated by the term $-Ka/b$,
the lattice coupling constant $g_0$ is clearly negative, 
and the lattice model is attractive, as already pointed out in \cite{kitp}. 
By the usual variational argument, 
this shows that the ground state energy of the zero range interacting gas is below
the one of the ideal gas, for the same trapping potential and atom numbers $N_\sigma$:
\be
\label{eq:bounde}
E_0 \leq E_0^{\rm ideal}.
\ee
Similarly, at thermal equilibrium in the canonical ensemble, the free energy of the interacting  gas is below
the one of the ideal gas:
\be
\label{eq:boundf}
F \leq F^{\rm ideal}.
\ee
As in \cite{Blaizot} one indeed introduces the free-energy functional of the (here lattice model) interacting gas, $\mathcal{F}[\hat{\rho}]=
\mathrm{Tr}[H\hat{\rho}] +k_B T \mathrm{Tr}[\hat{\rho}\ln\hat{\rho}]$, 
where $\hat{\rho}$ is any unit trace system density operator. Then
\be
\label{eq:ident}
\mathcal{F}[\hat{\rho}_{\rm th}^{\rm ideal}] = F^{\rm ideal}(b) + \mathrm{Tr}[\hat{\rho}_{\rm th}^{\rm ideal}V],
\ee
where $\hat{\rho}_{\rm th}^{\rm ideal}$ is the thermal equilibrium density operator
of the ideal gas in the lattice model, and  $V$
is the interaction contribution to the $N$-body Hamiltonian.
Since the minimal value of $\mathcal{F}[\hat{\rho}]$ over $\hat{\rho}$ 
is equal to the interacting gas lattice model free energy $F(b)$, 
the left hand side of Eq.(\ref{eq:ident}) is larger than $F(b)$.
Since the operator $V$ is negative for small $b$, because $g_0 <0$, the right hand side of Eq.(\ref{eq:ident}) is smaller than $F^{\rm ideal}(b)$.
Finally taking the limit $b\to 0$, one obtains the desired inequality.
The same reasoning can be performed in the grand canonical ensemble, showing
that the interacting gas grand potential is below the one of the ideal gas, for the same temperature and
chemical potentials $\mu_\sigma$:
\be
\Omega \leq \Omega^{\rm ideal}.
\ee
In \cite{ChevyNature}, for the unpolarized unitary gas, this last inequality was checked to be obeyed by the experimental
results, but it was shown, surprisingly, to be violated by some of the Quantum Monte Carlo results of
\cite{Burovski}.
For the particular case of the spatially homogeneous unitary gas, the above reasonings
imply that $\xi\leq 1$ in Eq.(\ref{eq:e0}), so that the unitary gas is attractive (in the ground branch,
see subsection \ref{subsec:aor}). Using the BCS variational ansatz in the lattice model
\footnote{One may check, e.g.\ in the sector $N_\uparrow=N_\downarrow=2$, 
that the BCS variational wavefunction, which is a condensate of pairs in some pair wavefunction,
does not obey the Wigner-Bethe-Peierls boundary conditions
even if the pair wavefunction does, so it looses its variational character in the zero-range
model.} \cite{Varenna06}
one obtains the more stringent upper bound \cite{Randeria2}:
\be
\xi \leq \xi_{\rm BCS} = 0.5906\ldots
\ee

\subsubsection{Finite-range corrections}
\begin{figure}[b]
\sidecaption
\includegraphics[width=7cm,clip=]{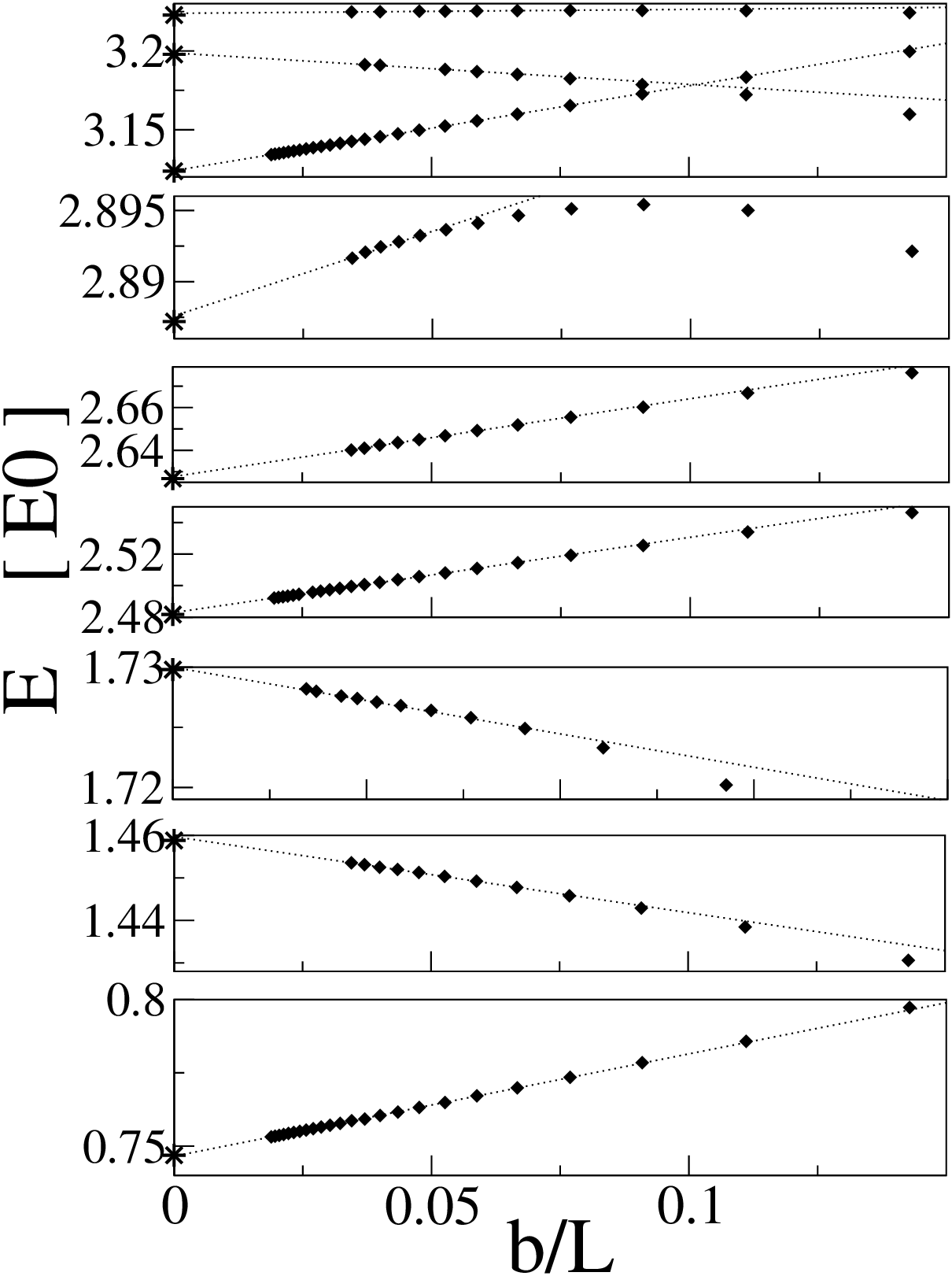}
\caption{
Diamonds: The first low eigenenergies for three $(\uparrow\uparrow\downarrow)$ 
fermions in a cubic box with a lattice model, as
functions of the lattice constant $b$ \cite{LudoVerif}. 
The box size is $L$, with periodic boundary conditions, the scattering length is
infinite, the dispersion relation is parabolic Eq.(\ref{eq:pdr}).
The unit of energy is $E_0=(2\pi\hbar)^2/2mL^2$.
Straight lines: Linear fits performed on the data over the range ${b/L\leq 1/15}$,
except for the energy branch $E\simeq 2.89 E_0$ which is linear on a smaller range.
Stars in $b=0$: Eigenenergies predicted by the zero-range model.
}
\label{fig:3piab}
\end{figure}

For the parabolic dispersion relation, the expectation Eq.(\ref{eq:lat_to_bp}) was checked analytically for
two opposite spin particles: For $b\to 0$, in free space the scattering amplitude (\ref{eq:fklm}),
and in a box the lattice energy spectrum, converge to the predictions of the zero-range model \cite{LudoVerif}.
It was also checked numerically for $N=3$ particles in a box, with two $\uparrow$ particles
and one $\downarrow$ particle: As shown in Fig.~\ref{fig:3piab}, for the first low energy eigenstates
with zero total momentum, a convergence of the lattice eigenenergies to the Wigner-Bethe-Peierls ones is observed,
in a way that is eventually linear in $b$ for small enough values of $b$.
As discussed in \cite{Tangen}, 
this asymptotic linear dependence in $b$ is expected
for Galilean invariant continuous space models, 
and the first order deviations of the eigenergies 
from their zero range values are linear in the effective range $r_e$ 
of the interaction potential,
as defined in Eq.(\ref{eq:lke}), with model-independent coefficients:
\be
\label{eq:deriv_univ}
\frac{dE_i}{d r_e}(b\to 0) \ \ \ \mbox{is model-independent}.
\ee
However, for lattice models, Galilean invariance is broken and the scattering between two particles 
depends on their center-of-mass momentum; this leads to a breakdown of
the universal relation~(\ref{eq:deriv_univ}), 
while preserving the linear dependence of the energy with $b$
at low $b$~\cite{BurovskiNJP}.

A procedure to calculate $r_e$ in the lattice model 
for a general dispersion relation
$\epsk$ in presented in Appendix~1. For the parabolic dispersion relation Eq.(\ref{eq:pdr}),
its value was given in \cite{Varenna06} in numerical form. With the technique exposed in Appendix~1,
we have now the analytical value:
\be
\label{eq:rep}
r_e^{\rm parab} = b\, \frac{12\sqrt{2}}{\pi^3} \arcsin\frac{1}{\sqrt{3}} =
0.336\ 868 \ 47\ldots b.
\ee
The usual Hubbard model, whose rich many-body physics is reviewed in \cite{AntoineVarenna},
was also considered in \cite{Varenna06}:
It is defined in terms of the tunneling
amplitude between neighboring lattice sites, here $t=-\hbar^2/(2mb^2)<0$, and of the on-site
interaction $U=g_0/b^3$. The dispersion relation is then
\be
\epsk = \frac{\hbar^2}{mb^2} \sum_{\alpha=x,y,z} \left[1-\cos(k_\alpha b)\right]
\label{eq:disphub}
\ee
where the summation is over the three dimensions of space. It reproduces the free space
dispersion relation only in a vicinity of $\kk=\mathbf{0}$.
The explicit version of Eq.(\ref{eq:g0_gen}) is 
obtained from Eq.(\ref{eq:g0_exp}) by replacing the numerical constant 
$K$ by $K^{\rm Hub}=3.175 911 \ldots$. In the zero range limit this leads for $a\neq 0$ to 
$U/|t| \to -7.913 552\ldots$, corresponding as expected to an {\sl attractive} Hubbard model,
lending itself to a Quantum Monte Carlo analysis for equal
spin populations with no sign problem \cite{Burovski,Bulgac}.
The effective range of the Hubbard model, calculated as in Appendix~1, remarkably is negative \cite{Varenna06}:
\be
\label{eq:reh}
r_e^{\rm Hub} \simeq -0.305\, 718 b.
\ee
It becomes thus apparent that an {\sl ad hoc} tuning of the dispersion relation $\epsk$
may lead to a lattice model with a zero effective range. 
As an example, we consider a dispersion relation
\be
\label{eq:mixte}
\epsk = \frac{\hbar^2 k^2}{2m} [ 1- C (kb/\pi)^2],
\ee
where $C$ is a numerical constant less than $1/3$.
From Appendix~1 we then find that
\be
\label{eq:magic}
r_e = 0 \ \ \ \ \mbox{for} \ \ \ \ C=0.2570224\ldots
\ee
The corresponding value of $g_0$ is given by Eq.(\ref{eq:g0_exp}) with $K=2.899952\ldots$.

As pointed out in \cite{BurovskiNJP},
 additionally fine-tuning the dispersion relation to cancel not only $r_e$ but also another coefficient (denoted by $B$ in \cite{BurovskiNJP})
 may have some practical interest for Quantum Monte Carlo calculations that are performed
with a non-zero $b$, by canceling the undesired linear dependence of thermodynamical quantities and of the critical temperature
$T_c$ on $b$.

\subsubsection{Energy functional,
tail of the momentum distribution
and pair correlation function at short distances}
\label{subsubsec:tail}

A quite ubiquitous quantity in the short-range or large-momentum
physics of gases with zero range interactions
is the so-called ``contact'', which, restricting here for simplicity to thermal equilibrium in the canonical ensemble,
can be defined by
\be
C\equiv\frac{4\pi m}{\hbar^2}\left(\frac{dE}{d(-1/a)}\right)_{\!S}
=\frac{4\pi m}{\hbar^2}\left(\frac{dF}{d(-1/a)}\right)_{\!T}.
\label{eq:defC}
\ee
For zero-range interactions, this quantity $C$ determines the
large-$k$ tail of the momentum distribution
\be
n_\sigma(\kk)\underset{k\to\infty}{\sim}\frac{C}{k^4}
\label{eq:C_nk}
\ee
as well as the
short-distance behavior of the 
pair distribution function
\be
\int d^3R\ g^{(2)}_{\up\down}\left(\RR+\frac{\rr}{2},\RR-\frac{r}{2}\right)\underset{r\to0}{\sim}\frac{C}{(4\pi r)^2}.
\label{eq:C_g2}
\ee
Here the spin-$\sigma$ momentum distribution $n_\sigma(\kk)$ is 
normalised as  $\int \frac{d^3k}{(2\pi)^3}n_\sigma(\kk)=N_\sigma$.
The 
relations~(\ref{eq:defC},\ref{eq:C_nk},\ref{eq:C_g2}) were obtained in
\cite{Tan1,Tan2}.
Historically, analogous relations were first established
for one-dimensional bosonic systems \cite{Lieb,Olshanii} with techniques
that may be straightforwardly extended to two dimensions and three
dimensions \cite{Tangen}.
Another relation derived in~\cite{Tan1} for the zero-range model
expresses the energy as a functional of the one-body density matrix:
\be
E=
\sum_{\sigma=\up,\down}\int\frac{d^3k}{(2\pi)^3} \frac{\hbar^2 k^2}{2m}\left[n_\sigma(\kk)-\frac{C}{k^4}\right]
+\frac{\hbar^2 C}{4\pi m a}
+\sum_{\sigma=\up,\down}\int d^3r\,U(\rr) \rho_\sigma(\rr)
\label{eq:fonctionelle}
\ee
where $\rho_\sigma(\rr)$ is the spatial number density.

One usually uses (\ref{eq:C_nk}) to define $C$, and then derives (\ref{eq:defC}).
Here we rather take (\ref{eq:defC}) as the definition of $C$. This choice is convenient both for the two-channel model discussed in Section~\ref{subsec:tcm} and for the rederivation of
 (\ref{eq:C_nk},\ref{eq:C_g2},\ref{eq:fonctionelle}) that we shall now present, where we use a lattice model before taking the zero-range limit.
 
From the Hellmann-Feynman theorem (that was already put forward in
\cite{Lieb}), the interaction energy $E_{\rm int}$ is equal to $g_0 (dE/dg_0)_{\!S}$. Since we have $d(1/g_0)/d(1/g)=1$ [see the relation (\ref{eq:g0_gen}) between $g_0$ and $g$], this can be rewritten as
\be
E_{\rm int}=\frac{\hbar^4}{m^2} \frac{C}{g_0}.
\label{eq:E_int}
\ee
Expressing $1/g_0$ in terms of $1/g$ using once again (\ref{eq:g0_gen}), adding the kinetic energy, and taking the zero-range limit, we immediately get the relation (\ref{eq:fonctionelle}). For the integral over momentum to be convergent, (\ref{eq:C_nk}) must hold (in the absence of mathematical pathologies).

To derive (\ref{eq:C_g2}), we again use (\ref{eq:E_int}), which implies that the relation
\be
\sum_\RR b^3\,g^{(2)}_{\up\down}\left(\RR+\rr/2,\RR-\rr/2 \right)=\frac{C}{(4\pi)^2}|\phi(\rr)|^2
\label{eq:g2_phi2}
\ee
holds for $\rr=\vn$, were $\phi(\rr)$ is the zero-energy two-body scattering wavefunction, normalised in such a way that 
\be
\phi(\rr)\simeq \frac{1}{r}-\frac{1}{a} {\rm\ \ for\ \ } r\gg b
\label{eq:phi_norm}
\ee
[see~\cite{Tangen} for the straightforward calculation of $\phi(\vn)$]. Moreover, in the regime where $r$ is much smaller than the typical interatomic distances and than the thermal de~Broglie wavelength
(but not necessarily smaller than $b$), it is generally expected that the $\rr$-dependence of $g^{(2)}_{\up\down}(\RR+\rr/2,\RR-\rr/2)$ is proportional to $|\phi(\rr)|^2$, so that (\ref{eq:g2_phi2}) remains asymptotically valid. Taking the limits $b\to0$ and then $r\to0$ gives the desired (\ref{eq:C_g2}).

Alternatively, the link (\ref{eq:C_nk},\ref{eq:C_g2}) between short-range pair correlations and large-$k$ tail of the momentum distribution can be directly deduced from the 
short-distance singularity of the wavefunction
coming from the contact condition~(\ref{eq:bpN})
and the corresponding tail in Fourier space~\cite{Tangen}, similarly to the original derivation in 1D~\cite{Olshanii}.
Thus this link remains true for a generic out-of-equilibrium statistical mixture of states satisfying the contact condition~\cite{Tan1,Tangen}. 

\subsubsection{Absence of simple collapse}

To conclude this subsection on lattice models, we try to address the question of the advantage of lattice models
as compared to the standard continuous space model with a binary interaction potential $V(\rr)$ between
opposite spin fermions.  Apart from practical advantages, due to the separable nature of the interaction
in analytical calculations, or to the absence of sign problem in the Quantum Monte Carlo
methods, is there a true physical advantage in using lattice models~?

One may argue for example that everywhere non-positive interaction potentials may be used
in continuous space, such as a square well potential, with a range dependent depth $V_0(b)$ adjusted
to have a fixed non-zero scattering and no two-body bound states. E.g.\ for a square well
potential $V(\rr)=-V_0 \theta(b-r)$, where $\theta(x)$ is the Heaviside function, one simply
has to take
\be
V_0 = \frac{\hbar^2}{m b^2} \left(\frac{\pi}{2}\right)^2
\ee
to have an infinite scattering length. 
For such an attractive interaction, it seems then that one can easily reproduce the reasonings leading
to the bounds Eqs.(\ref{eq:bounde},\ref{eq:boundf}). It is known however that there exists
a number of particles $N$, in the unpolarized case $N_\uparrow=N_\downarrow$,
such that this model in free space has a $N$-body bound state, necessarily of energy $\propto -\hbar^2/(m b^2)$ 
\cite{Blatt,Panda,Baym}.
In the thermodynamic limit, the unitary gas is thus not the ground phase of the system, it is at most a metastable
phase, and this prevents a derivation of the bounds Eqs.(\ref{eq:bounde},\ref{eq:boundf}).
This catastrophe is easy to predict variationally, taking as a trial wavefunction the ground state
of the ideal Fermi gas enclosed in a fictitious cubic hard wall
cavity of size $b/\sqrt{3}$ \cite{theseFelix}. 
In the large $N$ limit, the kinetic energy in the trial wavefunction is then $(3N/5) \hbar^2 k_F^2/(2m)$,
see Eq.(\ref{eq:e0}), where the Fermi wavevector is given by Eq.(\ref{eq:kf}) with
a density $\rho=N/(b/\sqrt{3})^3$, so that 
\be
E_{\rm kin} \propto N^{5/3}\frac{\hbar^2}{m b^2}.
\ee
Since all particles are separated by a distance less than $b$,
the interaction energy is exactly
\be
E_{\rm int} = -V_0 (N/2)^2
\ee
and wins over the kinetic energy for $N$ large enough, $2800 \lesssim N$ for the considered ansatz.
Obviously, a similar reasoning leads to the same conclusion for an everywhere negative, non-necessarily square
well interaction potential \footnote{In fixed node calculations, an everywhere negative interaction
potential is used \cite{Carlson,Panda,Giorgini}. It is unknown if $N$ in these simulations
exceeds the minimal value required to have a bound state. Note that the imposed nodal wavefunction in
the fixed node method, usually the one of the Hartree-Fock or BCS state, would be however quite different from
the one of the bound state.}.
One could imagine to suppress this problem by introducing a hard core repulsion, in which case
however the purely attractive nature of $V$ would be lost, ruining our simple derivation
of Eqs.(\ref{eq:bounde},\ref{eq:boundf}).

The lattice models
are immune to this catastrophic variational
argument, since one cannot put more than two spin $1/2$ fermions ``inside" the interaction potential, that is
on the same lattice site. Still they preserve the purely attractive nature of the interaction.
This does not prove however that their spectrum is bounded from below in the zero range limit, as pointed
out in the introduction of this section.


\subsection{Zero-range model, scale invariance and virial theorem}
\label{subsec:bp}

\subsubsection{The zero-range model} \label{subsubsec:ZRM}
The interactions are here replaced with contact conditions on the $N$-body
wavefunction.  In the two-body case, the model, introduced
already by Eq.(\ref{eq:bp}), is discussed in details in the literature,
see e.g.\ \cite{HouchesCastin99} in free space
where the scattering amplitude $f_k$
is calculated and the existence for $a>0$ of a dimer of energy
$-\hbar^2/(2\mu a^2)$ and wavefunction $\phi_0(r) =
(4\pi a)^{-1/2} \exp(-r/a)/r$ is discussed, $\mu$ being the reduced mass
of the two particles.
The two-body trapped case, solved in \cite{Wilkens}, was already
presented in subsection \ref{subsec:aor}. Here we present the model
for an arbitrary value of $N$.

For simplicity, we consider in first quantized form
the case of a fixed number $N_\uparrow$  of
fermions in spin state $\uparrow$ and a fixed number $N_\downarrow$ of
fermions in spin state $\downarrow$, assuming that the Hamiltonian
cannot change the spin state. We project the $N$-body 
state vector $|\Psi\rangle$ onto the non-symmetrized spin state with the 
$N_\uparrow$ first particles in spin state $\uparrow$
and the $N_\downarrow$ remaining particles in spin state $\downarrow$,
to define a scalar $N$-body wavefunction:
\be
\label{eq:def_psi}
\psi(\XX) \equiv \left(\frac{N!}{N_\uparrow! N_\downarrow!}\right)^{1/2}
\langle \uparrow, \rr_1| \otimes
\ldots \langle \uparrow, \rr_{N_\uparrow}|
\langle \downarrow, \rr_{N_\uparrow+1}| \otimes \ldots
\langle \downarrow, \rr_{N}| \Psi\rangle
\ee
where $\XX=(\rr_1,\ldots,\rr_N)$ is the set of all coordinates, 
and the normalization factor ensures that $\psi$ is normalized to unity
\footnote{
The inverse formula giving the full state vector
in terms of $\psi(\XX)$ is
$|\Psi\rangle = \left(\frac{N!}{N_\uparrow!N_{\downarrow}!}\right)^{1/2}
A |\uparrow\rangle^{N_\uparrow} 
|\downarrow\rangle^{N_\downarrow} |\psi\rangle$,
where the projector $A$ is the usual antisymmetrizing operator
$A=(1/N!) \sum_{\sigma\in S_N} \epsilon(\sigma) P_\sigma$.
}.
The fermionic symmetry of the state vector allows to express the wavefunction
on another spin state (with any different order of $\uparrow$
and $\downarrow$ factors) in terms of $\psi$.  For the considered
spin state, this fermionic symmetry imposes that $\psi$ is odd
under any permutation of the first $N_{\uparrow}$ positions
$\rr_1,\ldots,\rr_{N_\uparrow}$,
and also odd under any permutation of the last $N_{\downarrow}$ 
positions $\rr_{N_\uparrow+1},\ldots,\rr_{N}$.

In the Wigner-Bethe-Peierls model, that we also call zero-range model,
the Hamiltonian for the wavefunction $\psi$ is simply represented
by the same partial differential operator as for the ideal gas case:
\be
H =\sum_{i=1}^{N} \left[ -\frac{\hbar^2}{2m} \Delta_{\rr_i}
+U(\rr_i)\right],
\label{eq:hamil}
\ee
where $U$ is the external trapping potential supposed for 
simplicity to be spin state independent.
As is however well emphasized in the mathematics of operators
on Hilbert spaces \cite{analyse_spectrale}, an operator is defined
not only by a partial differential operator, but also by the choice
of its so-called {\sl domain} $D(H)$. A naive presentation
of this concept of domain is given in the Appendix~2. 
Here the domain does not coincide with the ideal gas one.
It includes the following Wigner-Bethe-Peierls contact conditions:
For any pair of particles $i,j$, when $r_{ij}\equiv |\rr_i-\rr_j|\to 0$
for a fixed position of their centroid $\RR_{ij}=(\rr_i+\rr_j)/2$,
there exists a function $A_{ij}$ such that
\be
\label{eq:bpN}
\psi(\XX) = A_{ij} (\RR_{ij}; (\rr_k)_{k\neq i,j})
(r_{ij}^{-1}-a^{-1}) + O(r_{ij}).
\ee
These conditions are imposed for all values of $\RR_{ij}$ different
from the positions of the other particles $\rr_k$, $k$ different from
$i$ and $j$. If the fermionic particles $i$ and $j$ are in the same
spin state, the fermionic symmetry imposes $\psi(\ldots, \rr_i=\rr_j,\ldots)=0$
and one has simply $A_{ij}\equiv 0$. 
For $i$ and $j$ in different spin states,
the unknown functions $A_{ij}$ have to be determined
from Schr\"odinger's equation, e.g.\ together with the
energy $E$ from the eigenvalue problem 
\be
\label{eq:ep}
H \psi = E \psi.
\ee
Note that in Eq.(\ref{eq:ep}) we have excluded the values of $\XX$
where two particle positions coincide. Since
$\Delta_{\rr_i} r_{ij}^{-1} = - 4 \pi \delta(\rr_i-\rr_j)$,
including these values
would require a calculation with distributions rather than with functions,
with regularized delta interaction pseudo-potential, which is
a compact and sometimes useful reformulation of the Wigner-Bethe-Peierls
contact conditions \cite{Petrov,HouchesCastin99,Huang,CRASCastin}.

As already pointed out below Eq.(\ref{eq:bpN}), $A_{ij}\equiv 0$ 
if $i$ and $j$ are fermions in the same spin state. One may wonder 
if solutions exist such that $A_{ij}\equiv 0$ even if $i$ and $j$
are in different spin states, in which case $\psi$ would
simply vanish when $r_{ij}\to 0$. 
These solutions would then be common eigenstates
to the interacting gas and to the ideal gas. They would correspond
in a real experiment to long lived eigenstates, protected from three-body
losses by the fact that $\psi$ vanishes when two particles or more
approach each other.
In a harmonic trap, one can easily  construct such ``non-interacting"
solutions, as for example
the famous Laughlin wavefunction of the Fractional Quantum Hall Effect.
``Non-interacting" solutions also exists for spinless bosons.
These non-interacting states actually dominate the ideal gas density of states at high energy
\cite{WernerPRL,theseFelix}.

\subsubsection{What is the kinetic energy?}
The fact that the Hamiltonian is the same as the ideal gas, apart from
the domain, may lead physically to some puzzles. E.g.\ the absence of
interaction term may give the impression that the energy $E$ is the
sum of trapping potential energy and kinetic energy only.
This is actually not so. The correct definition of the mean kinetic energy,
valid for general boundary conditions on the wavefunction, is
\be
E_{\rm kin} = \int d^{3N}X\, \frac{\hbar^2}{2m} 
|\partial_\XX \psi|^2.
\ee
This expression in particular guaranties that $E_{\rm kin}\geq 0$.
If $A_{ij}\neq 0$ in Eq.(\ref{eq:bpN}), one then sees that, although
$\psi$ is square integrable in a vicinity of $r_{ij}=0$
thanks to the Jacobian $\propto r_{ij}^2$ coming from three-dimensional
integration, the gradient of $\psi$ diverges as $1/r_{ij}^2$
and cannot be square integrable. Within the zero-range model
one then obtains an infinite kinetic energy
\be
E_{\rm kin}^{\rm WBP}=+\infty.
\ee
Multiplying Eq.(\ref{eq:ep}) by $\psi$ and integrating over $\XX$,
one realizes that the total energy is split as the trapping
potential energy,
\be
E_{\rm trap} = \int d^{3N}X\, |\psi(\XX)|^2 \sum_{i=1}^{N} U(\rr_i)
\ee
and as the sum of kinetic plus interaction energy:
\be
E_{\rm kin} + E_{\rm int} = -\int d^{3N}X\, \frac{\hbar^2}{2m}
\psi^* \Delta_\XX \psi.
\ee
This means that the interaction energy is $-\infty$ in the Wigner-Bethe-Peierls
model.  All this means is that, in reality, 
when the interaction has a non-zero range,
both the kinetic energy and the interaction energy
of interacting particles depend on the interaction range $b$,
and diverge for $b\to 0$, in such a way however that the sum 
$E_{\rm kin} + E_{\rm int}$ has a finite limit given by the Wigner-Bethe-Peierls
model.  
We have seen more precisely how this happens for lattice models in section \ref{subsubsec:tail},
see~the expression~(\ref{eq:E_int}) of $E_{\rm int}$ and the subsequent derivation of~(\ref{eq:fonctionelle}).\footnote{For a continuous-space model with an interaction potential $V(r)$,
we have~\cite{ZhangLeggett,Tangen} $E_{\rm int}=\frac{C}{(4\pi)^2}\int d^3r\,V(r) |\phi(r)|^2$ where $C$ is still defined by~(\ref{eq:defC}) and $\phi(r)$ still denotes the zero-energy two-body scattering state normalised according to~(\ref{eq:phi_norm}).}

\subsubsection{Scale invariance and virial theorem}

In the case of the unitary gas, the scattering length is infinite,
so that one sets $1/a=0$ in Eq.~(\ref{eq:bpN}). The domain of the Hamiltonian
is then imposed to be invariant by any isotropic rescaling Eq.(\ref{eq:sca_psi})
of the particle positions. To be precise, we define
for any scaling factor $\lambda >0$:
\be
\label{eq:sctr}
\psi_\lambda(\XX) \equiv \frac{\psi(\XX/\lambda)}{\lambda^{3N/2}},
\ee
and we impose that $\psi_\lambda\in D(H)$ for all $\psi\in D(H)$.
This is the precise mathematical definition of the scale invariance
loosely introduced in subsection \ref{subsec:ssfp}.
In particular, it is apparent in Eq.(\ref{eq:bpN})
that, for $1/a=0$, $\psi_\lambda$ obeys the Wigner-Bethe-Peierls contact 
conditions if $\psi$ does. On the contrary, if $\psi$ obeys 
the contact conditions for a finite scattering length $a$, $\psi_\lambda$ obeys
the contact condition for a different, fictitious
scattering length $a_\lambda =\lambda a \neq a$
and $D(H)$ cannot be scaling invariant.

There are several consequences of the scale invariance of the domain of the Hamiltonian $D(H)$
for the unitary gas. Some of them were presented in subsection
\ref{subsec:ssfp}, other ones will be derived in section \ref{sec:ds}.
Here we present another application, the derivation of a virial
theorem for the unitary gas. This is a first step towards 
the introduction of a SO(2,1) Lie algebra in section \ref{sec:ds}.
To this end, we introduce the infinitesimal generator $D$
of the scaling transform Eq.(\ref{eq:sctr}), such that
\footnote{
The scaling transform (\ref{eq:sctr}) defines
a unitary operator $T(\lambda)$ such that $\psi_\lambda = T(\lambda) \psi$.
One has $T(\lambda_1) T(\lambda_2) = T(\lambda_1 \lambda_2)$. 
To recover the usual additive structure as for the group of spatial
translations, one sets $\lambda=\exp\theta$, so that $T(\theta_1)
T(\theta_2)=T(\theta_1+\theta_2)$ and $T(\theta)=\exp(-i\theta D)$
where $D$ is the generator. This is why $\ln\lambda$ appears in
Eq.(\ref{eq:itod}).
}
\be
\label{eq:itod}
\psi_\lambda(\XX) = e^{-i D \ln \lambda} \psi(\XX).
\ee
Taking the derivative of Eq.(\ref{eq:sctr}) with respect to
$\lambda$ in $\lambda=1$, one obtains the hermitian operator
\be
\label{eq:defD}
D = \frac{1}{2i} (\XX\cdot \partial_\XX + \partial_\XX \cdot \XX)=
\frac{3N}{2i} -i \XX\cdot \partial_\XX.
\ee
The commutator of $D$ with the Hamiltonian is readily obtained.
From the relation
$\Delta_{\XX} \psi_\lambda(\XX) = \lambda^{-2}
(\Delta \psi)(\XX/\lambda)$, one has
\be
e^{iD\ln \lambda} (H-H_{\rm trap}) e^{-iD\ln \lambda} =
\frac{1}{\lambda^2} (H-H_{\rm trap})
\ee
where $H_{\rm trap}=\sum_{i=1}^{N} U(\rr_i)$ is the trapping potential
part of the Hamiltonian. It remains to take the derivative in $\lambda=1$ to
obtain
\be
\label{eq:comm1}
i [D, H-H_{\rm trap}] = - 2 (H-H_{\rm trap}).
\ee
The commutator of $D$ with the trapping potential is evaluated directly from
Eq.(\ref{eq:defD}):
\be
\label{eq:comm2}
i[D, H_{\rm trap}] = \sum_{i=1}^{N} \rr_i\cdot \partial_{\rr_i}U(\rr_i).
\ee
This gives finally
\be
\label{eq:comm}
i [D,H] = - 2 (H-H_{\rm trap}) + \sum_{i=1}^{N} \rr_i\cdot \partial_{\rr_i}U(\rr_i).
\ee
The standard way to derive the virial theorem in quantum mechanics
\cite{virial}, in a direct generalization of the one of classical
mechanics,  is then to take the expectation value of $[D,H]$ in an eigenstate
$\psi$ of $H$ of eigenenergy $E$. 
This works here for the unitary gas because the domain $D(H)$
is preserved by the action of $D$.
On one side, by having $H$ acting on $\psi$ from the right
or from the left, one trivially has $\langle [D,H]\rangle_\psi=0$.
On the other side, one has Eq.(\ref{eq:comm}), so that
\be
\label{eq:virth}
E = \sum_{i=1}^{N} \langle U(\rr_i) + 
\frac{1}{2} \rr_i\cdot \partial_{\rr_i}U(\rr_i) \rangle_\psi.
\ee
This relation was obtained with alternative derivations in the literature
(see \cite{Thomas_viriel_demo} and references therein).
One of its practical interests is that it gives access to the energy
from the gas density distribution \cite{ThomasVirial}.
As already mentioned, the scale invariance of the domain of $H$
is crucial to obtain this result. If $1/a$ is non zero,
a generalization of the virial relation can however
be obtained, that involves $dE/d(1/a)$, see \cite{TanVirial,WernerVirial}.

\subsection{Two-channel model and closed-channel fraction}
\label{subsec:tcm}

\subsubsection{The two-channel model}
\label{subsubsec:tcmdes}

The lattice models or the zero-range model are of course dramatic simplifications
of the real interaction between two alkali atoms.
At large interatomic distances, much larger than the radius of the electronic orbitals,
one may hope to realistically represent this interaction by a function $V(r)$ of the interatomic
distance, with a van der Waals attractive tail $V(r) \simeq -C_6/r^6$,
a simple formula that actually neglects retardation effects and long-range magnetic dipole-dipole
interactions.
As discussed below the gas phase condition Eq.(\ref{eq:gaz}), this allows to estimate
$b$ with the so-called van der Waals length, usually in the range of 1-10 nm.

At short interatomic distances, this simple picture of a scalar interaction potential $V(r)$
has to be abandoned. Following quantum chemistry or molecular physics methods, one has to
introduce the various Born-Oppenheimer potential curves obtained from the solution of the electronic
eigenvalue problem for fixed atomic nuclei positions. Restricting to one active electron of spin $1/2$
per atom, one immediately gets two ground potential curves, the singlet one
corresponding to the total spin $S=0$, 
and the triplet one corresponding to the total spin $S=1$. 
An external magnetic field $B$ is applied to activate the Feshbach resonance.
This magnetic field couples mainly to the total electronic spin and thus induces
{\sl different} Zeeman shifts for the singlet and triplet curves.
In reality, the problem is further complicated by the existence of the nuclear spin and the hyperfine
coupling, that couples the singlet channel to the triplet channel for nearby atoms, 
and that induces a hyperfine splitting within the ground electronic state for well separated atoms.

\begin{figure}[b]
\sidecaption
\includegraphics[width=7cm,clip=]{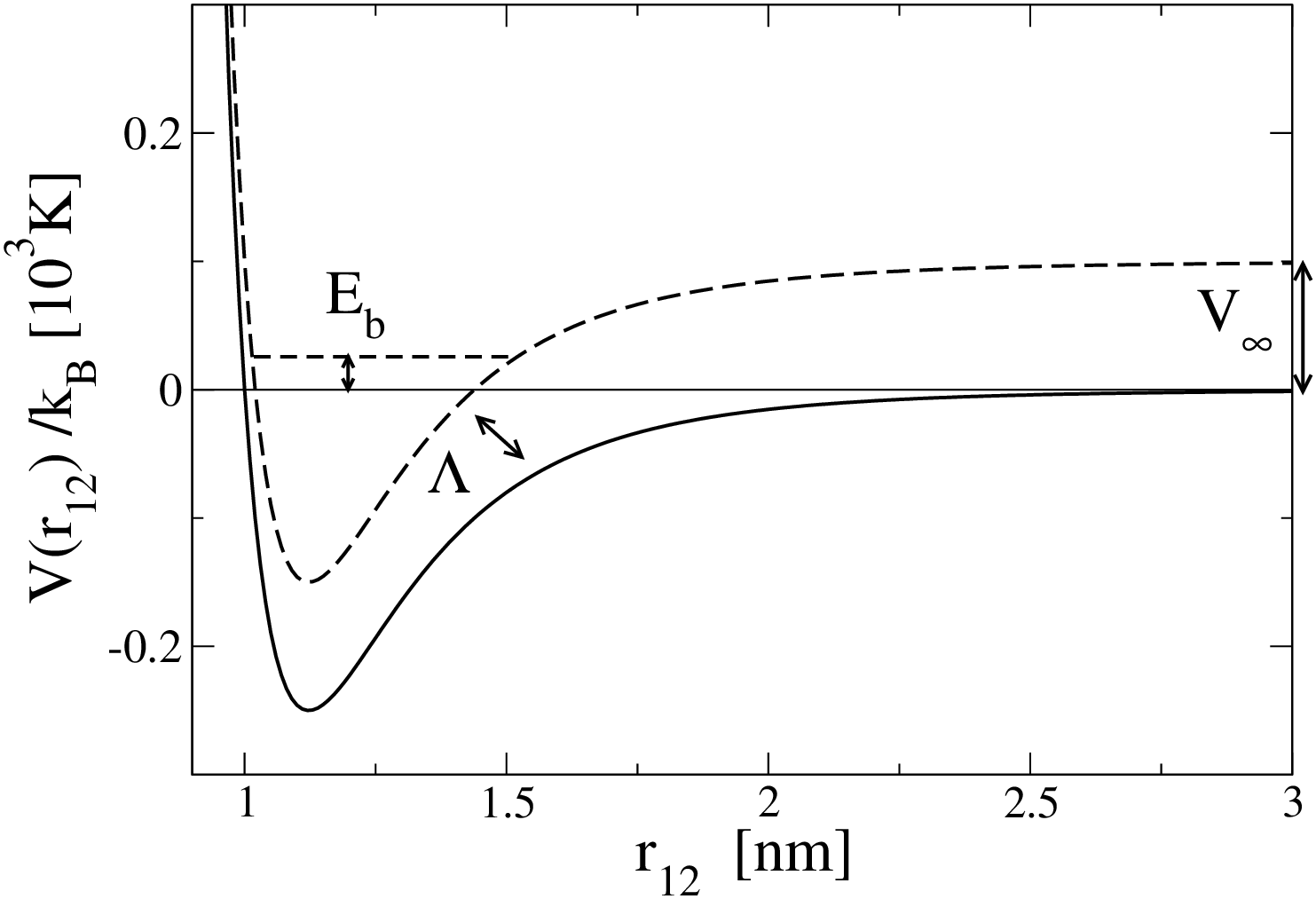}
\caption{Simple view of a Feshbach resonance.
The atomic interaction is described by two curves (solid line:
open channel, dashed line: closed channel).
When one neglects the interchannel coupling $\Lambda$, the closed channel
has a molecular state of energy $E_b$ close to the dissociation
limit of the open channel. The energy spacing $V_\infty$
was greatly exaggerated, for clarity.}
\label{fig:feshbach}
\end{figure}

A detailed discussion is given e.g. in \cite{revue_feshbach,Timm}. Here we take
the simplified view depicted in Fig.\ref{fig:feshbach}: The atoms
interact {\sl via} two potential curves, $V_{\rm open}(r)$ and
$V_{\rm closed}(r)$. 
At large distances, $V_{\rm open}(r)$ conventionally tends to zero,
whereas $V_{\rm closed}(r)$ tends to a positive value $V_\infty$, one of
the hyperfine energy level spacings for a single atom in the applied magnetic field.
In the two-body scattering problem, the atoms come from $r=+\infty$
in the internal state corresponding to $V_{\rm open}(r)$, the so-called open channel,
with a kinetic energy $E\ll V_\infty$. 
Due to a coupling between the two channels, 
the two interacting atoms can have access to the internal state corresponding
to the curve $V_{\rm closed}(r)$, but only at short distances; at long distances,
the external atomic wavefunction in this so-called closed channel
is an evanescent wave that decays exponentially with $r$ since $E<V_{\infty}$.

Now assume that, in the absence of coupling between the channels, the closed channel
supports a bound state of energy $E_b$, called in what follows {\sl the molecular state}, or {\sl the closed-channel molecule}.
Assume also that, by applying a judicious magnetic field, one tunes the energy of
this molecular state close to zero, that is to the dissociation limit of the
open channel. In this case one may expect that the scattering amplitude of
two atoms is strongly affected, by a resonance effect, given the non-zero coupling
between the two channels. This is in essence how the Feshbach resonance takes place.

The central postulate of the theory of quantum gases is that the short range details
of the interaction are unimportant, only the low-momentum scattering amplitude $f_k$ between
two atoms is relevant. As a consequence, any simplified model for the interaction, leading
to a different scattering amplitude $f_k^{\rm model}$, is acceptable provided that
\be
\label{eq:postu}
f_k^{\rm model} \simeq f_k
\ee
for the relevant values of the relative momentum $k$ populated in the gas.
We insist here that we impose similar scattering amplitudes over some momentum range, not just
equal scattering lengths $a$. For spin 1/2 fermions, typical values of $k$ can be
\be
k_{\rm typ}\in \{a^{-1}, k_F, \lambda_{\rm dB}^{-1}\}
\ee
where the Fermi momentum is defined in Eq.(\ref{eq:kf}) and the thermal de Broglie
wavelength in Eq.(\ref{eq:ldb}).
The appropriate value of $k_{\rm typ}$ depends on the physical
situation. The first choice $k_{\rm typ}\sim a^{-1}$ is well suited to the case
of a condensate of dimers ($a>0$) since it is the relative momentum of two atoms forming
the dimer. The second choice $k_{\rm typ}\sim k_F$ is well suited to a degenerate
Fermi gas of atoms (not dimers). The third choice $k_{\rm typ} \sim \lambda^{-1}$
is relevant for a non-degenerate Fermi gas.

The strategy is thus to perform an accurate calculation of the ``true" $f_k$,
to identify the validity conditions of the simple models and of the unitary regime
assumption Eq.(\ref{eq:fku}).
One needs a realistic, though analytically tractable, model of the Feshbach resonance.
This is provided by the so-called {\sl two-channel} models \cite{Timm,Holland0,Holland}.
We use here the version presented in \cite{Tarruell}, which is a particular case of the one used 
in~\cite{revue_feshbach,KoehlerBurnett} and Refs. therein:
The open channel part consists of the original gas of spin $1/2$ fermions that interact
{\sl via} a separable potential, that is in first quantized form for two opposite spin fermions,
in position space:
\be
\langle \rr_1, \rr_2 | V_{\rm sep} | \rr_1',\rr_2'\rangle =
\delta\left(\frac{\rr_1+\rr_2}{2}-\frac{\rr_1'+\rr_2'}{2}\right)
g_0 \chi(\rr_2-\rr_1)\chi(\rr_2'-\rr_1').
\ee
This potential does not affect the atomic center of mass, so it conserves total momentum
and respects Galilean invariance. Its matrix element involves the product of a function of the relative position
in the ket and of the same function of the relative position in the bra, hence the name {\sl separable}.
The separable potential is thus in general non local. As we shall take a function $\chi$ of width $\approx b$
this is clearly not an issue. The coupling constant $g_0$ of the separable potential
is well-defined by the normalization condition for $\chi$, $\int d^3r \, \chi(\rr)=1$.
In the presence of this open channel interaction only, the scattering length between fermions, 
the so-called background scattering length $a_{\rm bg}$, is usually small, of the order of the potential
range $b$, hence the necessity of the Feshbach resonance to reach the unitary limit.

In the closed channel part, a single two-particle state is kept, the one corresponding to the 
molecular state, of energy $E_b$ and of spatial range $\lesssim b$. The atoms thus exist in that channel
not in the form of spin $1/2$ fermions, but in the form of bosonic spinless molecules, of mass twice
the atomic mass.
The coupling between the two channels simply corresponds to the possibility for each boson
to decay in a pair of opposite spin fermions, or the inverse process that two opposite spin fermions
merge into a boson, in a way conserving the total momentum. This coherent Bose-Fermi conversion 
may take place only if the positions $\rr_1$ and $\rr_2$ of the two fermions are within a distance $b$, and is thus described
by a relative position dependent amplitude $\Lambda \chi(\rr_1-\rr_2)$, where for simplicity one takes
the same cut-off function $\chi$ as in the separable potential.
It is important to realize that the Bose-Fermi conversion effectively induces an interaction
between the fermions, which becomes resonant for the right tuning of $E_b$ and leads to
the diverging total scattering length $a$.

The model is best summarized in second quantized form \cite{Tarruell}, introducing the fermionic field operators
$\psi_\sigma(\rr)$, $\sigma=\uparrow,\downarrow$,
obeying the usual fermionic anticommutation relations, and the bosonic field operator 
$\psi_b(\rr)$ obeying the usual bosonic commutation relations:
\bea
\label{eq:definition_H2}
&& H = \int d^3r \left[\sum_{\sigma=\uparrow,\downarrow}
 \psi_\sigma^\dagger\left(-\frac{\hbar^2}{2m}\Delta_\rr + U\right) \psi_\sigma +
\psi_b^\dagger\left(E_b-\frac{\hbar^2}{4m}\Delta_\rr + U_b\right) \psi_b
\right]  \\
&& +
\Lambda \int d^3r_1 d^3r_2\, \chi(\rr_1-\rr_2)
\left\{
\psi_b^\dagger[(\rr_1+\rr_2)/2]
\psi_\downarrow(\rr_1) \psi_\uparrow(\rr_2)
+ {\rm h. c.}
\right\}
 \nonumber \\
&& + g_0 \int d^3\!R\, d^3\!r\, d^3\!r'
\chi(\rr) \chi (\rr') 
\psi_\uparrow^\dagger(\RR-\rr/2) \psi_\downarrow^\dagger(\RR+\rr/2)
\psi_\downarrow(\RR+\rr'/2) \psi_\uparrow(\RR-\rr'/2),
\nonumber
\eea
where $U(\rr)$ and $U_b(\rr)$ are the trapping potentials for the fermions and the bosons, respectively.

\subsubsection{Scattering amplitude and universal regime}
\label{subsubsec:fk_2canaux}

In free space, the scattering problem of two fermions 
is exactly solvable
for a Gaussian cut-off function $\chi(\rr)\propto \exp[-r^2/(2 b^2)]$~ \cite{revue_feshbach,Tarruell}. A variety of parameterizations are possible.
To make contact with typical notations, we assume that the energy $E_b$ of the molecule in the closed
channel is an affine function of the magnetic field $B$, a reasonable assumption close to the Feshbach resonance:
\be
\label{eq:reasas}
E_b(B) = E_b^0 + \mu_b (B-B_0)
\ee
where $B_0$ is the magnetic field value right on resonance and $\mu_b$ is the effective magnetic moment 
of the molecule. Then the scattering length for the model Eq.(\ref{eq:definition_H2}) can be exactly
written as the celebrated formula
\be
\label{eq:celeb}
a = a_{\rm bg} \left(1-\frac{\Delta B}{B-B_0}\right),
\ee
where $\Delta B$, such that $E_b^0 + \mu_b \Delta B = \Lambda^2/g_0$,
is the so-called width of the Feshbach resonance. As expected, for $|B-B_0|\gg |\Delta B|$, one finds
that $a$ tends to the background scattering length $a_{\rm bg}$ solely due to the open channel interaction.
With $\Delta B$ one forms a length $R_*$ \cite{PetrovBosons} which is always non-negative:
\be
\label{eq:Ret}
R_* \equiv \frac{\hbar^2}{m a_{\rm bg} \mu_b \Delta B} = 
\left(\frac{\Lambda}{2\pi b E_b^0}\right)^2,
\ee
where the factor $2\pi$ is specific to our choice of $\chi$.
Physically, the length $R_*$ is also directly related to the effective range on resonance:
\be
r_e^{\rm res} = - 2 R_* + \frac{4b}{\sqrt{\pi}},
\ee
where the numerical coefficient in the last term depends on the choice of $\chi$.
The final result for the scattering amplitude for the model Eq.(\ref{eq:definition_H2}) is
\be
\label{eq:lerpf}
-\frac{1}{f_{k}} = ik + \frac{e^{k^2b^2}}{a}
\left[1-\left(1-\frac{a}{a_{\rm bg}}\right)\frac{k^2}{k^2-Q^2}\right]
-i k\, \mathrm{erf}\,(-i k b)
\ee
where erf is the error function, that vanishes linearly in zero, and the wavevector $Q$, such that
\be
Q^2 \equiv \frac{m}{\hbar^2 g_0} \left(g_0 E_b-\Lambda^2\right) 
= \frac{-1}{a_{\rm bg} R_*(1-a_{\rm bg}/a)},
\ee
may be real or purely imaginary.

The  unitary limit assumption Eq.(\ref{eq:fku}) implies that all the terms in the right hand side
of Eq.(\ref{eq:lerpf}) are negligible, except for the first one.
We now discuss this assumption, restricting for simplicity to an infinite scattering length  $a^{-1}=0$ (i.e. a magnetic field sufficiently close to resonance)
and a typical relative momentum $k_{\rm typ}=k_F$ (i.e. a degenerate gas).
To satisfy Eq.(\ref{eq:postu}), with $f_k^{\rm model}=-1/(ik)$, one should then have,
in addition to the gas phase requirement $k_F b\ll 1$,  that
\be
\label{eq:cond_valid}
\frac{k R_*}{|1+k^2 a_{\rm bg} R_*|} \ll 1 \ \ \forall k\in [0,k_F].
\ee
Table~\ref{tab:valid} summarizes the corresponding conditions to reach the unitary limit.\footnote{We discarded for simplicity the  rather peculiar case where $k_F \sqrt{|a_{\rm bg}| R_*}$ is $\le1$ but not $\ll1$.} \footnote{
An additional condition actually has to be imposed
to
have a universal gas, as
we will see after Eq.(\ref{eq:Nb_unitaire}).}
Remarkably, the condition $k_F |r_e^{\rm res}|\ll 1$ obtained in Eq.(\ref{eq:zrl})
from the expansion of $1/f_k$ to order $k^2$ is not 
 the end of the story.
In particular,
 if $a_{\rm bg} < 0$, $Q_{\rm res}^2\equiv-1/(a_{\rm bg} R_*)$ is positive and $1/f_k$ diverges 
for $k=Q_{\rm res}$; if the location of this divergence is within the Fermi sea, the unitary
limit is not reachable. This funny case however requires huge values of $R_* a_{\rm bg}$,
that is extremely small values of the resonance width $\Delta B$: 
\be
\label{eq:narrow}
| \mu_b \Delta B | \lesssim \frac{\hbar^2 k_F ^2}{2m}.
\ee
This corresponds to 
very narrow Feshbach resonances \cite{Gurarie}, 
whose experimental use requires a good control of the magnetic
field homogeneity and is more delicate.
Current experiments rather use broad Feshbach resonances such as on lithium 6,
where $r_e^{\rm res}=4.7$nm \cite{Strinati},
$a_{\rm bg}=-74$ nm, $R_*=0.027$nm \cite{Grimm}, leading to $1/(|a_{\rm bg}| R_*)^{1/2}
=700 (\mu\mathrm{m})^{-1}$ much larger than $k_F \approx$ a few $(\mu\mathrm{m})^{-1}$, so that the unitary
limit is indeed well reached.

\begin{table}
\caption{In the two-channel model, conditions deduced from Eq.(\ref{eq:cond_valid}) (supplementary to the gas phase
condition $k_F b \ll 1$) to reach the unitary limit
for a degenerate gas of spin 1/2 fermions of Fermi momentum $k_F$. It is assumed
that the magnetic field is tuned right on resonance, so that the scattering length is infinite.
The last column corresponds to narrow Feshbach resonances satisfying Eq.(\ref{eq:narrow}).}
\label{tab:valid}  
\begin{tabular}{p{3.1cm}p{4.1cm}p{4.1cm}}
\hline\noalign{\smallskip}
                  & 
                  $k_F \sqrt{|a_{\rm bg}| R_*} \ll 1$  &
                  $k_F \sqrt{|a_{\rm bg}| R_*}> 1$ \\
\noalign{\smallskip}\svhline\noalign{\smallskip}
$a_{\rm bg} > 0$  & $k_F R_* \ll 1$ & 
$(R_*/a_{\rm bg})^{1/2} \ll 1$
\\
$a_{\rm bg} < 0$  & $k_F R_* \ll 1$ & unreachable \\
\noalign{\smallskip}\hline\noalign{\smallskip}
\end{tabular}
\end{table}

\subsubsection{Relation between number of closed channel molecules and ``contact''}
\label{subsubsec:nccm}

The fact that the two-channel model includes the underlying atomic physics of the Feshbach resonance
allows to consider an observable that is simply absent from
 single channel models, 
namely
the number of molecules in the closed channel, represented by the operator:
\be
N_b \equiv \int d^3\rr\ \psi_b^\dagger(\rr) \psi_b(\rr)
\ee
where $\psi_b$ is the molecular field operator. The mean number $\langle N_b\rangle$ of closed
channel molecules was recently measured by laser molecular excitation techniques \cite{Hulet_mol}.

This mean number can be calculated from a two-channel model by a direct application of
the Hellmann-Feynman theorem~\cite{BraatenLong,Tarruell}
(see also~\cite{ZhangLeggett}). The key point is that the only quantity depending on the magnetic
field in the Hamiltonian Eq.(\ref{eq:definition_H2}) is the internal energy $E_b(B)$ of a closed
channel molecule. 
At thermal equilibrium in the canonical ensemble,
 we thus have
\be
\label{eq:hell}
\left(\frac{dE}{dB}\right)_{\!S}= \langle N_b\rangle \frac{dE_b}{dB}.
\ee
Close to the Feshbach resonance, we may assume that $E_b$ is an affine function of $B$, see Eq.(\ref{eq:reasas}),
so that the scattering length $a$ depends on the magnetic field as in Eq.(\ref{eq:celeb}).
Parameterizing $E$ in terms of the inverse scattering length rather than $B$, we can replace
$dE/dB$ by $dE/d(1/a)$ times $d(1/a)/dB$.
The latter can be calculated explicitly from (\ref{eq:celeb}). Thus
\be
\langle N_b\rangle = \frac{C}{4\pi} R_* \left(1-\frac{a_{\rm bg}}{a}\right)^2,
\label{eq:CvsNb}
\ee
where 
$C$
is the contact defined in~Eq.(\ref{eq:defC}),
and
we introduced the length $R_*$ defined in Eq.(\ref{eq:Ret}).

If the interacting gas is in the universal zero range regime, its energy $E$ depends on the interactions only via the scattering length,
independently of the microscopic details of the atomic interactions, and its dependence with $1/a$ may
be calculated by any convenient model. Then, at zero temperature, for the unpolarized case $N_\uparrow
=N_\downarrow$,
the equation of state of the homogeneous gas can be expressed as
\be
e_0 = e_0^{\rm ideal} f\left(\frac{1}{k_F a}\right),
\label{eq:EOS_T=0}
\ee
where $e_0$ and $e_0^{\rm ideal}$ are the ground state energy per particle for the interacting gas
and for the ideal gas with the same density,
and the Fermi wavevector $k_F$ was defined in Eq.(\ref{eq:kf}).
In particular, $f(0)=\xi$, where the number
$\xi$ was introduced in Eq.(\ref{eq:e0}).
Setting $\zeta\equiv -f'(0)$,
we have for the homogeneous unitary gas
\be
\frac{C^{\rm hom}}{V}  =  \zeta\frac{2}{5\pi}k_F^4,
\label{eq:C_hom_vs_zeta}
\ee
so that
\be
\frac{\langle N_b\rangle^{\rm hom}}{N} = \frac{3}{10} k_F R_* \zeta.
\label{eq:Nb_unitaire}
\ee
This expression is valid for a universal gas consisting mainly of fermionic atoms, which requires that $\langle N_b\rangle^{\rm hom}/N\ll1$, i. e. $k_F R* \ll 1$. This condition was already obtained in \S\ref{subsubsec:fk_2canaux}
for the broad resonances of the left column of Table~\ref{tab:valid}.
In the more exotic case of the narrow resonances of the second column of Table~\ref{tab:valid}, this condition has to be imposed in addition to the ones of Table~\ref{tab:valid}.

\subsubsection{Application of general relations: Various measurements of the contact}

\begin{figure}[b]
\centerline{\includegraphics[width=9cm,clip=]{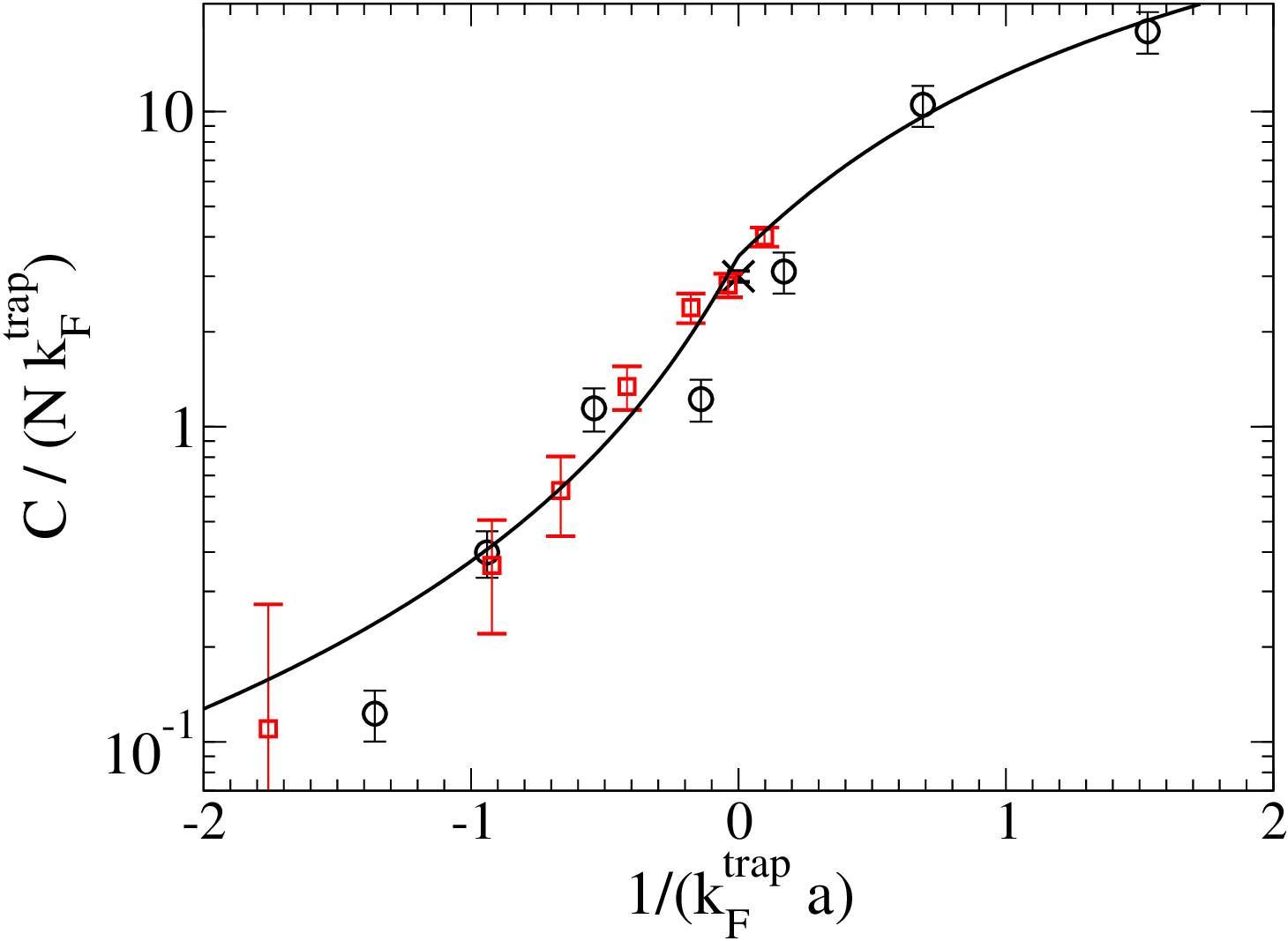}}
\caption{The contact $C=\frac{dE}{d(-1/a)}\frac{4\pi m}{\hbar^2}$ of a trapped unpolarized Fermi gas.
The circles are obtained from the measurements of
$\langle N_b\rangle$ in \cite{Hulet_mol}, combined with the two channel model theory linking
$\langle N_b\rangle$ to $C$ [Eq.(\ref{eq:CvsNb})].
The cross was obtained in~\cite{Australiens} by measuring the structure factor.
The squares were obtained in~\cite{Jin_tail} by measuring the momentum distribution.
Solid line: zero-temperature theoretical prediction extracted from \cite{Giorgini}
as detailed in \cite{Tarruell}. 
Here the Fermi wavevector $k_F^{\rm trap}$ of the trapped gas is defined by
$\hbar^2 (k_F^{\rm trap})^2/(2m)=(3N)^{1/3}\hbar \bar{\omega}$,
with $\bar{\omega}$ the geometric mean of the three oscillation frequencies
$\omega_\alpha$ and $N$ the total atom number.
}
\label{fig:hulet}
\end{figure}

The relation (\ref{eq:CvsNb}) allowed us to extract in~\cite{Tarruell} the contact $C$
of the trapped gas
[related to the derivative of the total energy of the trapped gas {\it via}~(\ref{eq:defC})]
from the values of $N_b$ measured in~\cite{Hulet_mol}.
The result is shown in Fig.\ref{fig:hulet}, together with a theoretical zero-temperature curve resulting from 
the local density approximation
in the harmonically trapped case 
where $U(\rr)=\frac{1}{2} m\sum_\alpha \omega_\alpha^2 x_\alpha^2$,
the function $f$ of~(\ref{eq:EOS_T=0}) being obtained by interpolating between
the fixed-node Monte-Carlo data of~\cite{Giorgini, LoboGiorgini}
and the known asymptotic expressions in the BCS and BEC limits\footnote{See~\cite{Tarruell} for details. The cusp at unitarity is of course an artefact of this 
interpolation procedure.}.

While
this is the 
first direct measurements of the contact in the BEC-BCS crossover,
it has also been measured more recently:
\begin{itemize}
\item
using Bragg scattering, {\it via} the
large-momentum tail of the structure factor, directly related
by Fourier transformation to the 
short-distance singularity Eq.(\ref{eq:C_g2}) of the pair correlation function~\cite{Australiens}, see the cross at unitarity in Fig.\ref{fig:hulet}
\item
{\it via} the tail of the momentum distribution Eq.(\ref{eq:C_nk}) measured by abruptly turning off both trapping potential and interactions~\cite{Jin_tail}, see the squares in Fig.\ref{fig:hulet}
\item
{\it via} 
(momentum resolved)
radio-frequency spectroscopy~\cite{Jin_tail,BraatenChap}. 
\end{itemize}

For the homogeneous unitary gas,
the contact is conveniently expressed in terms of the dimensionless parameter $\zeta$ [see (\ref{eq:C_hom_vs_zeta})].
 The experimental value
$\zeta=0.91(5)$ was obtained
by measuring the equation of state of the homogeneous gas
(with the technique proposed in \cite{Ho_pascal})
and taking the derivative of the energy with respect to the inverse scattering length [Eq.(\ref{eq:defC})]~(see 
\cite{NavonEOS} and the contribution 
of F. Chevy and C. Salomon to this volume
).
From the fixed-node Monte-Carlo calculations, one gets
$\zeta\simeq 1$ by taking a derivative of the data of~\cite{Giorgini} for the function $f$, 
while the data of~\cite{LoboGiorgini} for the pair correlation function together with the relation~(\ref{eq:C_g2}) give $\zeta\simeq0.95$.\footnote{This value is also compatible with the data of~\cite{LoboGiorgini} for the one-body density matrix, whose short-range singular part is related by Fourier transformation to the large-$k$ tail of the momentum distribution~\cite{Tangen}.}

At unitarity, 
the contact $C$ of the trapped gas is directly related to the contact of the homogeneous gas, i.e. to $\zeta$:
Indeed the average over the trap can be done analytically within 
the local density approximation,
yielding \cite{Tarruell}
\be
\frac{C}{N k_F^{\rm trap}}=\frac{512}{175}\,\frac{\zeta}{\xi^{1/4}}.
\ee

In conclusion, the smallness of the interaction range leads to mathematical singularities; at first sight this may seem to complicate the problem as compared to other strongly interacting systems;
however these singularities are well understood and have a useful consequence:
the existence of exact relations resulting from the Hellmann-Feynman theorem
\cite{Lieb} and from properties of the Fourier transform \cite{Olshanii}.
In particular this provides a ``useful check on mutual consistency of various experiments'', as foreseen in~\cite{Leggett_IHP}.

\section{Dynamical symmetry of the unitary gas 
}
\label{sec:ds}

In this section, we present some remarkable properties
of the unitary gas,
 derived
from the zero-range model.
The starting point is that the time evolution of the gas
in a 
time dependent
isotropic harmonic trap
 may be expressed exactly
in terms of a gauge and scaling transform, see subsection \ref{subsec:tds}.
This implies the existence of a SO(2,1) dynamical (or hidden)
symmetry of the system, a formal property that we shall link to concrete
consequences, such as the existence of an exactly decoupled
bosonic degree of freedom (the breathing mode of the gas),
see subsection \ref{subsec:so21},
or the separability of the $N$-body wavefunction in hyperspherical
coordinates, see subsection \ref{subsec:separability},
 which holds both in an isotropic harmonic trap and in free space and
  has several important consequences
such as the analytical solution of the trapped three-body
problem, see subsection \ref{subsec:pcots}.
In subsec.~\ref{subsec:viscosity} we
use the existence of the undamped breathing mode
to rederive a remarkable property of the homogeneous unitary gas: its bulk viscosity vanishes.
Subsection~\ref{subsec:scaling_laws} concerns short-range scaling laws, which are related to the separability in hyperspherical coordinates, but hold for any scattering length and external potential.

\subsection{Scaling solution in a time-dependent trap}
\label{subsec:tds}

In this section, we shall assume that the trapping potential
$U(\rr)$ introduced in Eq.(\ref{eq:hamil}) is an isotropic harmonic
potential. Whereas the hypothesis of harmonicity
may be a good approximation in present experiments for small enough
atomic clouds, the isotropy is not granted and requires some 
experimental tuning that, to our knowledge, remains to be done.
On the other hand, we allow a general time dependence of the trap
curvature, so that Schr\"odinger's equation for the $N$-body
wavefunction defined in Eq.(\ref{eq:def_psi}) is
\be
\label{eq:tdse}
i\hbar \partial_t \psi(\XX,t) =
\left[-\frac{\hbar^2}{2m} \Delta_\XX + \frac{1}{2} m \omega^2(t) X^2
\right] \psi(\XX,t),
\ee
where we recall that $\XX$ is the set of all particle coordinates,
and $\omega(t)$ is the instantaneous angular oscillation frequency.
The interaction between particles is described by the contact 
conditions Eq.(\ref{eq:bpN}), written here for the unitary
gas, that is for $a^{-1}=0$:
\be
\label{eq:bpu}
\psi(\XX) = \frac{A_{ij} (\RR_{ij}; (\rr_k)_{k\neq i,j})}{r_{ij}}+O(r_{ij}).
\ee

Let us consider the particular case, quite relevant experimentally,
where the gas is initially at equilibrium in a static trap
$\omega(t=0)=\omega$. The gas is then in a statistical mixture of
stationary states, so we can assume that the initial $N$-body
wavefunction is an eigenstate of the Hamiltonian with energy $E$.
At $t>0$, the trap curvature is varied, which leads to an arbitrary time
dependent $\omega(t)$. In typical experiments, one either sets abruptly
$\omega(t)$ to zero, to perform a time of flight measurement, 
or one modulates $\omega(t)$ at some frequency to study 
the gas collective modes. Can we predict the evolution of the system~?
As shown in \cite{CRASCastin}, the answer is yes, as we now explain.

In the absence of interactions, it is well known \cite{scaling_ideal}
that $\psi(\XX,t)$ is deduced from the $t=0$ wavefunction by
a simple gauge plus scaling ansatz:
\be
\label{eq:ansatz}
\psi(\XX,t) = \frac{e^{i\theta(t)}}{\lambda^{3N/2}(t)}
\exp\left[\frac{im\dot\lambda(t)}{2\hbar\lambda(t)} X^2\right]
\psi(\XX/\lambda(t),0),
\ee
where $\dot\lambda(t)=d\lambda(t)/dt$.
At time $t=0$, one clearly has $\theta(0)=0$,
\be
\label{eq:inco}
\lambda(0)  =  1 \ \ \ \ \mbox{and} \ \ \ \dot\lambda(0) = 0.
\ee
Inserting this ansatz into Schr\"odinger's equation (\ref{eq:tdse}),
we obtain a Newton like equation of motion for $\lambda$:
\be
\label{eq:russe}
\ddot\lambda(t) = \frac{\omega^2}{\lambda^3(t)} -\omega^2(t) \lambda(t)
\ee
to be solved with the initial conditions (\ref{eq:inco}).
We recall that $\omega$ stands for the {\sl initial} angular
oscillation frequency.
The equation (\ref{eq:russe}) is well studied in the literature,
under the name of the Ermakov equation
\cite{eq_russe}, and is in particular amenable to a linear form:
One recognizes an equation for the distance to the origin
for a two-dimensional harmonic oscillator
of angular frequency $\omega(t)$, as 
obtained from Newton's equation and from the law of equal areas. 
In particular, if $\omega(t)=\omega_{\rm ct}$ is a constant over some time 
interval, $\lambda(t)$ oscillates
with a period $\pi/\omega_{\rm ct}$ over that time interval.

The global phase $\theta(t)$ is given by
\be
\label{eq:tmod}
\theta(t) = -\frac{E}{\hbar} \int_0^{t} \frac{dt'}{\lambda^2(t')}.
\ee
This suggests that $\theta$ still evolves at the stationary pace
$-E/\hbar$ provided that one introduces a modified time,
as done in \cite{Shlyapnikov} in a bosonic mean field context:
\be
\label{eq:modt}
\tau(t) = \int_0^{t} \frac{dt'}{\lambda^2(t')}.
\ee
We shall come back to this point below.

In presence of interactions, one has to check that the ansatz
(\ref{eq:ansatz}) obeys the contact conditions (\ref{eq:bpu}).
First, the ansatz includes a scaling transform. As discussed
in subsection \ref{subsec:bp}, this preserves the contact
conditions and the domain of the Hamiltonian
for the unitary gas. Second, the ansatz includes a quadratic gauge
transform. Turning back to the definition of the contact conditions,
we select an arbitrary pair of particles
$i$ and $j$ and we take the limit $r_{ij}\to 0$ for a fixed centroid
position $\RR_{ij}=(\rr_i+\rr_j)/2$. In the gauge factor,
the quantity $X^2 = \sum_{k=1}^{N} r_k^2$ appears.
The positions $\rr_k$ of the particles other than $i$ and $j$
are fixed. What matters is thus $r_i^2 + r_j^2$ that we rewrite as
\be
r_i^2 + r_j^2 = 2 R_{ij}^2 + \frac{1}{2} r_{ij}^2.
\ee
$R_{ij}$ is fixed. $r_{ij}$ varies but it appears squared in the gauge
transform, so that
\be
\label{eq:chrcbp}
\exp\left[\frac{im\dot\lambda(t)}{2\hbar\lambda(t)} r_{ij}^2/2\right]
\left[\frac{1}{r_{ij}} + O(r_{ij})\right]  = \frac{1}{r_{ij}} + O(r_{ij})
\ee
and the contact conditions are preserved by the gauge
transform, even if the scattering length $a$ was finite.

We thus conclude that the ansatz (\ref{eq:ansatz}) gives
the solution also for the unitary gas. This has interesting practical
consequences. For measurements in position space,
one has simple scaling relations,
not only for the mean density $\rho_\sigma(\rr,t)$
in each spin component $\sigma$:
\be
\rho_\sigma(\rr,t) = \frac{1}{\lambda^{3}(t)}\, \rho_\sigma(\rr/\lambda(t),0)
\ee
but also for higher order density correlation functions:
For example, the second order density correlation function
defined in terms of the fermionic field operators as
\be
g^{(2)}_{\sigma\sigma'}(\rr,\rr') \equiv
\langle \psi_\sigma^\dagger(\rr) \psi_{\sigma'}^\dagger(\rr')
\psi_{\sigma'}(\rr') \psi_\sigma(\rr)\rangle,
\ee
evolves in time according to the scaling
\be
g^{(2)}_{\sigma\sigma'}(\rr,\rr',t) = \frac{1}{\lambda^6(t)}
g^{(2)}_{\sigma\sigma'}(\rr/\lambda(t),\rr'/\lambda(t),0).
\ee
As a consequence, if one abruptly switches off the trapping potential
at $t=0^+$, the gas experiences a ballistic expansion with a 
scaling factor
\be
\lambda(t) = [1+\omega^2 t^2]^{1/2},
\ee
which acts as a perfect magnifying lens on the density distribution. 

For non-diagonal observables in position space, some information
is also obtained, with the gauge transform now contributing. 
E.g.\ the first order coherence function
\be
g^{(1)}_{\sigma\sigma} (\rr,\rr') \equiv
\langle \psi_{\sigma}^\dagger(\rr') \psi_\sigma(\rr) \rangle,
\ee
which is simply the matrix element of the one-body density operator
between $\langle \rr,\sigma|$ and $|\rr',\sigma\rangle$,
evolves according to
\be
g^{(1)}_{\sigma\sigma} (\rr,\rr',t)=
\frac{1}{\lambda^3(t)} 
\exp\left[\frac{im\dot\lambda(t)}{2\hbar\lambda(t)} (r^2-r'^{2})\right]
g^{(1)}_{\sigma\sigma}(\rr/\lambda(t),\rr'/\lambda(t),0).
\ee
The momentum distribution $n_\sigma(\kk)$ in the spin component $\sigma$
is the Fourier transform over $\rr-\rr'$ and the integral over
$(\rr+\rr')/2$ of the first order coherence function.
For a ballistic expansion, directly transposing to three dimensions
the result obtained in \cite{Gangardt} from a time dependent scaling
solution for the one-dimensional gas of impenetrable bosons,
one has that the momentum distribution 
of the ballistically expanding unitary gas is asymptotically
homothetic to the gas initial spatial density profile:
\be
\lim_{t\to +\infty} n_\sigma(\kk,t) = 
\left(\frac{2\pi\hbar}{m\omega}\right)^{3}
\rho_\sigma\left(\rr=\frac{\hbar\kk}{m\omega},0\right).
\ee

We emphasize that the above results hold for an arbitrary gas polarization,
that is for arbitrary numbers of particles in each of the two
spin states $\sigma=\uparrow,\downarrow$. If the initial state is
thermal, they hold whatever the value of the temperature, larger
or smaller than the critical temperature $T_c$.
These results however require the unitary limit (in particular
$|a|=+\infty$) and a perfect isotropy of the harmonic trap.
If the experimental goal is simply to have the ballistic expansion
as a perfect magnifying lens, these two requirements
remarkably may be removed, as shown in \cite{Lobo}, if one is ready
to impose an appropriate time dependence to the scattering length
$a(t)$ and to the trap aspect ratio, in which case
the ansatz (\ref{eq:ansatz}) holds at all times. In the particular case of
an isotropic trap, the procedure of \cite{Lobo} is straightforward
to explain: If $\psi(t=0)$ obeys the contact conditions with a
finite scattering length $a$, the ansatz
(\ref{eq:ansatz}) obeys the contact conditions for a scattering
length $\lambda(t)a$ so one simply has to adjust the actual scattering
length in a time dependent way:
\be
a(t) = \lambda(t) a
\ee
where $\lambda$ evolves according to Eq.(\ref{eq:russe}).
As shown in the next subsection,  
the time dependent solution in the unitary case,
apart from providing convenient scaling relations on the density,
is connected to several interesting intrinsic properties of the system, 
whereas the procedure of \cite{Lobo} does not imply such properties.

To be complete, we finally address the general case where
the initial wavefunction of the unitary gas is not necessarily
a stationary state but is arbitrary \cite{WernerPRA}.
Then the observables of the gas have in general a non-trivial time dependence,
even for a fixed trap curvature.
If the trap curvature is time dependent, we modify
the gauge plus scaling ansatz as follows:
\be
\psi(\XX,t) = \frac{1}{\lambda^{3N/2}(t)}
\exp\left[\frac{im\dot\lambda(t)}{2\hbar\lambda(t)} X^2\right]
\tilde{\psi}(\XX/\lambda(t),\tau(t)),
\ee
where $\tau(t)$ is the modified time introduced in Eq.(\ref{eq:tmod}),
$\lambda(t)$ evolves according to Eq.(\ref{eq:russe}) with the initial
conditions (\ref{eq:inco}),
and the time-dependent wavefunction $\tilde{\psi}$ coincides
with $\psi$ at time $t=\tau=0$ and obeys the unitary gas contact
conditions. Then this ansatz obeys the contact conditions.
When inserted in the time dependent Schr\"odinger equation
(\ref{eq:tdse}), it leads to a Schr\"odinger equation
for $\tilde{\psi}$ in the time independent external potential
fixed to the $t=0$ trap:
\be
i\hbar \partial_\tau \tilde{\psi}(\XX,\tau)=
\left[-\frac{\hbar^2}{2m} \Delta_\XX + \frac{1}{2} m \omega^2 X^2
\right] \tilde{\psi}(\XX,\tau).
\ee
The gauge plus scaling transform, and the redefinition of time,
have then totally cancelled the time dependence of the trap.
If the initial wavefunction is an eigenstate of energy $E$,
as was previously the case, one simple has
$\tilde{\psi}(\tau) = \exp(-i E\tau/\hbar) \psi(t=0)$
and one recovers the global phase factor in Eq.(\ref{eq:ansatz}).

\subsection{SO(2,1) dynamical symmetry and the decoupled breathing mode}
\label{subsec:so21}

As shown in \cite{Rosch} for a two-dimensional Bose gas with $1/r^2$ interactions, 
the existence of a scaling solution such as Eq.(\ref{eq:ansatz})
reflects a hidden symmetry of the Hamiltonian, the SO(2,1)
dynamical symmetry. Following \cite{WernerPRA},
we construct explicitly this dynamical symmetry
for the unitary gas and we show that it has interesting consequences
for the energy spectrum in a static isotropic harmonic trap.

Let us consider a gedankenexperiment: Starting from the unitary gas
is an energy eigenstate $\psi$, we modify in an infinitesimal
way the trap curvature during the time interval $[0,t_f]$, and for $t>t_f$ 
we restore the initial trap curvature, $\omega(t)=\omega(0)=\omega$.
Linearizing Eq.(\ref{eq:russe}) around $\lambda=1$ for $t>t_f$, 
we see that the resulting change in the scaling parameter $\lambda$ is
\be
\lambda(t)-1 = \epsilon \, e^{-2i\omega t} + \epsilon^* \ e^{2i\omega t}
+O(\epsilon^2)
\label{eq:mode_castin}
\ee
where $\epsilon$ is proportional to the infinitesimal curvature change.
Since $\lambda$ oscillates indefinitely at frequency $2\omega$, 
this shows the existence of an undamped mode of frequency $2\omega$.
This conclusion actually extends
to excitations during $[0,t_f]$ of arbitrarily large amplitudes, as noted
below Eq.(\ref{eq:russe}) \cite{Rosch}.

\begin{figure}[b]
\sidecaption
\includegraphics[width=4cm,clip=]{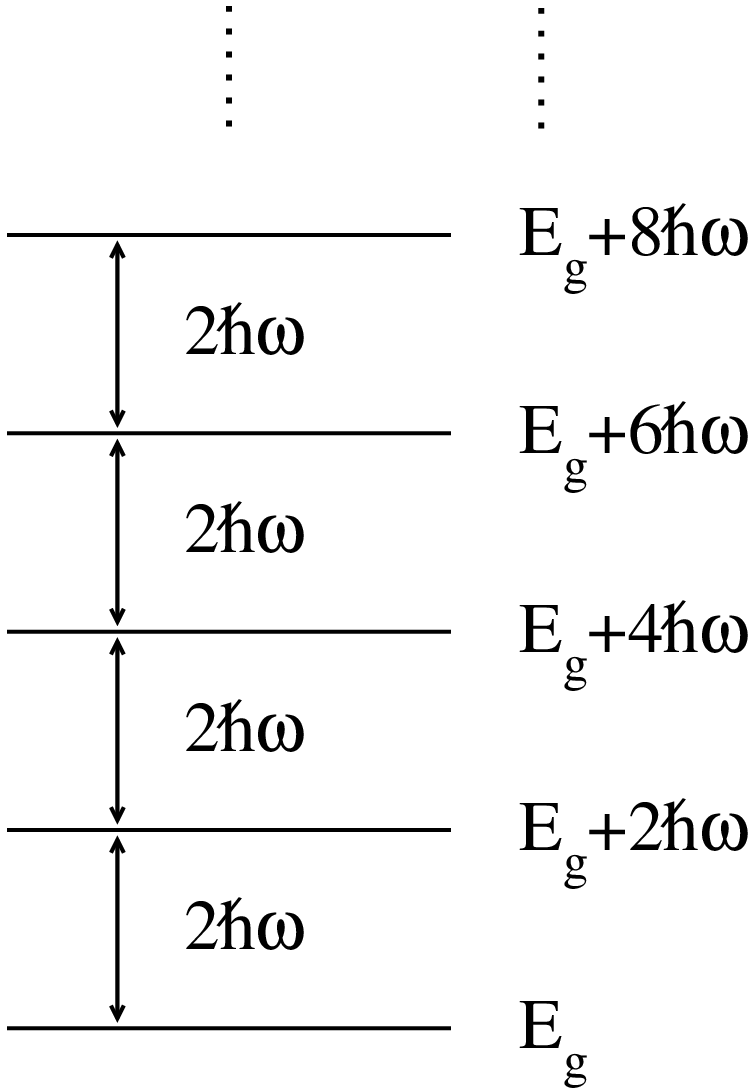} 
\caption{
The energy spectrum of the unitary gas in an isotropic harmonic trap
is a collection of semi-infinite ladders such as the one depicted
in the figure, with various ground step energies $E_g$. 
This structure is related to the existence of
a decoupled bosonic mode, and holds whatever the numbers of fermions in each
of the two spin components.
\label{fig:ladder}
}
\end{figure}

We calculate the resulting change in the $N$-body wavefunction,
expanding Eq.(\ref{eq:ansatz}) to first order in $\epsilon$,
putting in evidence the components that oscillate with Bohr
frequencies $\pm 2\omega$:
\be
\label{eq:cwf}
\psi(\XX,t) = e^{i\alpha}\left[e^{-iEt/\hbar}-\epsilon e^{-i(E+2\hbar\omega)t/\hbar}
L_+ + \epsilon^* e^{-i(E-2\hbar\omega)t/\hbar} L_-\right] 
\psi(\XX,0) + O(\epsilon^2).
\ee
The time independent phase $\alpha$ depends on the details of the excitation procedure.
We have introduced the operators
\be
L_{\pm} = \pm i D + \frac{H}{\hbar\omega} - \frac{m\omega}{\hbar} X^2
\ee
where $D$ is the generator of the scaling transforms, as defined
in Eq.(\ref{eq:defD}), and $L_+=L_-^\dagger$.
We then read on Eq.(\ref{eq:cwf}) the remarkable property that the 
action of $L_+$ on an energy eigenstate $\psi$ of energy $E$ produces
an energy eigenstate of energy $E+2 \hbar \omega$ \footnote
{As shown in \cite{WernerPRA}, $L_+ \psi$ cannot be zero.}.
Similarly, the action of $L_-$ on $\psi$ produces an energy eigenstate
of energy $E-2\hbar \omega$, or eventually gives zero since
the spectrum is bounded from below by $E\ge 0$ according to the
virial theorem (\ref{eq:virth}) applied to $U(\rr)=\frac{1}{2} m\omega^2 r^2$.
We see that the spectrum has thus a very simple structure, it is a
collection of semi-infinite ladders, each ladder being made
of equidistant energy levels separated by $2\hbar \omega$,
see Fig.~\ref{fig:ladder}, and $L_\pm$ acting respectively as
a raising/lowering operator in that structure.
Within each ladder, we call $\psi_g$ the wavefunction corresponding
to the ground step of that ladder, such that
\be
\label{eq:innocent}
L_- \psi_g = 0.
\ee

As shown in \cite{Rosch}, this structure implies a dynamical SO(2,1) symmetry,
meaning that the Hamiltonian $H$ is part of the SO(2,1) Lie algebra.
One starts with the commutation relations:
\be
[H,L_{\pm}] = \pm 2 \hbar \omega L_{\pm} 
\ee
\be
[ {L}_{+},{L}_{-} ] = -4 H /(\hbar\omega).
\label{eq:comm_ut}
\ee
The first relation was expected from the raising/lowering nature
of $L_{\pm}$. Both relations can be checked from the commutation relations
Eqs.(\ref{eq:comm1},\ref{eq:comm2}) and from
\be
[\frac{1}{2}X^2, -\frac{1}{2}\Delta_\XX] =  i D.
\ee
We emphasize again the crucial point that the operators $L_\pm$
preserve the domain of the Hamiltonian in the present unitary case,
since $D$ and $X^2$ do. Obtaining the canonical commutation
relations among the generators $T_1,T_2$ and $T_3$ of the SO(2,1)
Lie algebra,
\be
\label{eq:la}
[T_1,T_2] \equiv -i T_3, \ \ \ [T_2,T_3]\equiv i T_1, \ \ \ 
[T_3,T_1] \equiv i T_2,
\ee
is then only a matter of rewriting:
\be
T_1 \pm i T_2 \equiv \frac{1}{2} L_\pm \ \ \ \ 
\mbox{and}\ \ \ \ T_3 = \frac{H}{2\hbar\omega}.
\ee
Note the sign difference in the first commutator of Eq.(\ref{eq:la})
with respect to the other two commutators, and with respect to the
more usual SO(3) or SU(2) Lie algebra.

Have we gained something in introducing the SO(2,1) Lie algebra,
or is it simply a formal rewriting of the ladder structure
already apparent in the simple minded approach Eq.(\ref{eq:cwf}),
may ask a reader unfamiliar with dynamical symmetries.
Well, an advantage is that we can immediately exhibit the so-called Casimir
operator $C$, 
\be
C \equiv -4 [T_1^2 + T_2^2 - T_3^2] = H^2-\frac{1}{2} (\hbar\omega)^2
(L_+ L_- + L_- L_+),
\ee
guaranteed to commute with all the elements $T_1$, $T_2$ and
$T_3$ of the algebra, so that $C$ is necessarily a scalar within each
ladder. Taking as a particularly simple case the expectation value
of $C$ within the ground step $\psi_g$ of the ladder of energy
$E_g$, and using
Eq.(\ref{eq:comm_ut}) to evaluate $\langle \psi_g| L_- L_+ |\psi_g\rangle$,
we obtain $C |\psi_g\rangle = E_g (E_g - 2 \hbar \omega) |\psi_g\rangle$.
Inverting this relation thanks to the property
$E_g \ge 3\hbar\omega/2$ 
\footnote{To obtain this inequality, one uses a virial theorem after 
separation of the center of mass motion \cite{WernerPRA}.}, 
we can define the ground energy step operator $H_g$: 
\be
H_g = \hbar\omega + [C+(\hbar\omega)^2]^{1/2},
\ee
which is scalar and equal to $E_g$ within each ladder.
A useful application of $H_g$ is to rescale the raising
and lowering operator $L_\pm$ to obtain simpler commutation
relations: It appears that
\be
b = [2(H+H_g)/(\hbar\omega)]^{-1/2} L_- 
\ee
is a bosonic annihilation operator, which obeys the usual
bosonic commutation relations, in particular with its hermitian conjugate
\be
[b,b^\dagger] = 1.
\ee
$b^\dagger $ and $b$ have the same raising/lowering properties
as $L_\pm$, and commute with $H_g$. They have the usual simple
matrix elements, e.g.\ $b^\dagger|n\rangle = (n+1)^{1/2} |n+1\rangle$
where $|n\rangle$ is the step number $n$ of a ladder, $n$ starting from
0.  They allow an illuminating rewriting of the Hamiltonian:
\be
H = H_g + 2\hbar\omega b^\dagger b
\ee
revealing that the unitary gas in a harmonic isotropic trap
has a fully decoupled bosonic degree of freedom.
This bosonic degree of freedom, physically, is simply
the undamped breathing mode of the gas of frequency $2\omega$,
identified for a different system in \cite{Rosch}.

We now give two simple applications of the above formalism \cite{WernerPRA}.
First, one can calculate the various moments of the trapping Hamiltonian
$H_{\rm trap}= \frac{1}{2} m \omega^2 X^2$, from the identity
\be
\label{eq:htrap}
H_{\rm trap} = \frac{1}{2} H -\frac{\hbar\omega}{4} (L_+ + L_-)
= \frac{\hbar\omega}{2} A^\dagger A
\ee
where $A=[b^\dagger b+H_g/(\hbar\omega)]^{1/2}-b$. 
Taking the expectation value of Eq.(\ref{eq:htrap}) within a given 
eigenstate of energy $E$, or within a statistical mixture of eigenstates,
immediately gives
\be
\langle H_{\rm trap}\rangle = \frac{1}{2} \langle H\rangle,
\ee
a particular case of the virial theorem Eq.(\ref{eq:virth}).
Taking the expectation value of $H_{\rm trap}^2$ for the thermal
equilibrium density operator gives
\be
4 \langle H_{\rm trap}^2 \rangle = \langle H^2 \rangle
+ \langle H\rangle \hbar\omega [2\langle b^\dagger b\rangle +1]
\ee
where we used $\langle H_g b^\dagger b\rangle= \langle H_g\rangle \langle
b^\dagger b\rangle$ for the thermal equilibrium.
From the Bose formula, one has also 
$\langle b^\dagger b\rangle=[\exp(2\beta\hbar\omega)-1]^{-1}$,
with $\beta=1/(k_B T)$ and $T$ is the temperature.

The second, more impressive,
application is to uncover a very interesting structure
of the $N$-body wavefunction $\psi_g(\XX)$ of the ground energy step
of an arbitrary ladder. We introduce hyperspherical coordinates
$(X,\nn=\XX/X)$, where $\nn$ is a unit vector is the space of
$3N$ real coordinates. The innocent equation (\ref{eq:innocent}) becomes
\be
\label{eq:innocent2}
\left[-\frac{3N}{2} - X\partial_X +\frac{E_g}{\hbar\omega} 
-\frac{m\omega}{\hbar} X^2 \right] \psi_g(\XX)=0.
\ee
This is readily integrated for a fixed hyperdirection $\nn$:
\be
\label{eq:fascinating}
\psi_g(\XX) = e^{-m\omega X^2/(2\hbar)} 
X^{\frac{E_g}{\hbar\omega} -\frac{3N}{2}}
f(\nn)
\ee
where $f(\nn)$ is an unknown function of the hyperdirection.
Eq.(\ref{eq:fascinating}) has fascinating consequences.

First, it shows that $\psi_g$, being the product of a function
of the modulus $X$ and of a function of the hyperdirection,
is {\sl separable} in hyperspherical coordinates.  The physical consequences
of this separability, in particular for the few-body problem,
are investigated in subsection \ref{subsec:pcots}.
Note that this separability holds for all the other steps of the ladder,
since $L_+$ only acts on the hyperradius.

Second, we take the limit $\omega\to 0$ in Eq.(\ref{eq:fascinating}):
According to Eq.(\ref{eq:simple}), $E_g/(\hbar\omega)$ is a constant,
and $E_g\to 0$, whereas the Gaussian factor tends to unity. 
$\nn$ is dimensionless, and we can take $f(\nn)$ to be $\omega$ independent
if we do not normalize $\psi_g$ to unity.
We thus obtain in this limit a zero energy eigenstate
of the free space problem,
\be
\label{eq:psifsi}
\psi^{\rm free}(\XX) = X^{\frac{E_g}{\hbar\omega} -\frac{3N}{2}} f(\nn)
\ee
which is independent of $\omega$. This zero energy eigenstate
is scaling invariant,  in the sense that
\be
\label{eq:sca_inv}
\psi_\lambda^{\rm free} (\XX) = \frac{1}{\lambda^\nu} \psi^{\rm free}(\XX)
\ \ \ \forall \lambda>0,
\ee
where $\psi_\lambda$ is defined in Eq.(\ref{eq:sctr}) and 
\be
\nu = \frac{E_g}{\hbar\omega}.
\ee
In summary, starting from the wavefunction $\psi_g$ of any ladder ground state of 
the trapped gas spectrum,
one gets a scaling-invariant zero-energy free-space eigenstate $\psi_\lambda^{\rm
free}$,
simply by removing the gaussian factor $e^{-m\omega X^2/(2\hbar)}$
in the expression (\ref{eq:fascinating}) of $\psi_g$.

Remarkably, the reverse property is true. Let us imagine that we know
a zero energy eigenstate $\psi^{\rm free}$ of the free space problem 
$H_{\rm free}=-\frac{\hbar^2}{2m} \Delta_\XX$,
\be
\label{eq:eselen}
\Delta_\XX \psi^{\rm free}(\XX) = 0,
\ee
that of course also obeys the Wigner-Bethe-Peierls contact conditions
for the unitary gas. Since $H_{\rm free}$
commutes with the generator $D$ of the scaling transforms,
we generally expect $\psi^{\rm free}$ to obey Eq.(\ref{eq:sca_inv})
with some exponent $\nu$, so that
\be
i D \psi^{\rm free} = \nu \psi^{\rm free}.
\label{eq:dpsi}
\ee
Since $\psi^{\rm free}$ is not square integrable, the hermiticity of
$D$ does not imply that $\nu\in i\mathbb{R}$; on the contrary,
we will see that $\nu\in \mathbb{R}$.
Let us multiply $\psi^{\rm free}$ with a Gaussian factor:
\be
\label{eq:mapping}
\psi(\XX) \equiv e^{-m\omega X^2/(2\hbar)} \psi^{\rm free}(\XX).
\ee
As we did for the gauge transform, see Eq.(\ref{eq:chrcbp}),
we can show that $\psi$ so defined obeys the Wigner-Bethe-Peierls
contact conditions.
Calculating the action on $\psi$ of the Hamiltonian $H$ of the trapped gas,
and using Eq.(\ref{eq:dpsi}), we directly obtain
\be
H \psi = \nu \hbar \omega \psi,
\ee
i.e.\ $\psi$ is indeed an eigenstate of the trapped gas with the
eigenenergy $\nu \hbar \omega$.
This $\psi$ corresponds to the ground energy step of a ladder.
Repeated action of $L_+$ will generate the other states of the ladder.

We have thus constructed a mapping between the trapped case
and the zero energy free space case, for the unitary gas in an
isotropic harmonic trap. A similar mapping
(restricting to the ground state) was constructed by Tan in
an unpublished work \cite{TanScaling}. 

\subsection{Separability in internal hyperspherical coordinates}
\label{subsec:separability}

As shown in subsection \ref{subsec:so21}, the SO(2,1) dynamical symmetry
of the unitary gas in an isotropic harmonic trap implies that the eigenstate
wavefunctions $\psi(\XX)$ may be written as the product of a function
of the modulus $X$ and of a function of the direction $\XX/X$.
Here, following \cite{WernerPRA},  we directly use this property at the level of the $N$-body Schr\"odinger
equation, for $N>2$, and we derive an effective Schr\"odinger equation for a
hyperradial wavefunction, with interesting consequences discussed
in subsection \ref{subsec:pcots}. The derivation is restricted here to the case of particles of
identical masses, as in  the previous sections, but the separability in internal spherical coordinates
may also hold for particles of different masses, as detailed in Appendix~3.

First, we introduce a refinement to the separability of subsection \ref{subsec:so21}: In a harmonic trap,
the center of mass of the system is totally decoupled from the internal
variables, that is from the relative coordinates $\rr_i-\rr_j$ of the particles.
This is quite straightforward in Heisenberg picture, for an interaction modeled by a potential
$V(|\rr_i-\rr_j|)$. The Heisenberg equations of motion for the center of mass position
\be
\CC \equiv \frac{1}{N} \sum_{i=1}^{N} \rr_i
\ee
and the center of mass momentum $\PP=\sum_{i=1}^{N} \pp_i$
are indeed coupled only among themselves, due in particular to the fact that the interaction
potential cannot change the total momentum $\PP$ of the system:
\bea
\frac{d}{dt} \hat{\PP}(t) &=& -N m \omega^2 \hat{\CC}(t) \\
\frac{d}{dt} \hat{\CC}(t) &=& \frac{\hat{\PP}(t)}{Nm}.
\eea
The center of mass of the system thus behaves as a fictitious particle of mass $Nm$
trapped in the harmonic potential $N m\omega^2 \CC^2/2$, with a Hamiltonian
\be
H_{\rm CM} = -\frac{\hbar^2}{2Nm} \Delta_{\CC} + \frac{1}{2} N m \omega^2 C^2.
\ee
The center of mass has of course the same
angular oscillation frequency as the individual particles.
This center of mass decoupling property clearly holds in the general harmonic {\sl anisotropic} case.
It persists in the zero range limit so it holds also for the zero-range model.

We can thus split the Hamiltonian Eq.(\ref{eq:hamil}) 
as the sum of the center of mass Hamiltonian  $H_{\rm CM}$ and the internal Hamiltonian
$H_{\rm internal} \equiv H - H_{\rm CM}$.
As a consequence, we introduce as new spatial coordinates the center of mass position
$\CC$ and the set of internal coordinates
\be
\RR \equiv (\rr_1-\CC,\ldots,\rr_N-\CC),
\ee
and we can seek eigenstates in the factorized form $\psi(\XX) = \psi_{\rm CM}(\CC) \psi_{\rm internal}(\RR)$.

The crucial step is then to define {\sl internal} hyperspherical coordinates, consisting in the
hyperradius
\be
R = \left[\sum_{i=1}^{N} (\rr_i-\CC)^2\right]^{1/2}
\ee
and a convenient parameterization of the set of dimensionless internal coordinates $\RR/R$. 
There is a technical subtlety due to the
fact that the coordinates of $\RR$ are not independent variables: Since the sum of the components
of $\RR$ along each spatial direction $x$, $y$ and $z$ is exactly zero, and since $\RR/R$ is a unit vector,
the vector $\RR/R$ contains actually only $3N-4$ independent dimensionless real variables.
We then use the following result, that may be obtained with the appropriate Jacobi coordinates 
\footnote{For particles of equal masses one introduces the Jacobi coordinates
$\uu_i=(\frac{N-i}{N+1-i})^{1/2} \left[\rr_i-(N-i)^{-1} \sum_{j=i+1}^{N}\rr_j\right]$
for $1\leq i\leq N-1$. Then $\Delta_\XX = N^{-1} \Delta_\CC + \sum_{i=1}^{N-1} \Delta_{\uu_i}$
and $R^2=X^2-N C^2 =\sum_{i=1}^{N-1} \uu_i^2$. The general case of arbitrary masses is detailed in
the Appendix~3.}
\cite{Jacobi}:
There exists a parameterization of $\RR/R$
by a set of $3N-4$ internal hyperangles that we call $\Omegab$,
such that the internal Hamiltonian takes the form
\be
\label{eq:hamil_interne}
H_{\rm internal} = -\frac{\hbar^2}{2m} \left[ \partial_R^2 + \frac{3N-4}{R} \partial_R 
+\frac{1}{R^2} \Delta_{\Omegab}\right] + \frac{1}{2} m \omega^2 R^2,
\ee
where $\Delta_{\Omegab}$ is the Laplacian on the unit sphere of dimension $3N-4$.
The expression between square brackets is the standard form for
the usual Laplacian in dimension $d=3N-3$, written in hyperspherical coordinates,
which justifies the name of ``internal hyperspherical coordinates".

The separability in internal spherical coordinates means that
the internal eigenstates in a trap can be written as products of a function of $R$ and a function of $\Omegab$.
This
basically results from the reasoning
below Eq.(\ref{eq:eselen}), with the little twist that one can further assume that
the zero-energy free space eigenstate $\psi^{\rm free}(\XX)$ has a zero total momentum,
i.\ e.\ it is independent of the center of mass position
\footnote{The reasoning below Eq.(\ref{eq:innocent2}) can also be adapted
by putting the center of mass in its ground state $\psi_{\rm CM}(\CC)\propto
\exp[-Nm\omega C^2/(2\hbar)]$ and by constructing purely internal raising and lowering operators
of an internal SO(2,1) dynamical symmetry, that do not excite the center of mass
motion contrarily to $L_+$ and $L_-$ \cite{WernerPRA}.}.
The scale invariance Eq.(\ref{eq:sca_inv}) or equivalently Eq.(\ref{eq:dpsi}) 
then implies 
\be
\label{eq:psifreeR}
\psi^{\rm free}(\XX) = R^{s-(3N-5)/2} \phi(\Omegab)
\ee
with some exponent $s$ shifted for convenience by $(3N-5)/2$.
The challenge is of course to determine the unknown function $\phi(\Omegab)$ and the corresponding value
of $s$.
 From Schr\"odinger's equation $\Delta_\XX \psi^{\rm free}=0$ and the expression
of the internal Laplacian in hyperspherical coordinates, see Eq.(\ref{eq:hamil_interne}),
one finds that $s^2$ solves the eigenvalue problem
\be
\label{eq:hypera}
\left[-\Delta_{\Omegab} + \left(\frac{3N-5}{2}\right)^2 \right] \phi(\Omegab) = s^2 \phi(\Omegab),
\ee
where $\phi(\Omegab)$ has to obey the Wigner-Bethe-Peierls contact conditions Eq.(\ref{eq:bpu})
reformulated in hyperangular coordinates.\footnote{These reformulated contact conditions are given explicitly in~\cite{theseFelix}, Eq.(1.38).} The merit of the shift $(3N-5)/2$ is thus
to reveal a symmetry $s\leftrightarrow -s$.

The generalization of the zero energy free space solution Eq.(\ref{eq:psifreeR})
to the finite energy trapped problem
is simply provided by the ansatz:
\be
\psi(\XX) = \psi_{\rm CM}(\CC) \phi(\Omegab) R^{-(3N-5)/2} F(R).
\label{eq:psisepargen}
\ee
Here $\psi_{\rm CM}(\CC)$ is any center of mass eigenstate wavefunction of energy $E_{\rm CM}$, 
$\phi(\Omegab)$ is any solution
of the eigenvalue problem Eq.(\ref{eq:hypera}).
 Injecting the ansatz into Schr\"o\-din\-ger's equation of eigenenergy $E$
and using Eq.(\ref{eq:hamil_interne}), one finds that
\be
E = E_{\rm CM} + E_{\rm internal},
\ee
where the hyperradial wavefunction $F(R)$ and the internal eigenenergy $E_{\rm internal}$ solve the eigenvalue problem:
\be
\label{eq:schr_F}
-\frac{\hbar^2}{2m} \left[F''(R) + \frac{1}{R} F'(R) \right] +
\left(\frac{\hbar^2 s^2}{2m R^2} + \frac{1}{2} m \omega^2 R^2\right) F(R) = 
E_{\rm internal} F(R).
\ee
We note that, as detailed in the Appendix~3,
 this separability remarkably
also holds in the case where the $N$ particles have different masses \cite{WernerPRA}, provided that they all have
the same angular oscillation frequency $\omega$ in the trap, and that the Wigner-Bethe-Peierls model still defines a self-adjoint Hamiltonian
for the considered mass ratios. The separability even holds when the Wigner-Bethe-Peierls model {\it supplemented with an additional boundary condition for $R\to0$ and fixed $\Omegab$} is self-adjoint, as is the case e.g. for $N=3$ bosons, see below; indeed such a boundary condition only affects the hyperradial problem.

In practice the explicit calculation of $s$ is possible
 for the few-body problem.
The most natural approach in general is to try to calculate the functions $A_{ij}$ in Eq.(\ref{eq:bpu}) in momentum space.
 From Eq.(\ref{eq:bpu}) it appears that $A_{ij}$ is scaling invariant with an exponent
$s+1-(3N-5)/2$. Its Fourier transform\footnote{Since the Fourier transform 
$\tilde{A}_{ij}(\KK)=\int d^{3(N-2)}Y\, e^{-i\KK\cdot\YY} A_{ij}(\YY)$ may lead to non-absolutely
converging integrals at infinity, the calculation has to be performed using the language of distributions,
with a regularizing factor $e^{-\eta Y}$, $\eta\to 0^+$.}
is then also scaling invariant, with an exponent given
by a simple power-counting argument: Since $A_{ij}$ is a function of $3(N-2)$ variables, 
if one takes into account the fact that it does not depend on the center of mass position $\mathbf{C}$, and
since one has $[s+1-(3N-5)/2]+3(N-2)=s+(3N-5)/2$, its Fourier transform $\tilde{A}_{ij}$ scales as
\be
\label{eq:Afs}
\tilde{A}_{ij}(\KK) = K^{-[s+(3N-5)/2]} f_{ij}(\KK/K)
\ee
where $\KK$ collects all the $3(N-2)$ variables of $\tilde{A}_{ij}$ 
and $f_{ij}$ denotes some functions to be determined.
Remarkably it is the same quantity $(3N-5)/2$ which appears in both Eqs.(\ref{eq:psifreeR},\ref{eq:Afs}).

This momentum space approach leads to integral equations.
For $N=3$, this integral equation was obtained in~\cite{TerMartirosian}; it was solved analytically in~\cite{Danilov},
the allowed values of $s$ being the solutions of a transcendental equation.
This transcendental equation was rederived from a direct analytical solution of (\ref{eq:hypera}) in position space in \cite{Efimov70},
and generalised to arbitrary angular momenta, masses and statistics in~\cite{Efimov70,Efimov73};
for equal masses it is conveniently expressed in the form~(\cite{Birse} and refs. therein):
\be
\frac{\Gamma\left( l+\frac{3}{2} \right)}
{\Gamma\left(\frac{l+1+s}{2}\right) \Gamma\left(\frac{l+1-s}{2}\right)}
=
\frac{\eta}{\sqrt{3\pi}(-2)^l}\ _2F_1\left(\frac{l+1+s}{2},\frac{l+1-s}{2};l+\frac{3}{2};\frac{1}{4}\right)
\label{eq:trans_hypergeo}
\ee
or alternatively~\cite{WernerPRL}
\begin{eqnarray*}
\Bigg[i^l\sum_{k=0}^l\frac{(-l)_k(l+1)_k}{k!}\frac{(1\!-\!s)_l}{(1\!-\!s)_k}
\Big(
    2^{-k}i(k-s)e^{is\frac{\pi}{2}}
+\eta (-1)^l\frac{4}{\sqrt{3}}e^{i\frac{\pi}{6}(2k+s)}
\Big) \Bigg]
\nonumber 
\end{eqnarray*}
\begin{eqnarray*}
-\Bigg[(-i)^l\sum_{k=0}^l\frac{(-l)_k(l+1)_k}{k!}\frac{(1\!-\!s)_l}{(1\!-\!s)_k}
\Big(
    2^{-k}(-i)(k-s)e^{-is\frac{\pi}{2}}
+\eta (-1)^l\frac{4}{\sqrt{3}}e^{-i\frac{\pi}{6}(2k+s)}
\Big) \Bigg]  \nonumber 
\end{eqnarray*}
\be
\ \ \ \ =0 
\label{eq:trans}
\ee
where $l$ is the total internal angular momentum quantum number,
$\eta$ is $-1$ for fermions ($N_\uparrow=2$, $N_\downarrow=1$) or $+2$ for spinless bosons,
$\,_2F_1$ is a hypergeometric function,
and 
$(x)_n\equiv x (x+1)\ldots (x+n-1)$ with $(x)_0\equiv 1$.
The equation (\ref{eq:trans_hypergeo}) has some spurious integer solutions
($l=0,s=2$ for fermions; $l=0,s=4$ and $l=1,s=3$ for bosons)
which must be eliminated.
For $N=4$ there is no known analytical solution of the integral equation.
Using the scale invariance of $\tilde{A}_{ij}(\KK)$ as in Eq.(\ref{eq:Afs}) and rotational
symmetry however brings it to a numerically tractable integral equation involving the exponent
$s$, that allowed to predict a four-body Efimov effect for three same-spin state
fermions interacting with a lighter particle \cite{CMP}.

\subsection{Physical consequences of the separability}
\label{subsec:pcots}

As seen in the previous section \ref{subsec:separability}, the solution of the $N$-body problem ($N>2$)
for the unitary gas in a harmonic isotropic trap boils down to (i) the calculation
of exponents $s$ from zero-energy free space solutions, and (ii) the solution of the hyperradial eigenvalue
problem Eq.(\ref{eq:schr_F}). Whereas (i) is the most challenging part on a practical point of
view, the step (ii) contains a rich physics that we now discuss.

Formally, the hyperradial problem Eq.(\ref{eq:schr_F}) is Schr\"odinger's equation for one (fictitious) particle 
moving in two dimensions with zero angular momentum in the (effective) potential
\be
U_{\rm eff}(R)=\frac{\hbar^2}{2m}\,\frac{s^2}{R^2}
+ \frac{1}{2}m\omega^2 R^2
.
\label{Ueff}
\ee
We will see that the nature of this problem is very different depending on the sign of $s^2$.
The case $s^2\ge0$, i.e. $s$ real, happens
for $N=3$ fermions ($N_\uparrow=2$,\,$N_\downarrow=1$),
not only for equal masses [as can be tested numerically from (\ref{eq:trans}) and even demonstrated analytically from the corresponding hyperangular eigenvalue problem~\cite{WernerPRL}] but also for unequal masses
provided $m_\uparrow/m_\downarrow$ is below the critical value $13.60\ldots$ where one of the $s$
(in the angular momentum $l=1$ channel) becomes imaginary~\cite{Efimov73}.
For $N=4$ fermions with ($N_\uparrow=3$, $N_\downarrow=1$),
the critical mass ratio above which one of the $s$ (in the angular momentum $l=1$ channel) becomes imaginary is slightly smaller, $m_\uparrow/m_\downarrow\simeq13.384$~\cite{CMP}.
In the physics literature,
$s$ is believed to be real for fermions for any $(N_\uparrow, N_\downarrow)$ for equal masses,
this belief being supported by numerical and experimental evidence.
For $3$ identical bosons, it is well-known that one of the values of $s$ (in the $l=0$ channel) is imaginary~\cite{Efimov70}, all other values being real.

\subsubsection{Universal case}
In this subsection we assume that $s$ is real and we can take the sign convention $s\ge0$.
We impose that the hyperradial wavefunction $F(R)$ is bounded for $R\to0$; indeed, allowing $F(R)$ to diverge would physically correspond to a $N$-body resonance
 (see Appendix~6).
The spectrum and
the corresponding hyperradial wavefunctions
then are~\cite{WernerPRA}
\be
E_{\rm internal} = (s+1+2 q)\hbar\,\omega,\ \ \ q\in\mathbb{N}
\label{eq:Eint}
\ee
\be
F(R)=
\sqrt{\frac{2\,q !}{\Gamma(s+1+q)}}\ 
\frac{R^s}{(a_{ho})^{s+1}}
 \, e^{-\left(\frac{R}{2 a_{ho}}\right)^2}
L_q^{(s)}\!\left( \left(\frac{R}{a_{ho}}\right)^2\right)
\label{eq:F(R)}
\ee
where $L_q^{(s)}$ is a generalised Laguerre polynomial of order $q$, $a_{ho}\equiv\sqrt{\frac{\hbar}{m\omega}}$ is the harmonic oscillator length, and the normalisation is such that 
$\int_0^\infty dR\, R\, F(R)^2=1$.
Eq.~(\ref{eq:Eint}) generalises to excited states the result obtained 
for the ground state
in~\cite{TanScaling}.

We thus recover the $2\hbar\omega$ spacing of the spectrum discussed in section~\ref{subsec:so21}. We can also reinterpret the scaling solution of section~\ref{subsec:tds} as a time-evolution of the hyperradial wavefunction with a time-independent hyperangular wavefunction; in particular, the undamped breathing mode corresponds to an oscillation of the fictitious particle in the effective potential (\ref{Ueff}).\footnote{Strictly speaking, such a time evolution of the wavefunction in {\it internal} hyperspherical coordinates corresponds to an internal scaling solution where the center of mass wavefunction is constant, whereas the scaling solution of~\ref{subsec:tds} corresponds to a hyperradial motion in the hyperspherical coordinates $(X,\nn)$.}

The expression (\ref{eq:F(R)}) of $F(R)$ immediately yields the probability distribution $P(R)$ of the hyperradius
 via $P(R)=F(R)^2\,R$. This analytical prediction is in good agreement with the numerical results obtained in~\cite{Blume} for up to $17$ fermions.
 


In the large $N$ limit (more precisely if $N_\uparrow$ and $N_\downarrow$ tend to infinity and their ratio goes to a constant),
the ground state energy of the trapped unitary gas is expected to be given in an asymptotically exact way by hydrostatics (also called local density approximation). 
Amusingly, this allows to predict the large-$N$ asymptotics of the smallest possible value of $s$.
For $N_\uparrow=N_\downarrow=N/2\to\infty$ this gives~\cite{TanScaling,Blume}
\be
s\sim\sqrt{\xi}\frac{(3N)^{4/3}}{4}
\ee
where $\xi$ appears in the expression
Eq.(\ref{eq:e0}) for the ground state energy of the homogeneous unitary gas.

For spin-1/2 fermions, Eqs. (\ref{eq:Eint},\ref{eq:F(R)}), combined with
the transcendental equation (\ref{eq:trans_hypergeo})
and the expression of the hyperangular wavefunctions~\cite{Efimov73}, provide the complete solution of the unitary $3$-body problem in an isotropic harmonic trap~\cite{WernerPRL} (for completeness, one also has to include the eigenstates which are common to the unitary and the non-interacting problem~\cite{WernerPRL}, mentioned at the end of subsection~\ref{subsubsec:ZRM}).
This was first realised for the ground state in~\cite{TanScaling}.
Remarkably, this $3$-body spectrum in a trap allows to compute the third virial coefficient of the homogeneous unitary gas~\cite{Hu}, whose value was confirmed experimentally (see~\cite{ChevyNature} and the contribution of F. Chevy and C. Salomon to this volume).

For spinless bosons, the unitary $3$-body problem in an isotropic harmonic trap has two families of eigenstates (apart from the aforementioned common eigenstates with the non-interacting problem)~\cite{Jonsell,WernerPRL}: the states corresponding to real solutions $s$ of the transcendental equation~(\ref{eq:trans_hypergeo}), which we call universal states; and the states corresponding to the imaginary solution for $s$, which we call efimovian. Eqs.~(\ref{eq:Eint},\ref{eq:F(R)}) apply to universal states. The efimovian states are discussed in the next subsection. 

\subsubsection{Efimovian case}
In this subsection we consider the case $s^2<0$,
i.e. $s$ is purely imaginary.
In this case, {\rm all} solutions of the Schr\"{o}dinger-like  equation (\ref{eq:schr_F}) are bounded and oscillate more and more rapidly when $R\to 0$.
In order to obtain a hermitian problem with a discrete spectrum, one has to impose the boundary condition~\cite{Danilov,theseFelix}:
\begin{equation}
\exists A/\ F(R) 
\sim
A\ \mbox{Im}\, \left[\left(\frac{R}{R_t}\right)^{s}\right]
\ \ \ {\rm for}\ R\to 0
,
\label{eq:CL_Rt}
\end{equation}
where $R_t$ is an additional 3-body parameter.~An equivalent form is:
\begin{equation}
\exists A'/\ F(R)
\sim
 A'\ \sin\left[|s|\ln\left(\frac{R}{R_t}\right)\right]
\ \ \ {\rm for}\ R\to 0
.
\label{eq:CL_sin}
\end{equation}
The corresponding hyperradial wavefunctions are
\begin{equation}
F(R) = R^{-1}\, W_{E/2,s/2}(R^2/a_{ho}^2)
\label{eq:Fefimovien}
\end{equation}
where $W$ is a Whittaker function, and the spectrum is given by the implicit equation
\begin{equation}
\mbox{arg}\, \Gamma\left[\frac{1+s-E/(\hbar\omega)}{2}\right] =
-|s| \ln (R_t/a_{ho})
 +\mbox{arg}\, \Gamma(1+s) \ \mbox{mod}\ \pi
\label{eq:Eefimovien}
\end{equation}
obtained in~\cite{Jonsell}, whose solutions form a discrete series, which is unbounded from below, and can be labeled by a quantum number $q\in\mathbb{Z}$.

In free space ($\omega=0$), there is a geometric series of bound states
\begin{equation}
E_q=-\frac{2\hbar^2}{mR_t^{\phantom{.}2}} \exp\left({-q \frac{2\pi}{|s|}+\frac{2}{|s|}{\rm arg\,}\Gamma(1+s)}\right),\ q\in\mathbb{Z}
\label{eq:Eefimov}
\end{equation}
\begin{equation}
F(R)=K_s\left(R\sqrt{2m|E|/\hbar^2}\right)
\label{eq:Fefimov}
\end{equation}
where $K$ is a Bessel function.
For $3$ particles this corresponds to the well-known series of Efimov $3$-body bound states \cite{Efimov70,Efimov73}.
This also applies to the $4$-body bound states in the aforementioned case of $(3+1)$ fermions with $m_\up/m_\down$ between $\simeq 13.384$ and $13.607\ldots$~\cite{CMP}.
As expected,
in the limit $E\to -\infty$, the spectrum of the efimovian states in the trap (\ref{eq:Eefimovien}) approaches the free space spectrum (\ref{eq:Eefimov}).
The unboundedness of the spectrum in the zero-range limit is a natural consequence of the Thomas effect and of the limit cycle behavior~\cite{theseFelix}.

\subsection{Vanishing bulk viscosity}
\label{subsec:viscosity}

In this subsection, we give a simple rederivation of the fact that the bulk viscosity of the unitary gas in the normal phase is zero.
This result was obtained in~\cite{SonBulkVisco}
(see also \cite{NishidaSonChap}).
It helps analysing e.~g. the ongoing experimental studies of the shear viscosity, whose value is of fundamental importance~(\cite{SchaeferLong2010} and refs. therein).
Although the superfluid regime was also treated in~\cite{SonBulkVisco},
we omit it here for simplicity.
In our rederivation we shall use the scaling solution 
and the existence of the undamped breathing mode.~\footnote{
In article~\cite{SonBulkVisco}, the vanishing of the bulk viscosity was deduced from the so-called general coordinate and conformal invariance,
the
scaling solution being
 unknown to its author of at the time of writing
(although it had been obtained in~\cite{CRASCastin}).
The scaling solution was recently rederived using this 
general coordinate and conformal invariance~\cite{NishidaSonChap}.
Several other results presented in subsections~\ref{subsec:so21}, \ref{subsec:separability} and \ref{subsec:pcots} were also rederived
using this field theoretical formalism (\cite{NishidaSonChap} and refs. therein).}

In the hydrodynamic theory for a normal compressible viscous fluid~\cite{LandauHydro,SonBulkVisco},
 the (coarse-grained) evolution of the gas in a trapping potential $U(\rr,t)$ is described by the atom number density $\rho(\rr,t)$, 
the  velocity vector field $\mathbf{v}(\rr,t)$, and the entropy per particle (in units of $k_B$) $s(\rr,t)$.
These $5$ scalar functions solve $5$ equations
which are given for completeness in Appendix~4 
although we will not directly use them here.
We will only need the equation for the increase of the total entropy $S=\int \rho s\, d^3r$ of the gas
\be
\frac{dS}{dt}=\int \frac{\kappa \|\mathbf{\nabla} T\|^2}{T^2} d^3r
+
\int \frac{\eta}{2T}\sum_{ik}\left(
\frac{\partial v_i}{\partial x_k}+\frac{\partial v_k}{\partial x_i}-\frac{2}{3}\delta_{ik}\mathbf{\nabla}\cdot\mathbf{v} \right)^2 d^3r
+
\int \frac{\zeta}{T}\|\mathbf{\nabla}\cdot\vv\|^2 d^3r
\label{eq:dSdt}
\ee
which follows from the hydrodynamic equations~(\ref{eq:entr_prod},\ref{eq:cte});
note that the thermal conductivity $\kappa$, the shear viscosity $\eta$ and the bulk viscosity $\zeta$ have to be $\geq 0$ so that $dS/dt\geq0$~\cite{LandauHydro}.
The hydrodynamic theory is expected to become exact in the limit where the length (resp. time) scales on which the above functions vary are much larger than microscopic length (resp. time) scales such as $1/k_F$ (resp. $\hbar/E_F$).

We consider
the following gedankenexperiment:
Starting with the gas at thermal equilibrium in a trap of frequency $\omega$,
 we suddenly switch the trapping frequency 
 at $t=0$ to a different value $\omega_+$.
As we have seen in subsection~\ref{subsec:tds}
and at the beginning of subsection~\ref{subsec:so21},
this excites an {\it undamped} breathing mode:
For $t>0$, the size of the gas oscillates {\it indefinitely}.
This rigorously periodic evolution of the system implies that 
the total entropy $S(t)$ is periodic, and since it cannot decrease, it has to be constant.
Thus each of the terms in the right-hand-side of~(\ref{eq:dSdt}), and in particular the last term, has to vanish.
Thus
$\zeta(\rr,t) \|\mathbf{\nabla}\cdot\vv(\rr,t)\|^2\equiv0$.
This implies that $\zeta$ is identically zero, as we now check.
From the scaling evolution~(\ref{eq:ansatz}) of each many-body eigenstate, one can deduce (using the quantum-mechanical expression for the particle flux) that
\be
\vv(\rr,t)=\frac{\dot{\lambda}}{\lambda}\rr,
\label{eq:v_echelle}
\ee
so that $\mathbf{\nabla}\cdot\vv=3\dot{\lambda}/\lambda$.
For $t$ approaching $0$ from above,
we have $\dot{\lambda}(t)\neq0$,
as is intuitively clear and can be checked from Eqs.~(\ref{eq:inco},\ref{eq:russe});
thus $\zeta(\rr,t)=0$ and by continuity $\zeta(\rr,t=0)=0$. Since the central density and temperature in the initial equilibrium state of the gas are arbitrary, 
we conclude that $\zeta(\rho,T)=0$ for all $\rho$ and $T$.
An alternative derivation of this result is presented in Appendix~5.

\subsection{Short-range scaling laws}
\label{subsec:scaling_laws}

As opposed to the previous subsections, we now consider an arbitrary scattering length and an arbitrary external potential, possibly with periodic boundary conditions.
Of all the particles $1,\ldots,N$, let us consider a subset $J\subset\{1,\ldots,N\}$ containing $n_\up$ particles of spin $\up$ and $n_\down$ particles of spin $\down$.
From the particle positions $(\rr_i)_{i\in J}$, we can define
a hyperradius $R_J$ and hyperangles $\Omegab_J$, and
a center of mass position $\CC_J$.
The positions of all particles that do not belong to $J$ are denoted by $\mathbf{\mathcal{R}}_J=(\rr_i)_{i\notin J}$.
In the absence of a $(n_\up+n_\down)$-body resonance
(see Appendix~6),
 one expects that, for any eigenstate,
in the limit $R_J\to0$ where all particles belonging to the subset $J$ approach each other while $(\Omegab_J,\CC_J,\mathbf{\mathcal{R}}_J)$ remain fixed, there exists a function $A_J$ such that
\be
\psi(\rr_1,\ldots,\rr_N) = R_J^\nu\,\phi(\Omegab_J)
\,A_J(\CC_j,\mathbf{\mathcal{R}}_J)
+o(R_J^\nu).
\label{eq:RJ->0}
\ee
Here, $\nu=s_{\rm min}(n_\up,n_\down)-\frac{3(n_\up+n_\down)-5}{2}$ with $s_{\rm min}(n_\up,n_\down)$ the smallest possible value of $s$ for the problem of $n_\up$ particles of spin $\up$ and $n_\down$ particles of spin $\down$ ($s$ being defined in Sec.~\ref{subsec:separability})
and
$\phi(\Omegab_J)$ is the corresponding hyperangular wavefunction (also defined in Sec.~\ref{subsec:separability}).
This statement is essentially contained in~\cite{PetrovShlyapSalomon,TanScaling}. It comes from the intuition that, in the limit where the $n_\up+n_\down$ particles approach each other, the $N$-body wavefunction should be proprtional to the $(n_\up+n_\down)$-body zero-energy free space wavefunction Eq.(\ref{eq:psifreeR}).
This was used in~\cite{PetrovShlyapSalomon} to predict that
 the formation rate of deeply bound molecules by three-body recombination, $\Gamma\equiv-\dot{N}/N$, behaves as $\hbar \Gamma/E_F\sim K\cdot(k_F b)^{2 s_{\rm min}(2,1)}$ in the low-density limit, with $b$ on the order of the van der Waals range
and $K$ a numerical prefactor which depends on short-range physics. 
The analytical solution of the hyperangular three-body problem [Eq.(\ref{eq:trans})] yields $s_{\rm min}(2,1)=1.772724\ldots$ (this value is reached in the angular momentum $l=1$ channel).
Experimentally, this scaling has not been checked, 
but the smallness of $\hbar\Gamma/E_F$ is one of the crucial ingredients which allow to realise the unitary gas.

\acknowledgement
We thank O. Goulko,
O. Juillet, 
T.~Sch\"afer,
D.~T. Son,
B.~Svistunov
and S.~Tan
for helpful discussions while writing this manuscript,
and the authors of~\cite{Jin_p} for their data.
F.~W. is supported by NSF grant PHY-0653183, Y.~C. is member of IFRAF and acknowledges support from the ERC project FERLODIM N.228177.

\section*{Appendix 1: Effective range in a lattice model}
\addcontentsline{toc}{section}{Appendix 1}
\label{app:re}

To calculate the effective range $r_e$
[defined by Eq.(\ref{eq:lke})]
 for the lattice model
of subsection \ref{subsec:lm}, 
it is convenient to perform in the expression~(\ref{eq:fklm}) of the scattering amplitude an analytic continuation
to purely imaginary incoming
wavevectors $k_0$, setting 
$k_0=i q_0$ with $q_0$ real and positive.
Eliminating $1/g_0$ thanks to Eq.(\ref{eq:g0_gen}) we obtain the useful expression:
\be
\label{eq:flu}
-\frac{1}{f_{k_0}} = \frac{1}{a} +4\pi
\int_{\mathcal{D}} \frac{d^3k}{(2\pi)^3}
\left[\frac{1}{q_0^2+2m\epsk/\hbar^2}-\frac{1}{2m\epsk/\hbar^2}\right].
\ee

We first treat the case of the parabolic dispersion relation Eq.(\ref{eq:pdr}).
A direct expansion of Eq.(\ref{eq:flu}) in powers of $q_0$ leads to an infrared
divergence. The trick is to use the fact that the integral over $\mathcal{D}$
in Eq.(\ref{eq:flu}) can be written as the integral of the same integrand over the whole
space minus the integral over the supplementary space $\mathbb{R}^3\setminus \mathcal{D}$.
The integral over the whole space may be performed exactly using
\be
\int_{\mathbb{R}^3} 
\frac{d^3k}{(2\pi)^3} \left[\frac{1}{q_0^2+k^2}-\frac{1}{k^2}\right] = -\frac{q_0}{4\pi}.
\ee
This leads to the transparent expression, where the term corresponding to $ik$ in Eq.(\ref{eq:u(k)}),
and which is non-analytic in the energy $E$, is now singled out:
\be
-\frac{1}{f_{k_0}^{\rm parab}} =  \frac{1}{a} - q_0 -4\pi \int_{\mathbb{R}^3\setminus\mathcal{D}} \frac{d^3k}{(2\pi)^3}
 \left[\frac{1}{q_0^2+k^2}-\frac{1}{k^2}\right].
\ee
This is now expandable in powers of $q_0^2$, leading to the effective range for the parabolic dispersion relation:
\be
\label{eq:repa}
r_e^{\rm parab} = \frac{1}{\pi^2} \int_{\mathbb{R}^3\setminus\mathcal{D}} \frac{d^3k}{k^4}.
\ee

We now turn back to the general case. The trick is to consider the difference between the inverse scattering amplitudes
of the general case and the parabolic case with a common value of the scattering length:
\be
\frac{1}{f_{k_0}^{\rm parab}} - \frac{1}{f_{k_0}} =
4\pi \int_{\mathcal{D}} \frac{d^3k}{(2\pi)^3}
\left[\frac{1}{q_0^2+2m\epsk/\hbar^2}-\frac{1}{q_0^2+k^2}
-\frac{1}{2m\epsk/\hbar^2} + \frac{1}{k^2}\right].
\ee
This is directly expandable to second order in $q_0$, leading to:
\be
r_e - r_e^{\rm parab} = 8\pi \int_{\mathcal{D}} \frac{d^3k}{(2\pi)^3} \left[
\frac{1}{k^4} - \left(\frac{\hbar^2}{2m\epsk}\right)^2
\right].
\ee
The numerical evaluation of this integral for the Hubbard dispersion relation
Eq.(\ref{eq:disphub}) leads to the Hubbard model effective range Eq.(\ref{eq:reh}).

Finally, we specialize the general formula to the parabolic plus quartic form
Eq.(\ref{eq:mixte}). Setting $\kk=(\pi/b)\qq$ and using Eq.(\ref{eq:repa}), we obtain
\be
\frac{\pi^3 r_e^{\rm mix}}{b} = \int_{\mathbb{R}^3\setminus[-1,1]^3} \frac{d^3q}{q^4}
+ \int_{[-1,1]^3} \frac{d^3q}{q^4} \left[1-\frac{1}{(1-Cq^2)^2}\right].
\ee
The trick is to split the cube $[-1,1]^3$ as the union of $B(0,1)$, the sphere of center $0$ and unit radius,
and of the set $X=[-1,1]^3\setminus B(0,1)$. One has also $\left(\mathbb{R}^3\setminus[-1,1]^3\right) \cup X=\mathbb{R}^3\setminus
B(0,1)$ so that
\be
\label{eq:interm}
\frac{\pi^3 r_e^{\rm mix}}{b} = \int_{\mathbb{R}^3\setminus B(0,1)} \frac{d^3q}{q^4}
+\int_{B(0,1)}  \frac{d^3q}{q^4} \left[1-\frac{1}{(1-Cq^2)^2}\right]
- \int_X \frac{d^3q}{q^4} \frac{1}{(1-Cq^2)^2}.
\ee
One then moves to spherical coordinates of axis $z$.
The first two terms in the right hand side may be calculated exactly. In particular, one introduces
a primitive of $q^{-2}(1-C q^2)^{-2}$, given by $C^{1/2} \Phi(C^{1/2}q)$ with
\be
\Phi(x) = \frac{x}{2(1-x^2)} + \frac{3}{2}\mbox{arctanh}\, x -\frac{1}{x}.
\ee
In the last term of Eq.(\ref{eq:interm}) one integrates over the modulus $q$ of $\qq$ for a fixed direction
$(\theta,\phi)$ where $\theta$ is the polar angle and $\phi$ the azimuthal angle.
One then finds that $q$ ranges from $1$ to some maximal value $Q(\theta,\phi)$, and the integral
over $q$ provides the difference $\Phi(C^{1/2}Q)-\Phi(C^{1/2})$. Remarkably, the term
$-\Phi(C^{1/2})$ cancels the contribution of the first two integrals in the right hand side
of Eq.(\ref{eq:interm}), so that
\be
\frac{r_e^{\rm mix}}{b} = -\frac{C^{1/2}}{\pi^3} \int_0^{2\pi} d\phi \int_{-1}^{1} du\, \Phi[C^{1/2}Q(\theta,\phi)]
\ee
where as usual we have set $u=\cos\theta$.
Using the symmetry under parity along each Cartesian axis, which adds a factor $8$, and restricting to the face
$q_x=1$ of the cube, which adds a factor $3$, the expression of $Q(\theta,\phi)$ is readily obtained,
leading to
\be
\label{eq:remi}
\frac{r_e^{\rm mix}}{b} = -\frac{24 C^{1/2}}{\pi^3} \int_0^{\pi/4} d\phi \int_0^{\frac{\cos\phi}{\sqrt{1+\cos^2\phi}}}
du \, \Phi\left(\frac{C^{1/2}}{\cos\phi \sqrt{1-u^2}}\right).
\ee
In the limit $C\to 0$, $r_e^{\rm mix}\to r_e^{\rm parab}$, and Eq.(\ref{eq:remi}) may be calculated analytically
with $\Phi(x)\sim -1/x$ and with an exchange of the order of integration: This leads to Eq.(\ref{eq:rep}).
For a general value of $C\in [0,1/3[$ we have calculated Eq.(\ref{eq:remi}) numerically,
and we have identified the magic value of $C$ leading to a zero effective range, see
Eq.(\ref{eq:magic}). With the same technique, we can calculate the value of $K$ appearing in Eq.(\ref{eq:g0_exp})
from the expression
\be
K = \frac{12}{\pi C^{1/2}} \int_0^{\pi/4} d\phi \int_0^{\frac{\cos\phi}{\sqrt{1+\cos^2\phi}}}
du \, \mbox{arctanh}\, \frac{C^{1/2}}{\cos\phi \sqrt{1-u^2}}.
\ee

\section*{Appendix 2: What is the domain of a Hamiltonian?}
\addcontentsline{toc}{section}{Appendix 2}
\label{app:domain}

Let us consider a Hamiltonian $H$ represented by a differential
operator also called $H$. 
A naive and practical definition 
of the domain $D(H)$ of $H$ is that it is the set
of wavefunctions over which the action of the Hamiltonian is
indeed represented by the considered differential operator.
In other words, if a wavefunction $\psi_{\rm bad}$ 
does not belong to $D(H)$, one should not calculate the action of $H$
on $\psi_{\rm bad}$ directly using the differential operator $H$.
If $H$ if self-adjoint, one should rather expand $\psi_{\rm bad}$
on the Hilbert basis of eigenstates of $H$ and calculate the 
action of $H$ in this basis.

For example, for a single particle
in one dimension in a box with infinite walls in $x=0$ 
and $x=1$, so that $0\leq x \leq 1$, one has the Hamiltonian
\be
\label{eq:dfa}
H = -\frac{1}{2} \frac{d^2}{dx^2},
\ee
with the boundary conditions on the wavefunction
\be
\label{eq:bca}
\psi(0)=\psi(1)=0
\ee
representing the effect of the box.
To be in the domain, a wavefunction $\psi(x)$ should be twice
differentiable for $0<x<1$  
and should obey the boundary conditions (\ref{eq:bca}).
An example of a wavefunction which is not in the domain is the constant
wavefunction $\psi(x)=1.$
An example of wavefunction in the domain is
\be
\label{eq:ind}
\psi(x) = 30^{1/2} x (1-x).
\ee
If one is not careful, one may obtain wrong results. Let us calculate
the mean energy and the second moment of the energy for $\psi$
given by (\ref{eq:ind}). By repeated action of $H$ onto $\psi$,
and calculation of elementary integrals, one obtains
\bea
\label{eq:m1}
\langle H\rangle_\psi &=& 5 \\
\langle H^2\rangle_\psi &=& 0 ?!
\label{eq:m2}
\eea
Eq.(\ref{eq:m1}) is correct, but Eq.(\ref{eq:m2}) is wrong (it would lead
to a negative variance of the energy) because
$H\psi$ is not in $D(H)$ and the subsequent illicit 
action of $H$ as the differential
operator (\ref{eq:dfa}) gives zero.

How to calculate the right value of $\langle H^2\rangle_\psi$~?
One introduces the orthonormal Hilbert basis of eigenstates of $H$,
\be
\psi_n(x) = 2^{1/2} \sin [\pi(n+1)x], \ \ \
n\in \mathbb{N},
\ee
with the eigenenergy $\epsilon_n=\frac{\pi^2}{2} (n+1)^2$.
Then $\psi$ of Eq.(\ref{eq:ind}) may be expanded as
$\sum_n c_n \psi_n(x)$, and the $k^{\rm th}$ moment of the energy may be
defined as
\be
\langle H^k\rangle_\psi = \sum_{n\in \mathbb{N}} (\epsilon_n)^k 
|c_n|^2.
\ee
Since $c_n=4\sqrt{15}[1+(-1)^n]/[\pi(n+1)]^3$, one recovers
$\langle H\rangle_\psi =5$ and one obtains the correct value
$\langle H^2\rangle_\psi=30$, that leads to a positive energy
variance as it should be. Also $\langle H^k\rangle_\psi=+\infty$
for $k\geq 3$.

The trick of expanding $\psi$ in the eigenbasis of $H$ is thus quite powerful,
it allows to define the action of $H$ on any wavefunction $\psi$
in the Hilbert space (not belonging to the domain). It may be applied of
course only if $H$ is self-adjoint, as it is the case in our simple
example.

\section*{Appendix 3: Separability and Jacobi Coordinates for arbitrary masses}
\addcontentsline{toc}{section}{Appendix 3}

We here consider $N\geq 2$ harmonically trapped particles interacting in the unitary limit,
with possibly different masses $m_i$ but with the same isotropic angular
oscillation frequency
$\omega$. The Hamiltonian reads
\be
H = \sum_{i=1}^{N}
\left[ -\frac{\hbar^2}{2m_i} \Delta_{\rr_i} +
\frac{1}{2} m_i \omega^2 r_i^2
\right]
\label{eq:hamiljacob}
\ee
and the unitary interaction is described by the Wigner-Bethe-Peierls contact conditions on the $N$-body wavefunction:
For all pairs of particles $(i,j)$, in the limit $r_{ij}=|\rr_i-\rr_j|\to 0$ with a {\sl fixed} value of the centroid of the particles 
$i$ and $j$, $\RR_{ij}\equiv (m_i\rr_i+m_j \rr_j)/(m_i+m_j)$,
that differs from the positions $\rr_k$ of the other particles, $k\neq i,j$, there exists a function $A_{ij}$ such that
\be
\psi(\rr_1,\ldots,\rr_N) = \frac{A_{ij}(\RR_{ij};(\rr_k)_{k\neq i,j})}{r_{ij}} + O(r_{ij}).
\ee
As is well known and as we will explain below,
 the internal Hamiltonian
$H_{\rm internal}= H - H_{\rm CM}$, where $H_{\rm CM}=
-\frac{\hbar^2}{2M}\Delta_\CC + \frac{1}{2} M \omega^2 C^2$,
takes the form
\be
H_{\rm internal} = \sum_{i=1}^{N-1} 
\left[
-\frac{\hbar^2}{2\bar{m}} \Delta_{\uu_i}
+ \frac{1}{2} \bar{m} \omega^2 u_i^2
\right]
\label{eq:Hjacodreduit}
\ee
in suitably defined Jacobi coordinates [see Eqs.(\ref{eq:def_jac_y},\ref{eq:ui_vs_yi})].
Here $\CC=\sum_{i=1}^{N} m_i\rr_i/M$ is the center of mass position, $M=
\sum_{i=1}^{N} m_i$ is the total mass, and $\bar{m}$ is some arbitray
mass reference, for example the mean mass $M/N$.
Then it is straightforward to express Eq.(\ref{eq:Hjacodreduit}) 
in hyperspherical coordinates, the vector $(\uu_1,\ldots,\uu_{N-1})$ with $3N-3$ coordinates
being expressed in terms of its modulus $R$ and a set
of $3N-4$ hyperangles $\Omegab$, so that
\be
\label{eq:hamilseparjacob}
H_{\rm internal}  = -\frac{\hbar^2}{2\bar{m}}
\left[\partial_R^2 + \frac{3N-4}{R} \partial_R
+\frac{1}{R^2} \Delta_{\Omegab}
\right]  
+\frac{1}{2} \bar{m} \omega^2 R^2
\ee
where $\Delta_{\Omegab}$ is the Laplacian over the unit sphere of dimension $3N-4$.
As we shall see, the expression for the hyperradius is simply
\be
R^2  \equiv \sum_{i=1}^{N-1} u_i^2 = \frac{1}{\bar{m}} \sum_{i=1}^{N} m_i (\rr_i - \CC)^2.
\label{eq:Rjacob}
\ee
This form of the Hamiltonian is then useful to show the separability of Schr\"odinger's equation
for the unitary gas in hyperspherical coordinates \cite{theseFelix,WernerPRA} for $N\geq 3$ and arbitrary masses.
The separability Eq.(\ref{eq:psisepargen}) that was described for simplicity in the case of equal mass particles in
subsection \ref{subsec:separability} indeed still holds in the case of different masses, if the Wigner-Bethe-Peierls model defines a self-adjoint Hamiltonian.
\footnote{Strictly speaking, it is sufficient that the Laplacian on the unit sphere
together with the Wigner-Bethe-Peierls boundary conditions reexpressed in terms of hyperangles 
 is self-adjoint, as extensively
used in \cite{CMP}. This is less restrictive than having the full Hamiltonian self-adjoint, since it allows for example
to have a $N$-body Efimov effect while the $N-1$ zero-range model is perfectly well-defined and does not experience any Efimov effect.}
We recall here the various arguments. First, for zero energy free space eigenstates, the form Eq.(\ref{eq:psifreeR}) is expected from scale invariance,
if the Hamiltonian is self-adjoint \cite{WernerPRA}. Second, the form Eq.(\ref{eq:psisepargen}) for the general case, including non-zero energy
and an isotropic harmonic trap, is expected because (i) 
the Hamiltonian (\ref{eq:hamiljacob}), after separation of the center of mass,
has the separable form (\ref{eq:hamilseparjacob}) in hyperspherical coordinates, and (ii) Eq.(\ref{eq:psisepargen})
obeys the Wigner-Bethe-Peierls contact conditions if Eq.(\ref{eq:psifreeR}) does. 
This point (ii) results from the fact that the Wigner-Bethe-Peierls conditions are imposed, for each pair of particles $(i,j)$,
for $r_{ij}\to 0$ with a {\sl fixed} value of $\RR_{ij}$
that differs from the positions $\rr_k$ of the other particles, $k\neq i,j$. Using $\rr_i=\RR_{ij}+[m_j/(m_i+m_j)] \rr_{ij}$
and $\rr_j=\RR_{ij}-[m_i/(m_i+m_j)] \rr_{ij}$, with $\rr_{ij}\equiv \rr_i-\rr_j$, we indeed find that 
\be
\bar{m} R^2 = \frac{m_i m_j}{m_i+m_j} r_{ij}^2 + (m_i+m_j) (\RR_{ij}-\CC)^2 + \sum_{k\neq i,j} m_k (\rr_k-\CC)^2.
\ee
For $N\geq 3$, we see that $\lim_{r_{ij}\to 0} R^2 > 0$, so that $R$ varies only to second order in $r_{ij}$
in that limit. Provided that the function $F(R)$ in Eq.(\ref{eq:psisepargen}) has no singularity at non-zero $R$,
the Wigner-Bethe-Peierls contact conditions are preserved [similarly to the argument Eq.(\ref{eq:chrcbp})].
Third, bosonic or fermionic exchange symmetries imposed on the $N$-body wavefunction cannot break the separability 
in hyperspherical coordinates: Exchanging the positions of particles of same mass does not change the value of the hyperradius
$R$, it only affects the hyperangles and thus the eigenvalues $[(3N-5)/2]^2-s^2$ of the Laplacian on the unit sphere.

To derive the form Eq.(\ref{eq:Hjacodreduit}) of the internal Hamiltonian,
we introduce the usual Jacobi coordinates given for example in \cite{libro_italiano}:
\be
\yy_i \equiv \rr_i -\frac{\sum_{j=i+1}^{N} m_j \rr_j}{\sum_{j=i+1}^{N} m_j}
\ \ \ \mbox{for} \ 1\leq i \leq N-1.
\label{eq:def_jac_y}
\ee
We note that $\yy_i$ simply gives the relative coordinates
of particle $i$ with respect to the center of mass of the particles from
$i+1$ to $N$. To simplify notations, we also set $\yy_N \equiv \CC$.~\footnote{Alternatively, Eq.(\ref{eq:Hjacodreduit}) can be derived easily by recursion, see p.~63 of~\cite{theseFelix}.}
In compact form, the Jacobi change of variables corresponds to
setting
$\yy_i = \sum_{j=1}^{N} M_{ij} \rr_j$ for  $1\leq i \leq N$,
where the non-symmetric matrix $M$ is such that:
\begin{itemize}
\item In the case  $1\leq i<N$, one has: $M_{ij}=0$ for $1\leq j < i$,
$M_{ij}=1$ for $j=i$, and $M_{ij}=-m_j/(\sum_{k=i+1}^{N} m_k)$
for $i< j \leq N$.
\item $M_{Nj} = m_j/(\sum_{k=1}^{N} m_k)$ for $1 \leq j \leq N$.
\end{itemize}
From the formula giving the derivative of a composite function,
the kinetic energy operator writes
\be
H_{\rm kin} \equiv \sum_{i=1}^{N} -\frac{\hbar^2}{2m_i} \Delta_{\rr_i}
= -\frac{\hbar^2}{2} \sum_{j=1}^{N} \sum_{k=1}^{N} S_{jk} \mathrm{grad}_{\yy_j}
\cdot \mathrm{grad}_{\yy_{k}},
\ee
where the {\sl symmetric} matrix $S$ is defined as
$S_{jk} = \sum_{i=1}^{N}  M_{ji} M_{ki}/m_i.$
The explicit calculation of the matrix elements $S_{jk}$
is quite simple. Taking advantage of the fact that $S$ is symmetric,
one has to distinguish three cases,
(i) $1\leq j,k\leq N-1$, with $j=k$ and $j<k$ as subcases, 
(ii) $j=k=N$, and (iii) $j<N, k=N$.
One then finds that $S$ is purely diagonal, with $S_{ii}=1/\mu_i$
for $1\leq i\leq N-1$ and $S_{NN}=1/M$. Here $\mu_i$ is the reduced
mass for the particle $i$ and for a fictitious particle of mass
equal to the sum of the masses of the particles from $i+1$ to $N$:
\be
\frac{1}{\mu_i} = \frac{1}{m_i} + \frac{1}{\sum_{j=i+1}^{N} m_j} \ \ \ 
\mbox{for} \ \ 1\leq i\leq N-1.
\ee
This results in the following form
\be
H_{\rm kin} = -\frac{\hbar^2}{2M} \Delta_{\CC} -
\sum_{i=1}^{N-1} \frac{\hbar^2}{2\mu_i} \Delta_{\yy_i}.
\label{eq:kinjac}
\ee

The next step is to consider the trapping potential energy term.
Inspired by Eq.(\ref{eq:kinjac}) one may consider the guess
\be
H_{\rm trap} \equiv \sum_{i=1}^{N} \frac{1}{2} m_i \omega^2 r_i^2
\stackrel{?}{=} \frac{1}{2} M\omega^2 C^2 + \sum_{i=1}^{N-1} 
\frac{1}{2} \mu_i \omega^2 y_i^2.
\label{eq:guessjacob}
\ee
Replacing each $\yy_i$ by their expression in the guess gives
\be
M C^2 + \sum_{i=1}^{N-1} \mu_i y_i^2 = \sum_{j=1}^{N} \sum_{k=1}^{N}
Q_{jk} \rr_j \cdot \rr_k
\ee
where $Q$ is uniquely defined once it is imposed to be a
{\sl symmetric} matrix. Setting $M_i=\sum_{j=i+1}^{N} m_j$
for $0\leq i \leq N-1$, and $M_N=0$, we find for the off-diagonal
matrix elements
\be
Q_{jk} = 
-\frac{\mu_{\min(j,k)}m_{\max(j,k)}}{M_{\min(j,k)}} + \frac{m_j m_k}{M}
+m_j m_k \sum_{i=1}^{\min(j,k)-1} \frac{\mu_i}{M_i^2}
\ee
where $1\leq j,k\leq N$,
$\min(j,k)$ and $\max(j,k)$ respectively stand for the smallest
and for the largest of the two indices $j$ and $k$.
The key relation is then that
\be
\frac{\mu_i}{M_i^2} = \frac{1}{M_i} - \frac{1}{m_i+M_i} 
=\frac{1}{M_i} - \frac{1}{M_{i-1}}
\label{eq:trick1}
\ee
since $M_i + m_i = M_{i-1}$ for $1\leq i\leq N$. This allows to calculate
the sum over $i$ of $\mu_i/M_i^2$, as all except the border terms 
compensate by pairs. E.g. for $j < k$:
\be
\sum_{i=1}^{j-1} \frac{\mu_i}{M_i^2} 
= \frac{1}{M_{j-1}}- \frac{1}{M}
\label{eq:trick2}
\ee
since $M_0=M$. One then finds that the off-diagonal elements
of the matrix $Q$ vanish. The diagonal elements of $Q$ may
be calculated using the same tricks (\ref{eq:trick1},\ref{eq:trick2}),
one finds $Q_{ii}= m_i$ for $1\leq i\leq N$.
As a consequence, the guess was correct and the question mark
can be removed from Eq.(\ref{eq:guessjacob}).

The last step to obtain Eq.(\ref{eq:Hjacodreduit}) is to appropriately
rescale the usual Jacobi coordinates, setting
\be
\uu_i \equiv (\mu_i/\bar{m})^{1/2} \yy_i
\label{eq:ui_vs_yi}
\ee
where $\bar{m}$ is an arbitrarily chosen mass.
A useful identity is 
the expression
 for the square of the hyperradius, Eq.(\ref{eq:Rjacob}). Starting from the definition
[first identity in Eq.(\ref{eq:Rjacob})] we see that $R^2=\sum_{i=1}^{N-1} \frac{\mu_i}{\bar{m}} y_i ^2$.
Then the second identity in Eq.(\ref{eq:Rjacob}) results from the fact that the guess
in Eq.(\ref{eq:guessjacob}) is correct.

\section*{Appendix 4: Hydrodynamic equations}\label{app:hydro}
\addcontentsline{toc}{section}{Appendix 4}

The hydrodynamic equations for a normal compressible viscous fluid are~(see~\cite{SonBulkVisco}\footnote{There is a typo in Eq.(10) of~\cite{SonBulkVisco}: $\mathbf{\nabla}_i(\rho v^i \partial_i s)$ should be replaced by  $\mathbf{\nabla}_i(\rho v^i s)$.}, or \S15 and~\S49 in~\cite{LandauHydro}):
\begin{itemize}
\item
the continuity equation
\be
\frac{\partial \rho}{\partial t}+\mathbf{\nabla}\cdot(\rho\mathbf{v}) =0,
\label{eq:cte}
\ee
\item
the equation of motion
\bea
m \rho \left(\frac{\partial v_i}{\partial t}+\vv\cdot\gr v_i\right)&=&-\frac{\partial p}{\partial x_i}-\rho\frac{\partial U}{\partial x_i}+\sum_k\frac{\partial}{\partial x_k}\left[\eta\left(\frac{\partial v_i}{\partial x_k}+\frac{\partial v_k}{\partial x_i}-\frac{2}{3}\delta_{ik}\mathbf{\nabla}\cdot\mathbf{v}\right)\right]
\nonumber
\\ & &+\frac{\partial}{\partial x_i}\left(\zeta\, \mathbf{\nabla}\cdot\mathbf{v}\right)
\label{eq:motion}
\eea
where $m$ is the atomic mass,  $\eta$ is the shear viscosity, $\zeta$ is the bulk viscosity, and the pressure $p(\rr,t)$
[as well as the temperature $T(\rr,t)$ appearing in the next equation]
is as always expressible in terms of $\rho(\rr,t)$ and $s(\rr,t)$ {\it via} the equation of state,\footnote{If we would neglect the position-dependence of $\eta$ and $\zeta$, (\ref{eq:motion}) would reduce to the Navier-Stokes equation.}
\item
the entropy-production equation
\be
\rho T \left(\frac{\partial s}{\partial t}+\vv\cdot\mathbf{\nabla}s\right)
= \gr\cdot(\kappa\mathbf{\nabla}T)+\frac{\eta}{2}\sum_{i,k}\left(
\frac{\partial v_i}{\partial x_k}+\frac{\partial v_k}{\partial x_i}-\frac{2}{3}\delta_{ik}\mathbf{\nabla}\cdot\mathbf{v} \right)^2
+\zeta\|\mathbf{\nabla}\cdot\mathbf{v}\|^2
\label{eq:entr_prod}
\ee
where $\kappa$ is the thermal conductivity.
\end{itemize}

\section*{Appendix 5: Alternative derivation of the vanishing bulk viscosity}\label{app:hydro_alternatif}
\addcontentsline{toc}{section}{Appendix 5}

Consider the particular case of a unitary gas initially prepared at thermal
equilibrium in an isotropic harmonic trap at a temperature $T$ above
the critical temperature. When the harmonic trap becomes time dependent,
$U(\rr,t)=\frac{1}{2} m \omega^2(t) r^2$,
each many-body eigenstate of the statistical mixture evolves under
the combination Eq.(\ref{eq:ansatz}) of a time dependent gauge transform and a 
time dependent scaling transform of scaling factor $\lambda(t)$. 
The effect of the gauge transform is to shift the momentum
operator $\pp_i$ of each particle $i$ by the spatially slowly varying 
operator $m\rr_i \dot{\lambda}/\lambda$. In the hydrodynamic framework,
this is fully included by the velocity field Eq.(\ref{eq:v_echelle}).
\footnote{To formalize this statement, we consider
a small but still macroscopic element of the equilibrium
gas of volume $dV$ around
point $\bar{\rr}$, with $k_F^{-1} \ll dV^{1/3} \ll R$ where
$k_F$ is the Fermi momentum and $R$ the Thomas-Fermi radius of the gas.
We can define the density operator $\hat{\rho}_{\rm elem}$ of this element
by taking the trace of the full $N$-body density operator over
the spatial modes outside the element. Since the gauge transform
in Eq.(\ref{eq:ansatz}) is local in position space, $\hat{\rho}_{\rm elem}$
experiences the same unitary gauge transform. It would be tempting
to conclude from the general formula $dS = -k_B \mbox{Tr}[\hat{\rho}_{\rm elem} 
\ln \hat{\rho}_{\rm elem}]$ that the entropy $dS$ of the element 
is not changed by
the gauge transform. This is a valid conclusion however only if
the gauge transform does not bring $\hat{\rho}_{\rm elem}$ too far
from local thermal equilibrium.
To check this, we split the gauge transform for a single particle
of position $\rr$ as $m r^2 \dot{\lambda}/(2 \hbar \lambda) =
  m \dot{\lambda}/(2 \hbar \lambda) [\bar{r}^2 + 2 \bar{\rr}\cdot (\rr-
\bar{\rr}) + (\rr-\bar{\rr})^2].$ The first term is an innocuous
uniform phase shift. The second term performs a uniform shift 
in momentum space by the announced value $m\vv(\bar{\rr},t)$. 
Due to Galilean invariance, this has no effect on the thermodynamic
quantities of the small element, such as its temperature, its pressure,
its density, its entropy.
With the estimate $\dot{\lambda}/\lambda \sim \omega$, $\bar{r}\sim
R$, $m\omega R \sim \hbar k_F$, 
this second term is of order $k_F dV^{1/3}\gg 1$, not negligible.
The third term is of order $m\omega dV^{2/3}/\hbar 
\sim N^{-1/3} k_F^2 dV^{2/3}$, negligible in the thermodynamic limit.}
Using the macroscopic consequences of a spatial scaling 
Eqs.(\ref{eq:resca_macro_dens},\ref{eq:resca_macro_temp},\ref{eq:resca_macro_entropie_par_particule},\ref{eq:resca_macro_pression}),
one {\sl a priori} obtains a time dependent solution of the 
hydrodynamic equations:
\bea
T(\rr,t) &=& T(t=0)/\lambda^2(t) \\
\rho(\rr,t) &=& \rho(\rr/\lambda,0)/\lambda^3(t) \\
s(\rr,t) &=& s(\rr/\lambda,0) \\
p(\rr,t) &=& p(\rr/\lambda,0)/\lambda^5(t)  \\
v_i(\rr,t) &=& x_i \dot{\lambda}(t)/\lambda(t).
\eea
One then may {\sl a posteriori} check that Eq.(\ref{eq:cte}) is inconditionally satisfied, and that Eq.(\ref{eq:entr_prod}) is satisfied if $\zeta\equiv 0$. 
Setting $\zeta\equiv 0$ in Eq.(\ref{eq:motion}), and using the hydrostatic condition $\mathbf{\nabla}p=-\rho\mathbf{\nabla} U$
at time $t=0$, one finds that Eq.(\ref{eq:motion}) holds provided that $\lambda(t)$ solves 
Eq.(\ref{eq:russe}) as it should be.

\section*{Appendix 6: $n$-body resonances}
\addcontentsline{toc}{section}{Appendix 6}
\label{app:res_Ncorps}

Usually in quantum mechanics one takes the boundary condition that the wavefunction is bounded when two particles approach each other; in contrast, the Wigner-Bethe-Peierls boundary condition (\ref{eq:bpN}) expresses the existence of a $2$-body resonance.
If the interaction potential is fine-tuned not only to be close to a two-body resonance (i.e. to have $|a|\gg b$) but also to be close to a $n$-body resonance (meaning that a real or virtual $n$-body bound state consisting of $n_\up$ particles of spin $\up$ and $n_\down$ particles of spin $\down$ is close to threshold),
then one similarly expects that, in the zero-range limit, the interaction potential can be replaced by the
Wigner-Bethe-Peierls boundary condition,
{\it together with an additional boundary condition in the limit where any subset of $n_\up$ particles of spin $\up$ and $n_\down$ particles of spin $\down$ particles approach each other}.
Using the notations of Section~\ref{subsec:scaling_laws},
this additional boundary condition reads~\cite{PetrovBosons, WernerPRA,NishidaSonTan,theseFelix}:
\be
\psi(\rr_1,\ldots,\rr_N) = \left(R_J^{-s}-\frac{\epsilon}{l^{2s}} R_J^s\right)\,R_J^{-\frac{3 n-5}{2}}\,\phi(\Omegab_J)
\,A_J(\CC_j,\mathbf{\mathcal{R}}_J)
+o(R_J^\nu)
\label{eq:NbodyRes}
\ee
where $s=s_{\rm min}(n_\up,n_\down)$,
while $l>0$ and $\epsilon=\pm1$
are parameters of the model playing a role analogous to the absolute value and the sign of the two-body scattering length.
This approach is only possible if the wavefunction remains square integrable, 
 i.~e. if $0\le s<1$,
which we assume in what follows.
This condition is satisfied e.g. for $n_\up=2, n_\down=1$ for a mass ratio $m_\up/m_\down \in ]8.62\ldots;13.6\ldots]$~\cite{Efimov73}.
Moreover we are assuming for simplicity that $s\neq0$.

Let us now consider the particular case where the two-body scattering length is infinite, and the external potential is either harmonic isotropic, or absent.
Then the
separability in internal hyperspherical coordinates of Section~\ref{subsec:separability} still holds for $n=N$. Indeed, Eq.(\ref{eq:NbodyRes}) then translates into the boundary condition on the hyperradial wavefunction
\be
\exists A\in\mathbb{R}/\ F(R)\underset{R\to 0}{=} A \cdot \left( R^{-s} - \frac{\epsilon}{l^{2s}} R^s \right) + O\left(R^{s+2}\right)
\ee
and does not affect the hyperangular problem.
Consequently~\cite{theseFelix}, 
\begin{itemize}
\item
For the $n$-body bound state, which exists if $\epsilon=+1$:
\be
E=-\frac{2\hbar^2}{m\,l^2} \left[\frac{\Gamma(1+s)}{\Gamma(1-s)}\right]^{\frac{1}{s}},\ee
\be
F(R)=K_s\left(R \sqrt{-2 E \frac{m}{\hbar^2}}\right).
\ee
\item
For the eigenstates in a trap:
\be E {\rm\ solves:\ \ }
-\epsilon\cdot\left(\frac{\hbar}{m\omega\,l^2}\right)^s =
\frac{\Gamma\left(
\frac{1+s-E/(\hbar\omega)}{2}
\right) \Gamma(-s)}
{\Gamma\left(
\frac{1-s-E/(\hbar\omega)}{2}
\right) \Gamma(s)},
\ee
\be
F(R)=\frac{1}{R}\,W_{\frac{E}{2\hbar\omega},\frac{s}{2}}\left(R^2\frac{m\omega}{\hbar}\right).
\ee
\end{itemize}
In particular, for $l=\infty$, 
we are exactly at the $n$-body resonance, since
the energy of the $n$-body bound state vanishes. The spectrum in a trap then is $E=(-s+1+2 q)\hbar\omega$ with $q\in\mathbb{N}$.

Note that, most often, $s\geq1$, in which case one would have to use an approach similar to the one developped by Pricoupenko for the case of $2$-body resonances in non-zero angular momentum channels, and to introduce a modified scalar product~\cite{LudoPRL_ondeP,Ludo_onde_L}.

\input references.tex
\end{document}

%% file: abstract.txt
The physics of atomic quantum gases is currently taking advantage
of a powerful tool, the possibility to fully adjust the interaction
strength between atoms using a magnetically controlled Feshbach resonance.
For fermions with two internal states,
formally two opposite spin states $\uparrow$ and $\downarrow$,
this allows to prepare long lived
strongly interacting three-dimensional gases and 
to study the BEC-BCS crossover.
Of particular interest along the BEC-BCS crossover is the so-called
unitary gas, where the atomic interaction potential between
the opposite spin states has virtually an infinite scattering length
and a zero range. This unitary gas is the main subject of the
present chapter: It has fascinating symmetry properties, from a
simple scaling invariance, to a more subtle dynamical symmetry in an
isotropic harmonic trap, which is linked to a separability of the
$N$-body problem in hyperspherical coordinates.
Other analytical results, valid over the whole BEC-BCS crossover,
are presented, 
establishing a connection between three recently measured quantities, the tail of the momentum
distribution, the short range part of the pair distribution function
and the mean number of closed channel molecules.